\documentclass[english,twocolumn]{emulateapj}
\usepackage{amsmath}

\usepackage{color}

\shorttitle{}
\shortauthors{}

\begin{document}

\title{The Secular Dynamics of TNOs and Planet Nine Interactions}
\author{Gongjie Li \altaffilmark{1, 2}, Samuel Hadden\altaffilmark{1}, Matthew Payne \altaffilmark{1}, Matthew J. Holman \altaffilmark{1}}
\affil{$^1$ Harvard-Smithsonian Center for Astrophysics, The Institute for Theory and
Computation, \\60 Garden Street, Cambridge, MA 02138, USA}
\affil{$^2$ Center for Relativistic Astrophysics, School of Physics, Georgia Institute of Technology, Atlanta, GA 30332, USA }
\email{gongjie.li@physics.gatech.edu}

\begin{abstract}
The existence of Planet Nine has been suggested to explain the pericenter clustering of extreme trans-Neptunian objects (TNOs). However, the underlying dynamics involving Planet Nine, test particles and Neptune is rich, and it remains unclear which dynamical processes lead to the alignment and how they depends on the properties of Planet Nine. Here, we investigate the secular interactions between an eccentric outer perturber and TNOs starting in a near-coplanar configuration. We find that a large number of TNOs could survive outside of mean motion resonances at 4Gyr, which differs from previous results obtained in the exact coplanar case with Neptune being treated as a quadrupole potential. In addition, secular dynamics leads to the orbital clustering seen in N-body simulations. %
We find that a near coplanar Planet Nine can flip TNO orbital planes, and when this happens, the geometrical longitudes of pericenter of the TNOs librate around $180^\circ$ during the flip. 
Orbital precession caused by the inner giant planets can suppress the flips while keeping the longitude of pericenter librating when $30\lesssim r_p \lesssim 80$ AU $\&$ $a\gtrsim 250$ AU. This results in the alignment of the pericenter of the low inclination TNOs ($i \lesssim 40^\circ$). We find the anti-aligned population and the flipped orbits could be produced by an eccentric ($e_9 \gtrsim 0.4$) outer planet of $\sim 10M_{\oplus}$ in a wide $a_9 \gtrsim 400 \sim 800$ AU orbit. Future surveys on the high inclination TNOs will help further constrain the properties of possible outer planets.
\end{abstract}
\bigskip

\section{Introduction}
Recent observations of the vast expanse of the outer Solar System have revealed around a dozen distant ($a\gtrsim 150$ AU) trans-Neptunian objects (TNOs) in our Solar System with pericenter distances outside the orbit of Neptune \citep{Gladman02, Brown04, Chen13, Trujillo14}. 
The orbits of such objects exhibit interesting architectures. 
For instance, there seems to be a clustering in the orbital orientation of the TNOs \citep[e.g.,][]{Trujillo14}. 

Many studies have shown that the alignment of the orbits is not due to selection biases \citep{delaFuenteMarcos14, Gomes15, Sheppard16, Brown16, Brown17}, although \citet{Shankman17} demonstrate that the ``Outer  Solar  System  Origins  Survey''  (OSSOS: \citealt{Bannister16}) contains non-intuitive biases for the detection of TNOs that lead to apparent clustering of orbital angles in their data, and the angular elements of the distant TNOs are consistent with uniform distribution \citep{Bannister18}. Recent observations have suggested additional clustering features of the TNOs \citep{Sheppard16, Brown17}. 
Ongoing observational TNO surveys will provide a better understanding of the architecture of the outer Solar System and the details (if any) of the TNO clustering.

It has been suggested that the clustering of the TNO orbits can be explained by an undetected outer planet, ``Planet Nine,'' in our own Solar System \citep{Trujillo14, Batygin16}. 
The location of the putative Planet Nine has been constrained using dynamical simulations of TNOs orbiting under the gravitational influence of Planet Nine \citep{Brown16, delaFuenteMarcos16}, and by simulations of the tidal perturbation induced by Planet Nine on the relative distance between the Earth and the Cassini spacecraft \citep{Holman16} and Pluto and other TNOs \citep{Holman16b}. 
In addition, the formation mechanism for Planet Nine has been investigated, including scenarios for capturing Planet Nine from another star,  scattered giant planets originating within the Solar System \citep{Li16, Mustill16, Bromley16, Parker17}, pebble accretion in a large ($250-750$AU) ring of solids \citep{Kenyon16}, as well as circularization of Planet Nine with an extended cold planetesimal disk \citep{Eriksson18}.

The dynamics involved in the interactions between TNOs and the putative Planet Nine are rich, and the mechanism by which the clustering of TNO orbits arises due to interactions with Planet Nine and the four known giant planets is not well characterized. \citet{Batygin16} suggested that the origin of the clustering is produced by mean motion resonances. This is supported by \citet{Malhotra16}, who noted that some of the TNOs are likely to be in mean motion resonances with an exterior planet. Indeed, dynamical simulations of the detected TNOs with hypothetical orbits of Planet Nine show that the TNOs can move between different mean motion resonances with Planet Nine \citep{Millholland17, Becker17, Hadden18}.

On the other hand, secular interactions can also produce similar orbital alignment in the longitude of pericenter of the TNOs. This has been investigated in detail in the coplanar configuration, where the TNOs and Planet Nine all lie in the same plane. In particular, \citet{Beust16} investigated the secular interactions of the test particles with Planet Nine in the coplanar case, and found the alignment can be produced by secular effects. Recently, \citet{BatyginMorbi17} found from a detailed study of the coplanar case, that the effect of secular dynamics embedded in mean motion resonances is to regulate the clustering of the TNOs. In addition, \citet{Hadden18} noted that TNOs with pericenters initially aligned with that of Planet Nine follow secular trajectories, but are more likely to be ejected due to overlaps of mean motion resonances, if their pericenter distances become small. Thus, the clustering of the TNO orbits is likely due to a combined effect of MMR and secular interactions. 


It is likely that Planet Nine is not in an exact coplanar configuration with TNOs: with small inclinations, many TNOs could survive outside MMRs with Planet Nine. Thus, in this article, we focus on the secular interactions between TNOs and Planet Nine in a near-coplanar configuration. We use the orbital phase averaged Hamiltonian and generalize the dynamical analysis to higher dimensions, which allows the inclination of the TNOs to vary, and we consider the clustering of the orbits due to secular effects. Extending beyond the coplanar configurations, \citet{Saillenfest17} recently identified secular resonances of the TNO orbits using surface of sections. Here, we focus on the near-coplanar configuration and study the how the secular interactions lead to the clustering of the TNOs. The remainder of this article is organized as follows. 

In section \textsection \ref{s:sec} we analyze the secular effects following the orbital phase averaged Hamiltonian, then in section \textsection \ref{s:Nbd} perform full N-body simulations and to compare with the secular results from section \textsection \ref{s:sec}. 
In section \textsection \ref{s:incl}, we discuss the inclination distribution of the TNOs and the origin of the aligned orbits. 
Finally, we present our conclusions in section \textsection \ref{s:dis}.

\section{Secular interactions between TNOs and Planet Nine}
\label{s:sec}
For a large range of orbital parameters, TNO orbits can cross that of Planet Nine. Although orbit-crossing (outside of commensurability) will eventually lead to close encounters between Planet Nine and TNOs, which may result in the ejection of TNOs, \citealt{Gronchi99} and \citealt{Gronchi02} have shown that the averaging principle is a powerful tool to study the secular evolution for a long time span, with applications to asteroids whose orbits are planet-crossing. The secular interactions also play a key role in the dynamics of TNOs. In particular, it has been shown that in a coplanar configuration, secular effects guide the overall evolution of the TNOs, and cause the libration of the longitude of pericenter about an anti-aligned configuration with Planet Nine ($\Delta \varpi \sim 180^\circ$)\citep{Beust16, BatyginMorbi17}. In particular, \citet{Hadden18} showed that any clustering that initially exists near $\Delta \varpi \sim 0^\circ$ will be removed due to instabilities caused by overlaps of mean motion resonances, and that secular interactions shape the clustering near $\Delta \varpi \sim 180^\circ$. 

It is likely that TNOs are misaligned in inclination with that of the Planet Nine, and N-body simulations have demonstrated interesting clustering of the TNOs which start near coplanar with that of Planet Nine\citep{BatyginMorbi17, Khain18}. It is not clear how the secular dynamics contribute to the overall clustering of the TNOs in the misaligned configurations. Here, we focus on the pure secular analysis and extend the previous secular study to higher dimensions, allowing the inclination of the TNOs to vary, and characterize the secular effects in the observed clustering of the TNO orbital orientations.

\subsection{Hamiltonian Framework}
\label{s:HF}
\begin{figure}[ht]
\includegraphics[width=4in]{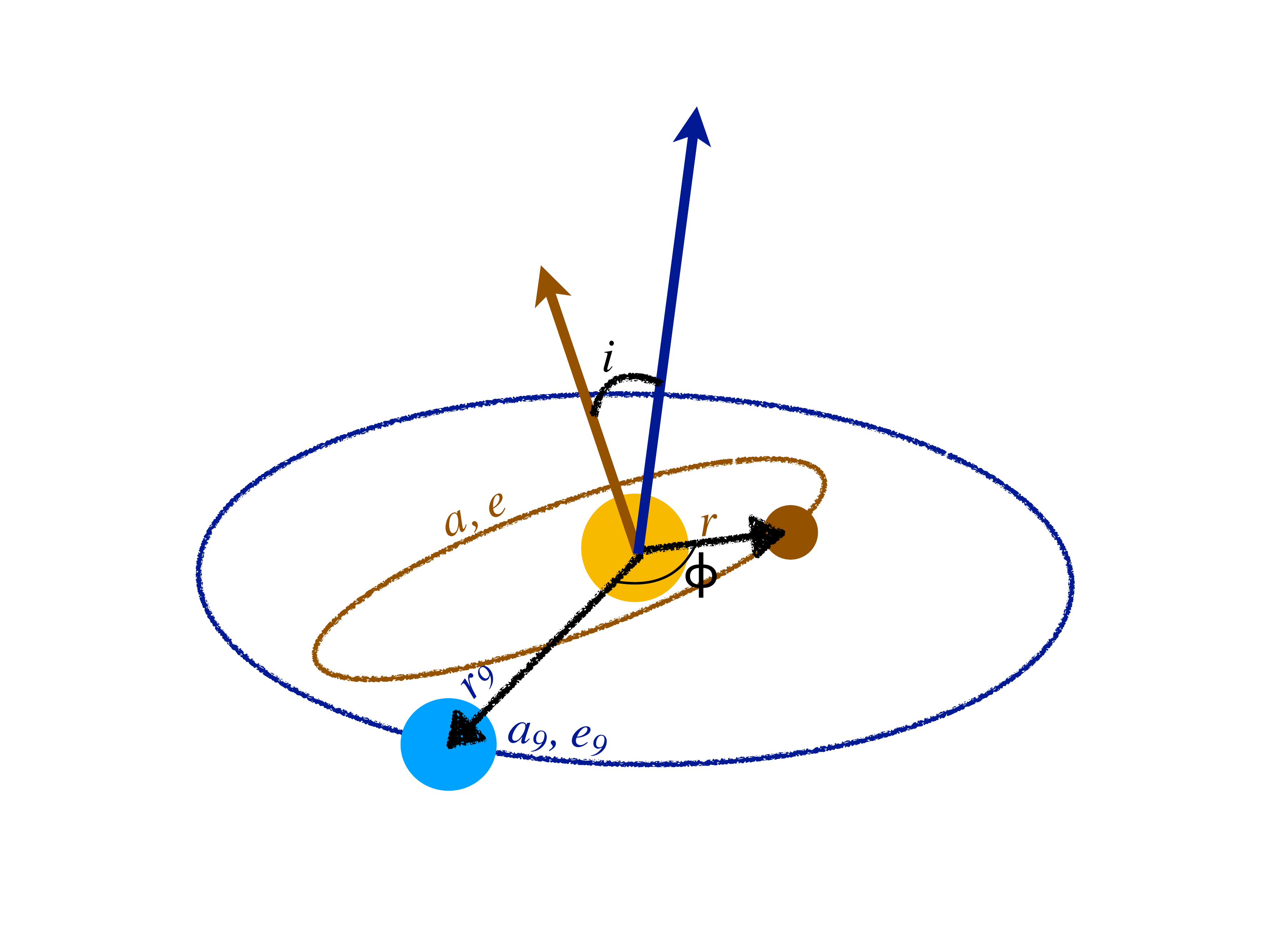}
\caption{The configuration of the system. The brown circle represents a TNO, and the blue circle represents the Planet Nine. The black arrows represent the position vectors of the TNO and the Planet Nine. On the other hand, the blue and brown arrows represent the angular momentum direction of the orbits of Planet Nine and TNO separately. 
\label{f:config}}
\vspace{0.1cm}
\end{figure}

We consider an outer planet (Planet Nine) and a TNO orbiting the Sun, illustrating the configuration of the system in Figure \ref{f:config}. 
The mass of the TNO is much smaller than Planet Nine and our Sun, and thus the TNO can be treated as a test particle. We allow the orbit of Planet Nine to be eccentric and to be misaligned with the inner orbit, different from the circular restricted case. 

In the non-hierarchical configuration, when $a/a_9 \gtrsim 0.1$, the usual expansion in the semi-major axes ratio is not a good approximation. However, in the case when the outer perturber is much less massive than that of the center body, the perturbation from the outer companion is not strong enough to destabilize the system, i.e., the semi-major axes of the orbits are almost constant. In this case, one can consider the long-term secular evolution of the system by the averaging out the fast varying orbital phase of the inner and the outer orbits.  

Specifically, the secular (averaged) Hamiltonian of the interaction energy can be expressed as the following:
\begin{equation}
\label{eq:En}
H_{sec, 0}=-\frac{G m_9}{4\pi^2}\int^{2\pi}_{0}\int^{2\pi}_{0}\Big(\frac{1}{|\vec{r}-\vec{r_9}|}-\frac{\vec{r}\cdot \vec{r_9}}{r_9^3}\Big)~dl~dl_9 \,
\end{equation}
where $r$ and $r_9$ is the distance to the test particle and the perturber from the central object, $m_9$ is the mass of the perturber as illustrated in Figure \ref{f:config}. $l$ and $l_9$ are the mean anomaly of the test particle and the perturber respectively. $\vec{r}-\vec{r_9}$ can be expressed as the following: 
\begin{equation}
|\vec{r}-\vec{r}_9| = \sqrt{r^2 +r_9^2-2 r r_9 \cos{\phi}}  ,
\end{equation}
and $r$, $r_9$ and $\phi$ can be expressed by orbital elements. 

The evolution of the TNO's orbit can be well described using the averaged Hamiltonian, which converges in most configurations even for crossing orbits, as discussed in \citet{Gronchi99, Gronchi02}. For illustration, we compare the secular evolution with N-body simulations in Figure \ref{f:secNbd} in Appendix \ref{APP:SEC1}.

TNOs in the outer Solar System undergo perturbations from Planet Nine as well as the known inner giant planets. The TNOs are quite far away from the inner giant planets, and thus, the secular effects of the inner giant planets on the TNOs can be well approximated based on the Hamiltonian to the second order in the ratio of the TNO to giant planet semi-major axes. Combining the secular Hamiltonian for the interaction energy of the TNO with a perturber (equation \ref{eq:En}), the secular Hamiltonian can be expressed as the following \citep[e.g.,][]{Kaula64, Murray99}: 
\begin{align}
\label{eq:sec1}
H_{sec, 1} = H_{sec, 0} - \frac{1}{8}\frac{GM}{a}\frac{(3\cos^2{i} - 1)}{(1-e^2)^{3/2}}\sum_{i=1}^4 \Big(\frac{m_{pi} a_{pi}^2}{Ma^2} \Big) ,
\end{align}
where $M$ is the mass of the Sun, $a$, $e$ and $i$ are the semi-major axis, eccentricity and inclination of the TNO, and $m_{pi}$ and $a_{pi}$ are the masses and semi-major axes of the inner giant planets. 
Here,we assume that the inner planets are co-planar with the orbit of planet Nine.

The orbit of Planet Nine is also perturbed by the inner giant planets, which causes precession. The precession rate of Planet Nine's orbit in the low inclination regime can be expressed as the following:
\begin{align}
\label{eqn:prec2}
\frac{d\omega_9}{dt} &= \frac{3}{2}\frac{\sqrt{GM}}{a_9^{3/2} (1 - e_9^2)^2 } \sum_{i=1}^4 \Big(\frac{m_{pi} a_{pi}^2}{Ma_9^2} \Big)\\
\frac{d\Omega_9}{dt} &= -\frac{3}{4} \frac{\sqrt{GM}}{a_9^{3/2} (1 - e_9^2)^2 } \sum_{i=1}^4 \Big(\frac{m_{pi} a_{pi}^2}{Ma_9^2} \Big) ,
\end{align}
where $m_9$, $a_9$, $e_9$, $\omega_9$ and $\Omega_9$ are the mass, semi-major axis, eccentricity, argument of pericenter and longitude of ascending node of Planet Nine separately.

The Hamiltonian of the TNOs (eqn \ref{eq:sec1}) is implicitly expressed in terms of the canonical variables of the TNOs, 
\begin{align}
q &= (M, \omega, \Omega) \\
p &= ( \sqrt{GMa}, \sqrt{GMa(1-e^2)}, \sqrt{GMa(1-e^2)}\cos{(i)} )  \nonumber ,
\end{align}  
where $M$ is the mean anomaly of the TNO, $J=\sqrt{GMa(1-e^2)}$ and $J_z=\sqrt{GMa(1-e^2)}\cos{(i)}$ are the angular momentum and z-component of angular momentum of the TNO, and $L=\sqrt{GMa}$ is a constant in the secular regime. 

To obtain the alignment of the orbit of the TNO relative to that of Planet Nine, we transfer the coordinates of the TNOs to the difference in the argument of pericenter and the longitude of ascending node between the TNO and Planet Nine. Using the  type-III generating function, $G_{3} = -(\Delta\omega+\omega_9)J-(\Delta\Omega+\Omega_9)J_z$, we transform to new canonical angles $\Delta\omega=\omega - \omega_9$ and $\Delta\Omega=\Omega-\Omega_9$.
Then, the new Hamiltonian becomes 
\begin{align}
\label{eq:Hsectot}
H_{sec} &= H_{sec, 1} - \dot{\omega}_9J -\dot{\Omega}_9J_z~. 
\end{align} 

\begin{figure}[ht]
\begin{center}
\includegraphics[width=\columnwidth]{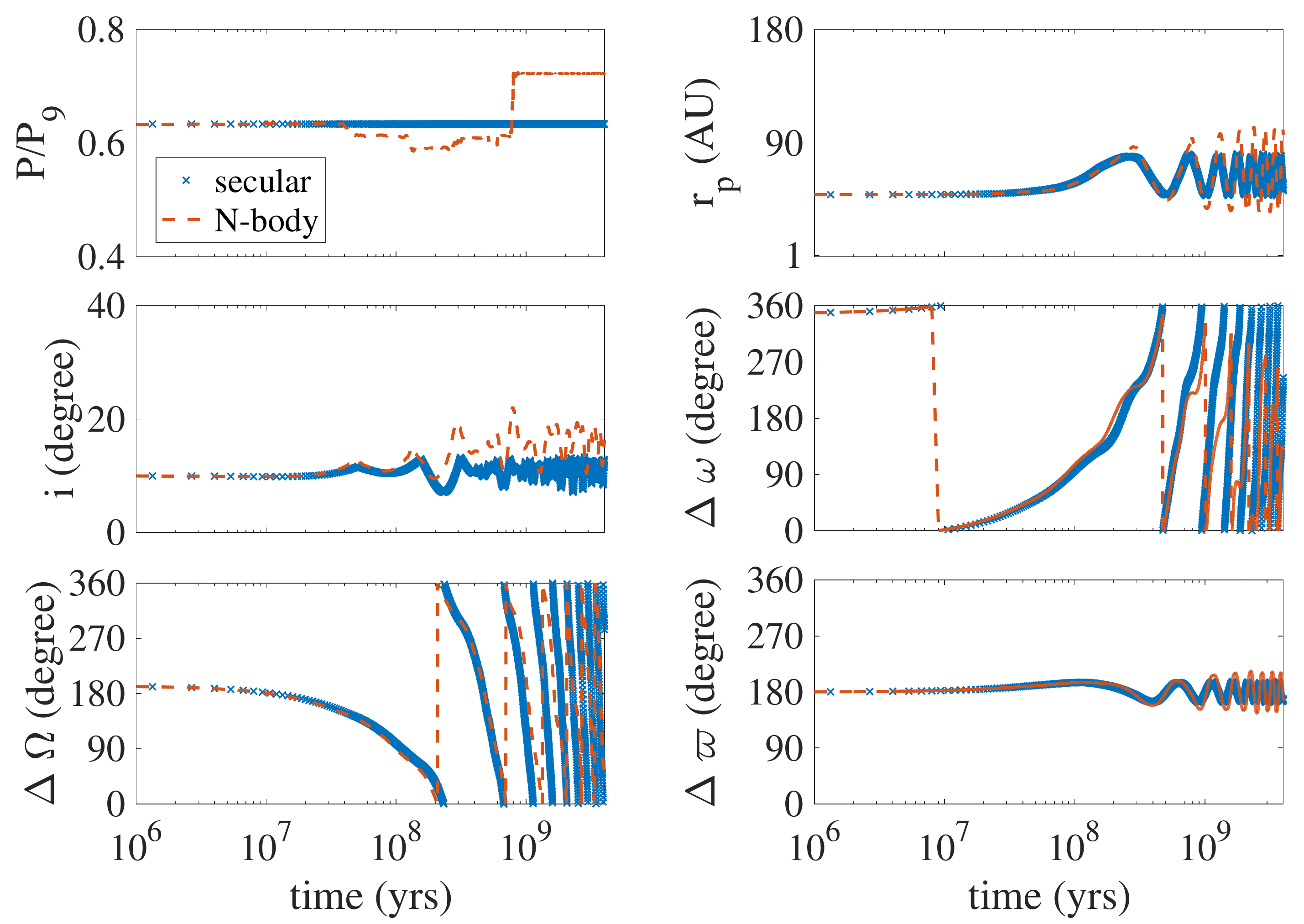} 
\caption{Orbital evolution of a TNO that starts with $P/P_9 = 0.633$ and ends with $P/P_9 = 0.722$. The blue crosses represent the secular results following the Hamiltonian (eqn \ref{eq:Hsectot}), and the dashed red lines represent the N-body results, with the inner giant planets substituted by a $J_2-$term as described in the text. The secular results are consistent with the N-body results until a close encounter at $\sim 50$ Myr with Planet Nine, and still reproduce the main important dynamical features.
\label{f:NbdSec}}
\end{center}
\end{figure}

The secular result following eqn (\ref{eq:Hsectot}) is a good approximation until close encounters occur between the TNO and Planet Nine or the inner giant planets.  For illustration, Figure \ref{f:NbdSec} compares the secular and N-body results, which shows the evolution of a TNO starting with $a = 368.75$ AU, $e = 0.867$,  and $\Delta\varpi=180^\circ$, and $i = 10^\circ$ relative to the ecliptic, and with Planet Nine and the inner four giant planets all co-planar in the ecliptic plane. The semi-major axis and eccentricity of Planet Nine is $a=500$ AU, and $e=0.6$. The blue crosses are the secular results and the red dashed lines are the N-body results. To isolate the effects of close encounters with Planet Nine, we substitute the inner giant planets (Jupiter, Saturn, Uranus and Neptune) by an equivalent $J_2$ term in the N-body simulation. We use the package Mercury for the N-body simulation, with the ``hybrid'' Wisdom-Holman/Bulirsch-Stoer integrator \citep{Wisdom92, Press92, Chambers99}, and a time-step of $dt = 3000$ days, which is roughly $5\%$ of Neptune's orbital period.   

Figure \ref{f:NbdSec} shows that the secular result is a good approximation up to $\sim 50$ Myr, at which point the TNO has a close encounter with Planet Nine, causing changes in the semi-major axis of the TNO. However, the secular effects are not immediately suppressed after the close encounter. In particular, the libration of $\Delta\varpi$ around $180^\circ$ shown in the secular results can still be observed in the N-body results after the close encounter until the end of simulation. Neither $\Delta\omega$ nor $\Delta\Omega$ librates for both the secular and N-body results. The secular results agree qualitatively with the N-body results after the change in the semi-major axis. This is due to the weak dependence of the secular interactions on the semi-major axis for TNOs with large semi-major axis ($\gtrsim 300$ AU), as shown in Figure \ref{f:Sec_ethetae} and in \citet{Hadden18}.

Note that the precession of the orbits increases the chance of close encounters between TNO and Planet Nine, which causes the secular approximation to deviate from the N-body results. Thus, the secular integration is a better approximation for the three-body interactions without $J_2-$precession, as illustrated in Appendix \ref{APP:SEC1}. For instance, the $J_2-$precession timescale due to the inner giants in $\Omega$ for Planet Nine is $13$ Gyr, and for the TNO is $276$ Myr.

\subsection{Secular Clustering of anti-aligned TNOs}
\label{s:secw180}
\subsubsection{Alignment in pericenter orientation}
It has previously been found (e.g., \citealt{Hadden18}, and our section \ref{s:Nbd}) that particles starting with anti-aligned pericenters are more likely to survive and play an important role in sculpting the overall orbital architecture of the TNOs. 
Thus, we start the analysis by focusing on the secular evolution of an (initially) anti-aligned population. 
We consider a set of $500$ test particles initialized with $150<a<550$ AU, $30<r_p<50$ AU, $i = 10^\circ$, $\varpi = 180^\circ$, and $\omega$ uniformly distributed between $0-360^\circ$. As in the illustrative example in Fig. \ref{f:NbdSec}, Planet Nine lies in the ecliptic plane together with the inner four giant planets, and we substitute the inner giant planets by the equivalent $J_2$ potential. The semi-major axis and eccentricity of Planet Nine is set to be $a_9=500$ AU, and $e_9=0.6$, and the longitude of pericenter of Planet Nine is $\varpi_9 = 0^\circ$. Thus, the pericenters of the test particles are initially anti-aligned with that of Planet Nine. We follow the particles' secular evolution by integrating equations of motion generated from the secular Hamiltonian, Equation \eqref{eq:Hsectot}. Details of our numerical method are given in Appendix \ref{APP:SEC1}.

\begin{figure*}[ht]
\begin{center}
\includegraphics[width=\textwidth]{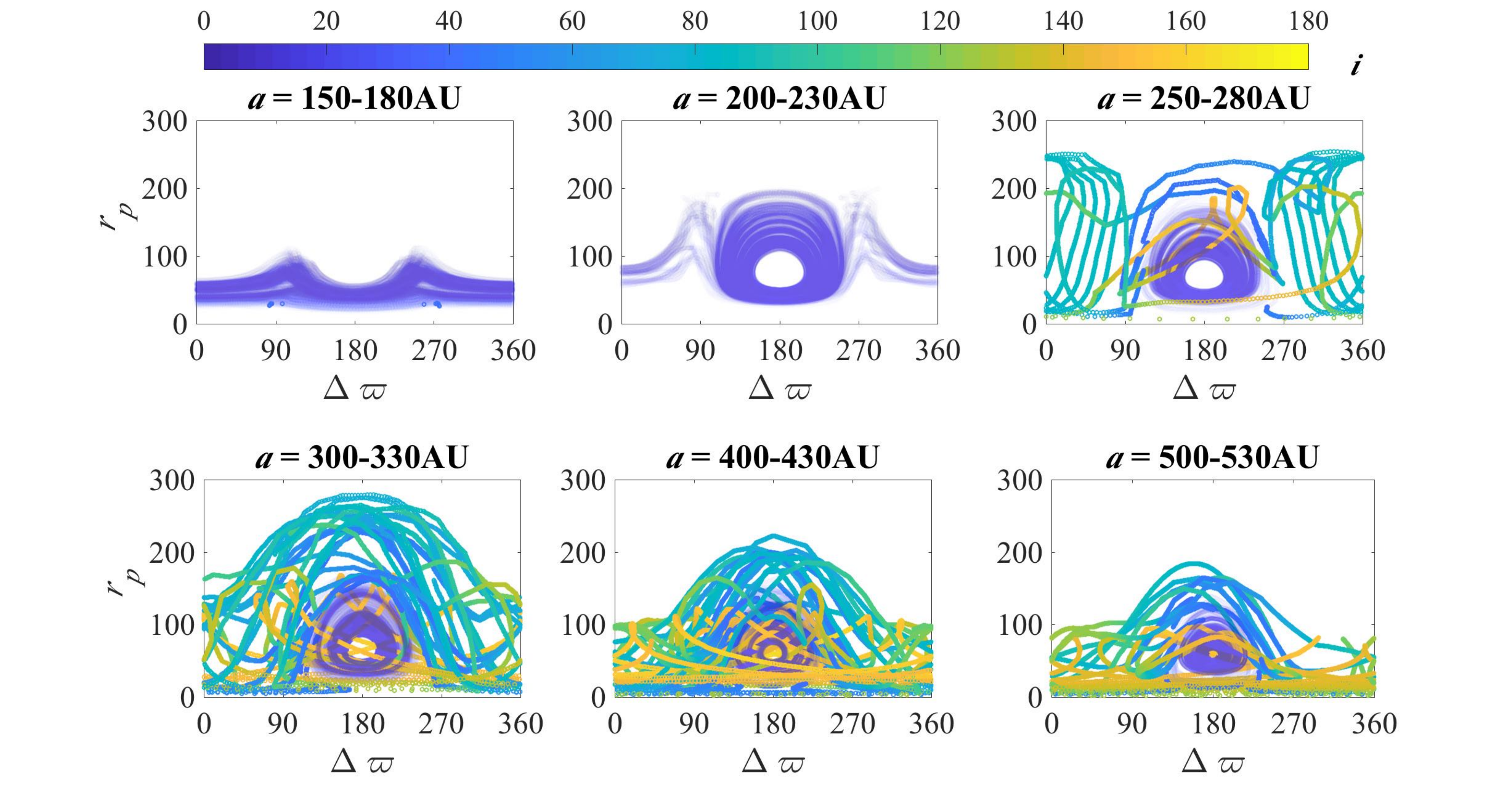} 
\caption{Secular evolution of TNOs in the plane of longitude of pericenter vs. pericenter distance for different semi-major axes, orbiting under the influence of a Planet Nine with $a=500$ AU, $e=0.6$ and $i=0^{\circ}$. 
The color represents the inclination of the TNOs. The part of the trajectories with low inclination ($<40^\circ$) are shown with a transparency parameter of $\alpha = 0.01$ to highlight the high inclination evolution.
Libration of $\Delta\varpi$ can be seen clearly for $a \gtrsim 200$ AU when the TNO inclination is low ($i \lesssim 40^\circ$), similar to the exact coplanar secular interactions \citep{Beust16,BatyginMorbi17}. When $a$ increases, the inclination of the TNOs can be excited. $\varpi$ no longer librates when the inclination becomes high. In addition, the libration region in $r_p$ shrinks when $a$ increases.  
\label{f:Sec_epo}
}
\end{center}
\end{figure*}

\begin{figure*}[ht]
\begin{center}
\includegraphics[width=\textwidth]{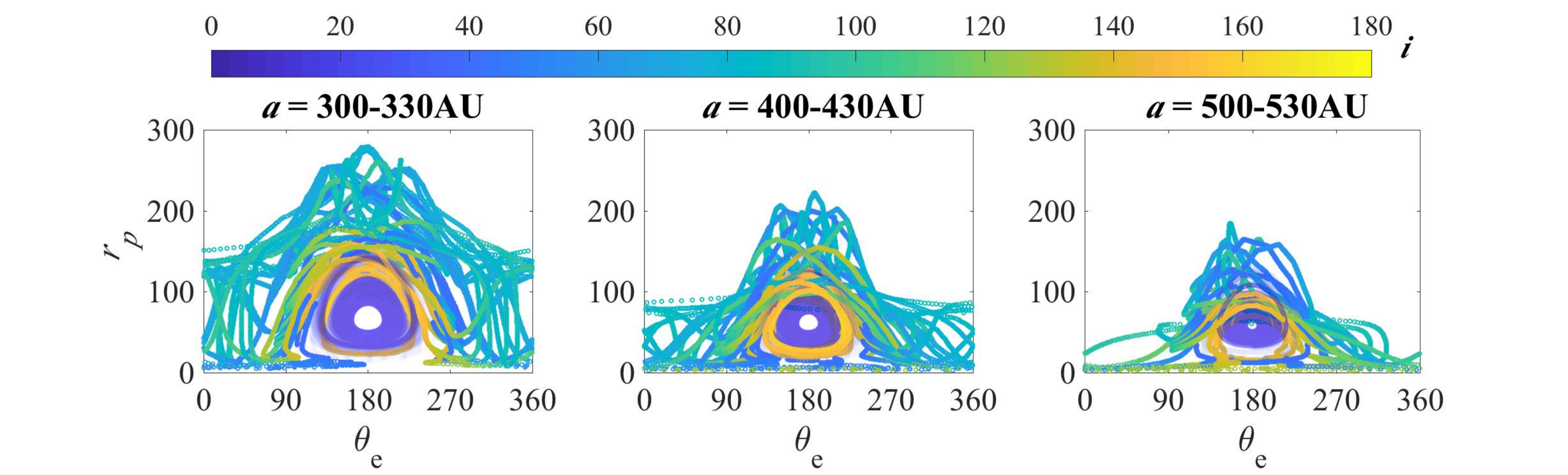} 
\caption{Similar to Figure \ref{f:Sec_epo}, except in the plane of geometrical longitude of pericenter ($\Delta\theta_e$) vs. pericenter distance for different semi-major axes. We only plot the TNOs with larger semi-major axes, which more likely allow high inclination excitations of TNOs (lower panels in Figure \ref{f:Sec_epo}). When inclination is low, $\theta_e \sim \varpi$ and the dynamical evolution in $\Delta \theta_e$ vs $r_p$ looks similar to that in $\Delta \varpi$ vs $r_p$. Different from $\Delta \varpi$, libration of $\Delta\theta_e$ can be seen also when the TNO inclination is high $i\gtrsim 140^\circ$.
\label{f:Sec_ethetae}
}
\end{center}
\end{figure*}

Figure \ref{f:Sec_epo} shows the secular evolution of the TNOs. Dividing the particles into semi-major axis regions, we notice that $\Delta\varpi$ circulates when TNO semi-major axes are low: $a\lesssim 200$ AU (upper left panel), where we illustrate the trajectories of a few representative particles. This is the region where the $J_2-$precession dominates. For instance, the precession caused by the $J_2$ term is roughly $\dot{\varpi} = 4200^\circ/$Gyr, calculated following equation \ref{eqn:prec2}. On the other hand, the change rate of $\varpi$ caused by Planet Nine is $\dot{\varpi} = -360^\circ/$Gyr, following the averaged Hamiltonian eqn (\ref{eq:En}) without the $J_2$ term in a coplanar configuration. Therefore, the $J_2$-precession leads to the circulation in $\varpi$ in the positive direction. 


When $a$ increases, the dynamical influence due to the inner giant planets becomes weaker and the perturbation due to the outer planet becomes stronger. Thus the $J_2$ term no longer dominates the evolution. As shown in the upper middle and left panel of Figure \ref{f:Sec_epo}, when $a$ increases to $\sim 210-300$ AU, $\Delta\varpi$ stops circulating and starts to librate around $180^\circ$. In the first row where $a<250$ AU, the inclination of the TNOs stay low for the entire 4 Gyr simulation. 

As $a$ further increases ($a \gtrsim 250$ AU), the initial eccentricities approach closer to unity, because we chose a fixed range of initial pericenter distances ($30<r_p<50$ AU). 
This is shown in  the lower panels in Figure \ref{f:Sec_epo}: as the eccentricity increases, the $J_2$ precession increases, while the precession due to Planet Nine decreases, leading to the circulation in $\Delta\varpi$ at high eccentricity.

Thus far, the secular dynamics of anti-aligned particles we have described are qualitatively the same as those seen in the strictly coplanar case previously studied by \citet{Beust16,BatyginMorbi17,Hadden18}. In particular, an island of librating trajectories appears at a critical semi-major axis where the apsidal precession induced by Planet Nine is able to balance that from the solar system giants.

Figure \ref{f:Sec_epo} shows that Planet Nine can also excite the inclinations of some TNOs; a dynamical effect that, by construction, is absent from previous coplanar studies. When TNOs reach very high eccentricity orbits, Planet Nine is able to excite extreme inclinations, often leading to orbital flips. We will study the dynamics of these orbital flips in more detail in Section~\ref{s:incl}. 

During the flips, the pericenter orientation of the TNO in the ecliptic plane ($\theta_e$, the geometrical longitude of pericenter\footnote{this was refered to as the ``pericenter longitude'' by \citet{Brown16}.}) librates (as illustrated in Section~\ref{s:incl}). This is similar to the coplanar flips shown in the hierarchical octupole limit \citep{Li14a}. In contrast to the nearly co-planar cluster, for high inclination orbits ($i>140^\circ$), $\Delta\varpi$ circulates, as can be seen in the lower panels of Figure \ref{f:Sec_epo}. This differs from Figure 10 in \citet{BatyginMorbi17}, where $\Delta\varpi$ remains confined as the inclination becomes larger. This is because the objects selected in \citet{BatyginMorbi17} only stay briefly in the high inclination region, before circulation is completed. We can see both types of objects in our secular and N-body simulations (e.g., some trajectories in Figure \ref{f:Sec_epo} and Figure \ref{f:highi_traj}), some of them stay only briefly in the retrograde stage and some of them stay much longer. The circulation of $\Delta\varpi$ when the TNO stays at high inclinations can also be seen in Figure \ref{f:secincl_traj3}. During the circulation of $\Delta\varpi$, the eccentricity of the TNO becomes high and the pericenter distance is reduced to the point where TNOs may be scattered and ejected by the inner giant planets, an effect that is not captured by our secular calculations. However, we do find instances in our our N-body simulations  where TNOs survive such phases of high eccentricity and inclination during which $\Delta \varpi$ rapidly circulates.  Figure \ref{f:highi_traj} illustrates two such examples, near $t \sim 1.5$ Gyr for the TNO in red and $t\sim3.5$ Gyr for the TNO in yellow.

In addition, we note that in the top-right panel of Figure \ref{f:Sec_epo}, where $250<a<280$ AU, there are two teal color objects displaying an interesting evolution, where their eccentricities vary with large amplitudes. 
There are only two teal color objects in this regime, one of them has $\varpi$ centered-in, and oscillating around, zero degrees.

To illustrate the libration of the geometrical longitude of pericenter in the ecliptic plane for some of the flipped orbits, we show in Figure \ref{f:Sec_ethetae} the evolution of TNO pericenter distances versus $\Delta\theta_e$ in the same semi-major axes ranges as those in the lower panels of Figure \ref{f:Sec_epo}, where the inclination of the TNOs can be excited. At high inclinations (e.g., $i\gtrsim 150^\circ$), $\Delta\theta_e$ librates, while $\Delta\varpi$ circulates. 
We note that $\theta_e \sim \varpi$ when inclinations are low ($i\sim 0^\circ$), and $\theta_e \sim 2\Omega-\varpi$ when TNOs counter-orbit w.r.t. Planet Nine ($i\sim 180^\circ$). The libration of $\theta_e$ is consistent with the libration of $2\Omega-\varpi$ noted by \citep{BatyginMorbi17} for high inclination orbits. 


\begin{figure*}[ht]
\begin{center}
\includegraphics[width=\textwidth]{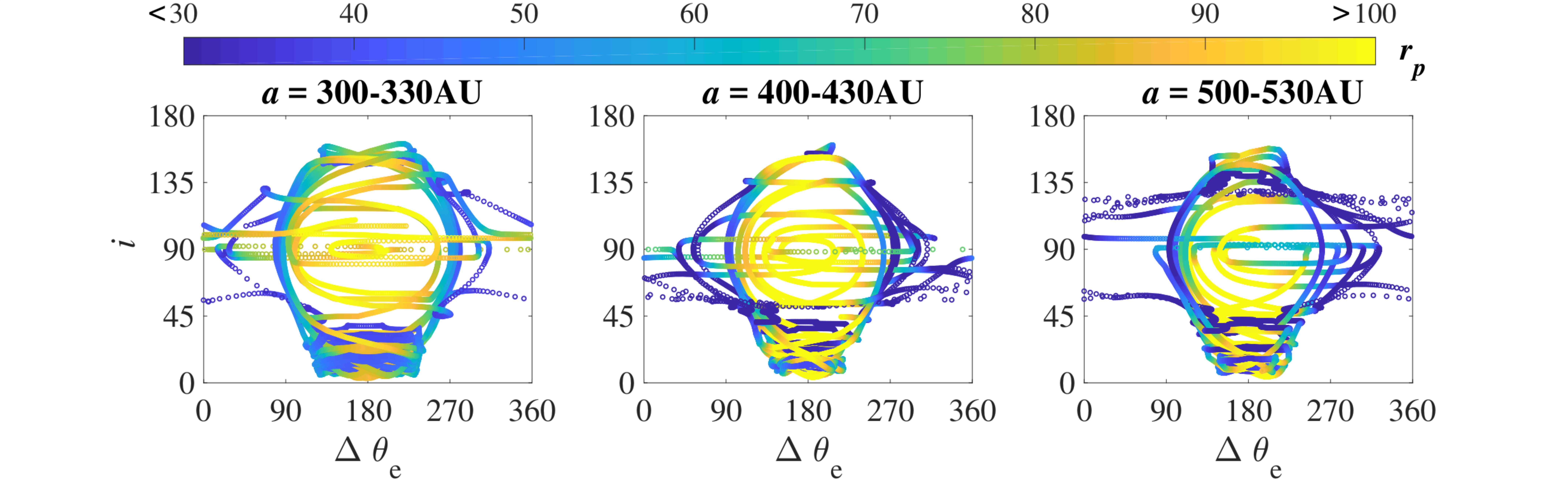} 
\caption{Secular evolution of TNOs for selected particles in the plane of geometrical longitude of pericenter ($\Delta\theta_e$) vs. inclination in different semi-major axes panels. The color represents the pericenter distance of the TNOs. For the high inclination objects ($40^\circ \lesssim i \lesssim 140^\circ$) with low pericenter distances ($r_p \lesssim 80$ AU), their pericenter orientation ($\Delta\theta_e$) is clustered around $\sim 90^\circ$ and $\sim 270^\circ$ in the low pericenter detection limit. When the inclination is $\lesssim 40^\circ$ or above $\sim 140^\circ$, their pericenter orientation clusters around $\sim 180^\circ$. As $a$ increases, the pericenter distances of the high inclination TNOs that clusters around $\theta_e\sim 180^\circ$ decrease.
\label{f:Sec_ipo}
}
\end{center}
\end{figure*}

105 out of the 500 particles have their inclination excited to retrograde configurations during the $4$ Gyr simulation. It is more likely for the inclination to flip if the particles start with large semi-major axes and small pericenter distances, but the detailed dependencies are more complicated as we demonstrate in section \ref{s:incl}.

To illustrate the clustering of high-inclination TNO orbits, Figure \ref{f:Sec_ipo} presents trajectories in the plane of $(\Delta\theta_e, i)$, color coded in the pericenter distance. 
There is a moderate clustering of trajectories around $\Delta\theta_e \sim 90^\circ$ and $\sim 270^\circ$ when the inclinations are high, $i \gtrsim 60-120^\circ$ and the pericenter distances are low $\lesssim80$ AU. 
Around $i \gtrsim 140^\circ$, $\Delta\theta_e$ is clustered around $150-210^\circ$. 

\begin{figure}[ht]
\begin{center}
\includegraphics[width=0.9\columnwidth]{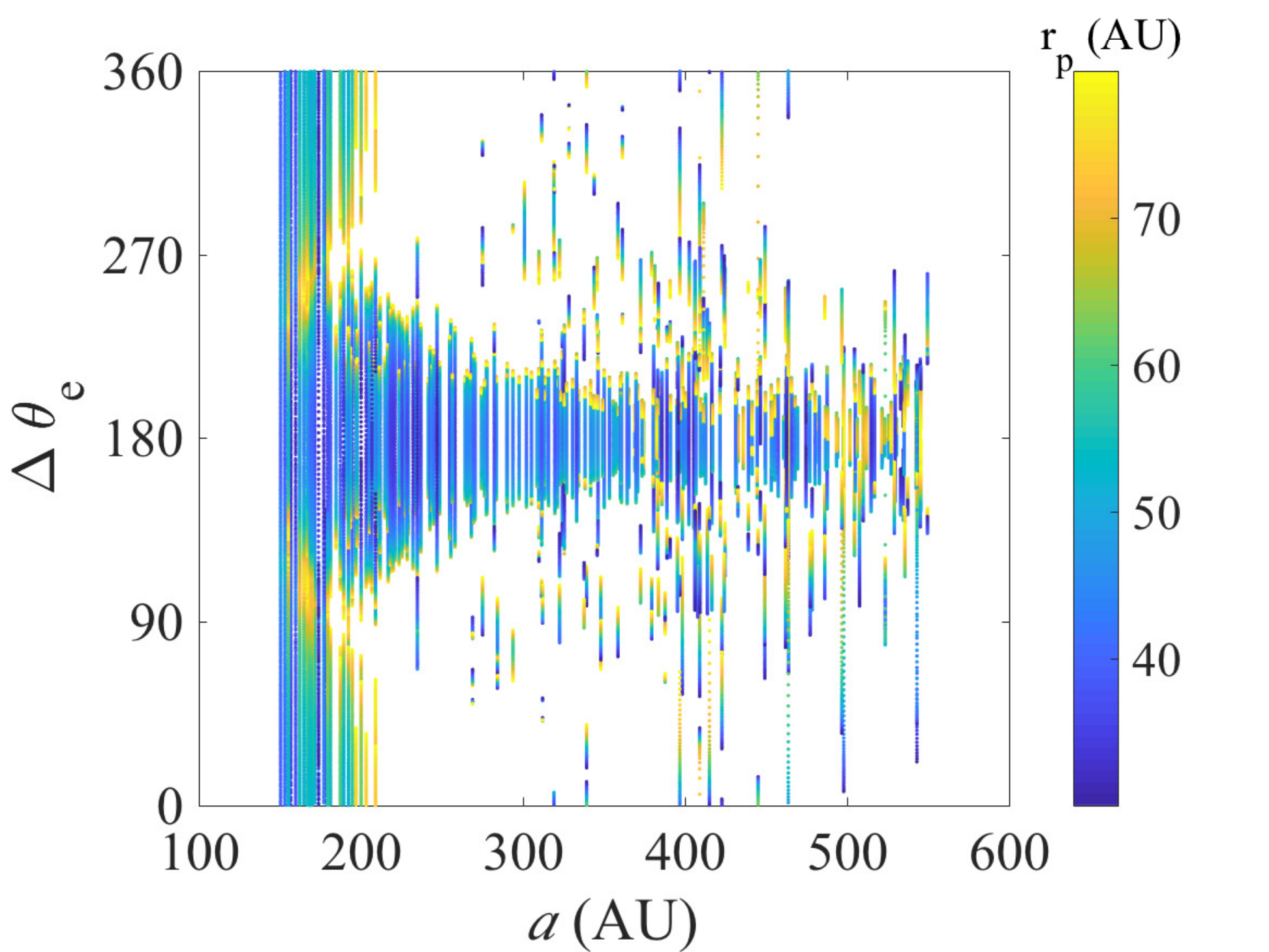}
\includegraphics[width=0.9\columnwidth]{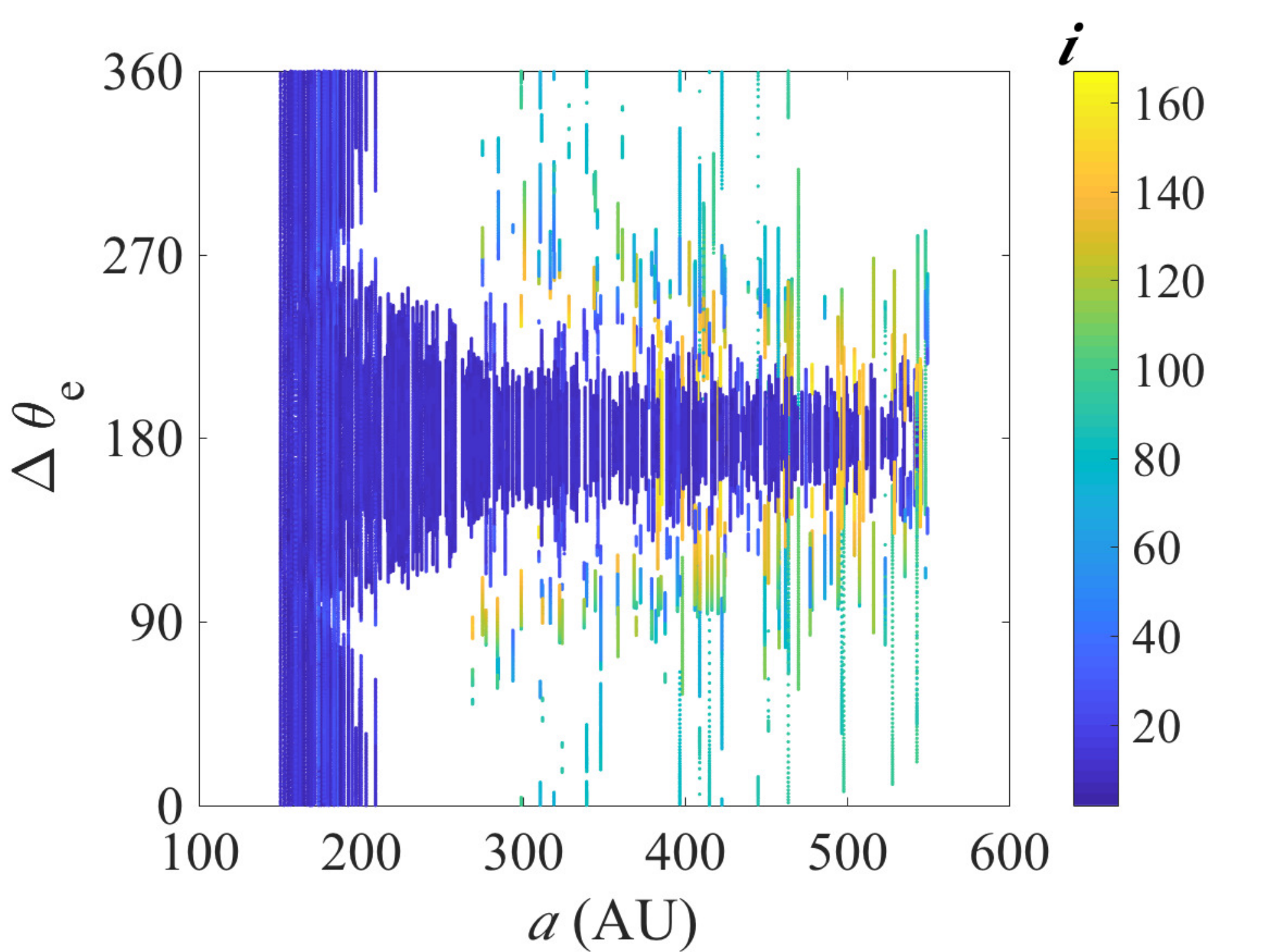} 
\caption{Results of secular simulations in which the particles are initially anti-aligned from the pericenter of Planet Nine ($\Delta \varpi = 180^\circ$). The top panel is color coded in pericenter distances and the lower panel is color coded in inclination. Particles with $t>3$ Gyr and $30<r_p<80$ AU are plotted (with time step of 1Myr). We find that there is a strong clustering at $\Delta\theta_e\sim 180^\circ$, when $30<r_p<80$ AU (top). In addition, there is a clustering in $\Delta\theta_e(\sim \varpi)\sim 180^\circ$ when inclination is low: $i\lesssim 40^\circ$, and there is a wider clustering in $\theta_e(\sim 2\Omega-\varpi)\sim 180^\circ$ when the inclination is high: $i\gtrsim 140^\circ$ (bottom). It shows that the pure secular effects can lead to the observed TNO orbital clusterings. There is no apparent clustering in $\Delta\omega$ or $\Delta\Omega$ (not plotted), except very mild ones which can be seen in the histogram in Figure \ref{f:hist_w180}. 
\label{f:scatter_w180}
}
\end{center}
\end{figure}

Combining all the particles in the secular simulations with different semi-major axes, we show the scatter in semi-major axes vs.~geometrical longitude of pericenter Figure \ref{f:scatter_w180}. The top panel is color coded by the pericenter distance, and the bottom panel is color coded by the inclination of the TNOs. To approximate detectable objects, we select only particles with $t>3$ Gyr. 
In addition, we select only particles with $30\text{ AU}<r_p<80$ AU to focus on the closer-in objects which are more likely to be detectable. 
All 500 particles are started with anti-aligned pericenter ($\Delta\varpi = 180^\circ$) and inclinations of $10^\circ$ from the ecliptic. Each point corresponds to a test particle, with snapshots taken at $1$ Myr timesteps. 
There is a clear alignment in $\Delta\theta_e$ that begins when the semi-major axes of the TNOs are around $\gtrsim 200-300$ AU, corresponding to the libration region in Figure \ref{f:Sec_epo}. In addition, when we consider only the lower inclination objects ($\lesssim 40^\circ$) the alignment near $\Delta\theta_e \sim \Delta\varpi\sim 180^\circ$ is stronger, particularly when $a\gtrsim 300$ AU. This shows that the observed clustering can be produced by pure secular effects alone, when starting with TNOs in the near coplanar configuration. We note that the clustering in $\Delta\theta_e$ is stronger than that in $\Delta\varpi$, because $\Delta \varpi$ no longer librates for the high inclination population.  This is consistent with what we find in Figures \ref{f:Sec_epo}, \ref{f:Sec_ethetae} and \ref{f:Sec_ipo}. The clustering in both $\Delta\omega$ and $\Delta\Omega$ is very weak and cannot be seen in the scatter plots (as shown in the histogram plots in Figure \ref{f:hist_w180}). Thus, we do not include them here.

\subsubsection{Alignment in argument of pericenter $\omega$ and longitude of node $\Omega$}

\begin{figure*}[ht]
\begin{center}
\includegraphics[width=6.in]{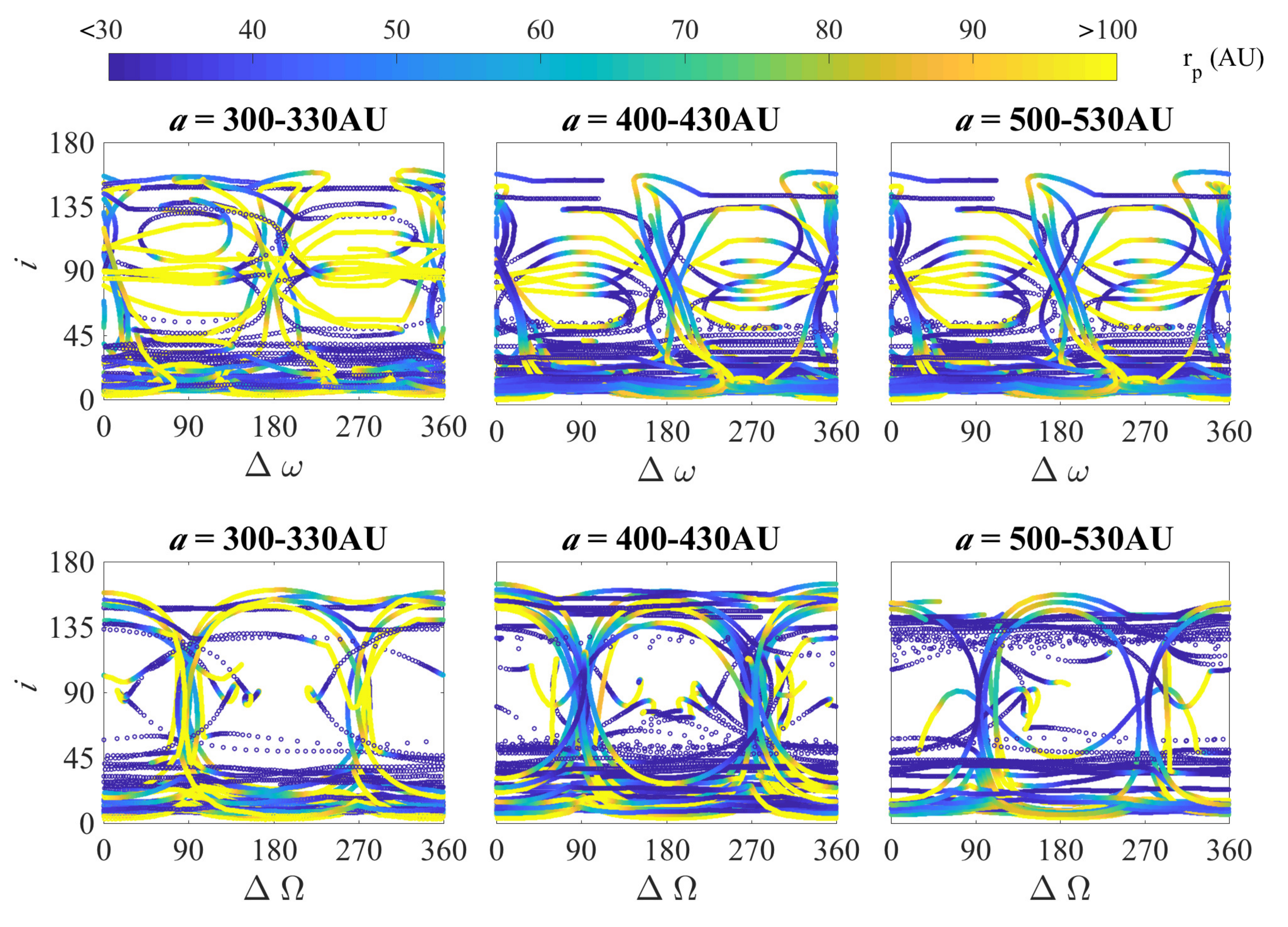} 
\caption{Secular evolution of TNOs, but now plotting results in the plane of argument of pericenter vs. inclination (upper row) and longitude of ascending node vs. inclination (lower row) in different semi-major axes panels. The color represents the pericenter distance of the TNOs. Clusters in $\Delta\omega$ and $\Delta\Omega$ can be seen when $60^\circ \lesssim i \lesssim 120^\circ$ for low pericenter distance $r_p\lesssim80$ AU. The libration in $\Delta \omega \sim 90^\circ$ and $\Delta \omega \sim 270^\circ$ are consistent with the secular resonances identified in \citet{Saillenfest17} using surface of sections that are analogous to Kozai resonances.
\label{f:Sec_iomega}
}
\end{center}
\end{figure*}

We now consider the alignment of the orbital plane in argument of pericenter and longitude of ascending node. 
We focus on TNOs with $a\gtrsim 300$ AU, which allows the inclination of the TNOs to be excited to large values. 
Figure \ref{f:Sec_iomega} presents the evolution of the TNOs in the plane of $(\Delta\omega,i)$ and $(\Delta\Omega,i)$, color-coded in the pericenter distance $r_p$. 
The figure illustrates ``looping'' trajectories followed by TNOs that reach high inclination, resulting in the clustering of $\omega$ near $\Delta \omega \sim 0^\circ$ and $\Delta \omega \sim 180^\circ$ and  $\Omega$ near $\Delta \Omega \sim 90^\circ$ and $\Delta \Omega \sim 270^\circ$ among the high-inclination TNOs when the pericenter distance is low $\lesssim 80$ AU.

In the top row of Figure \ref{f:Sec_iomega}, the dynamical region where the trajectories librate around $\Delta\omega \sim 90^\circ$ and $\sim 270^\circ$ for $40^\circ<i<90^\circ$ and $90^\circ<i<140^\circ$ are analogous to the quadrupole-order Kozai-Lidov resonant regions, which are seen for both interior and exterior test particles \citep[e.g.,][]{Kozai62, Lidov62, Naoz17, Vinson17}. These resonant regions are also shown in the secular study by \citet{Saillenfest17} on TNOs perturbed by Planet Nine using surface of sections. However, due to the close separation between the TNOs and the Planet, some of the hierarchical approximation breaks down as shown in \citet{Batygin16}. 
In the bottom row of Figure \ref{f:Sec_iomega} in the $(\Delta\Omega, i)$ plane, the region where the trajectories librate around $\Delta\Omega\sim180^\circ$ are analogous to the octupole Kozai-Lidov resonances, where $i$ can be excited from near zero inclination, and the cat-eye shaped regions centered around inclination $i\sim 90^\circ$ and $\Delta\Omega\sim 0^\circ \& \sim 180^\circ$ is also analogous to the octupole Kozai-Lidov  resononances, when the interaction energies are higher (e.g., Figure 4 of \citet{Li14b}, panel $H=-1$ and $H=-0.5$).  

\begin{figure}[ht]
\begin{center}
\includegraphics[width=3.2in]{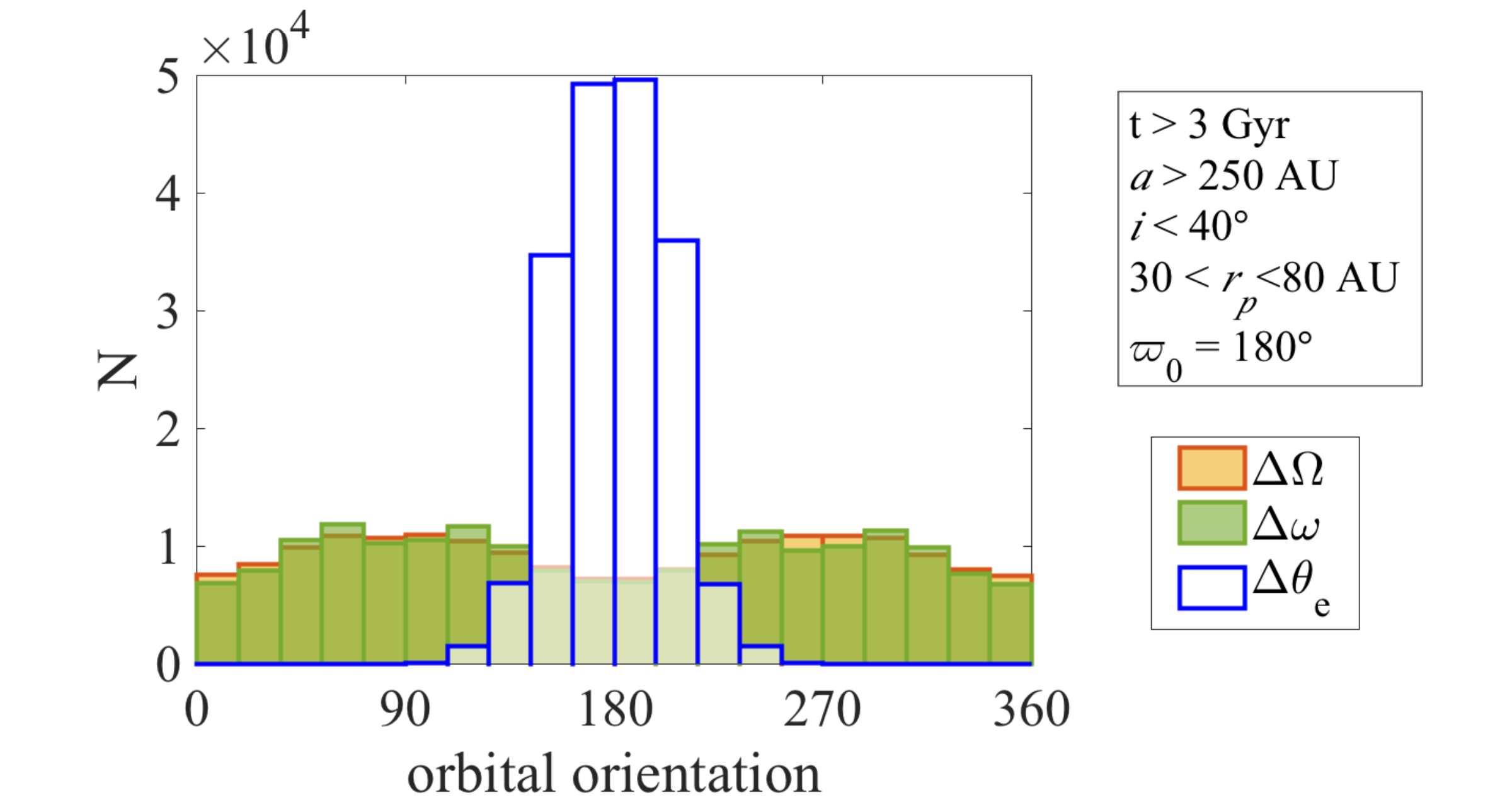} \\
\includegraphics[width=3.2in]{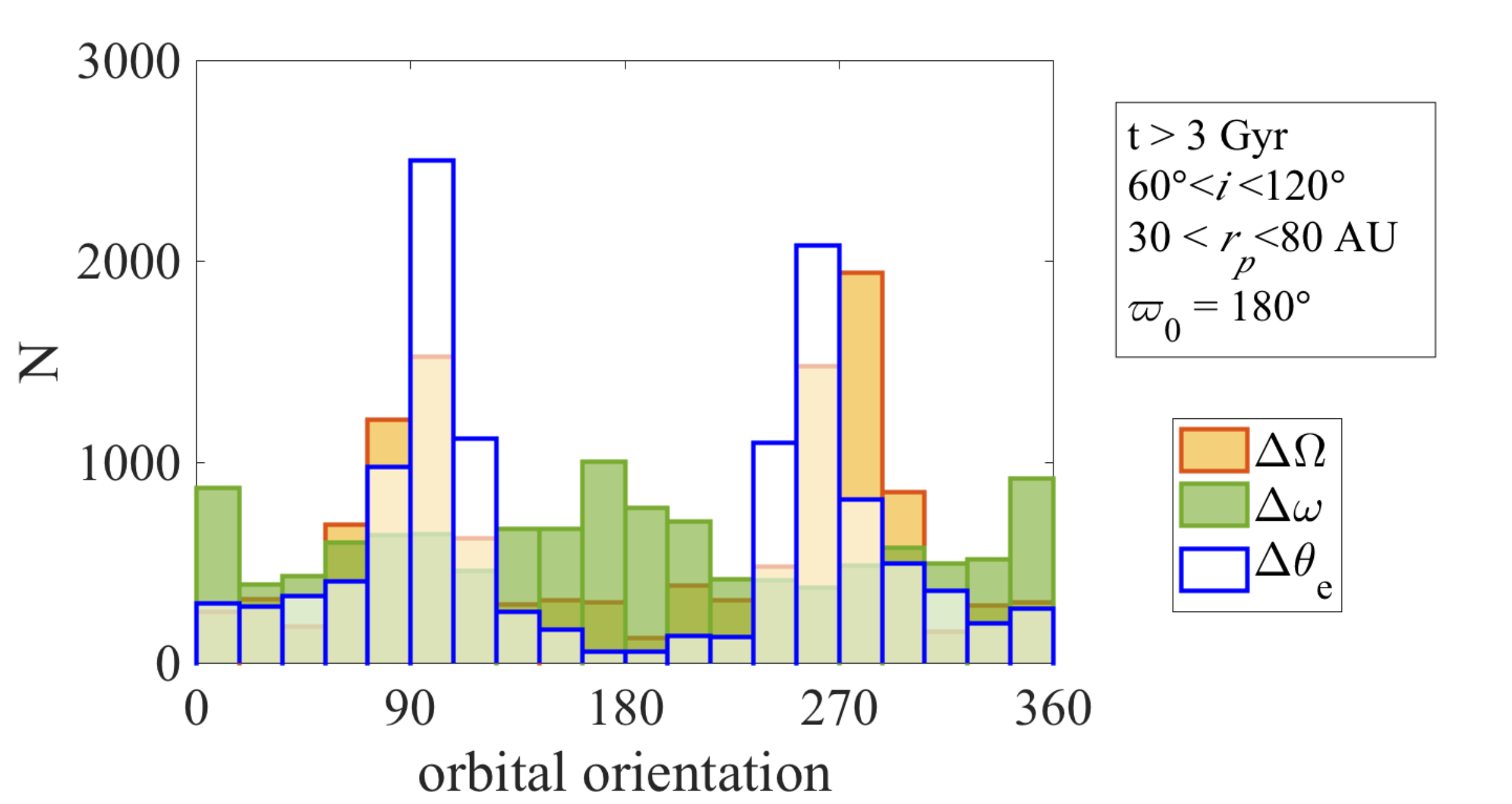} \\
\includegraphics[width=3.2in]{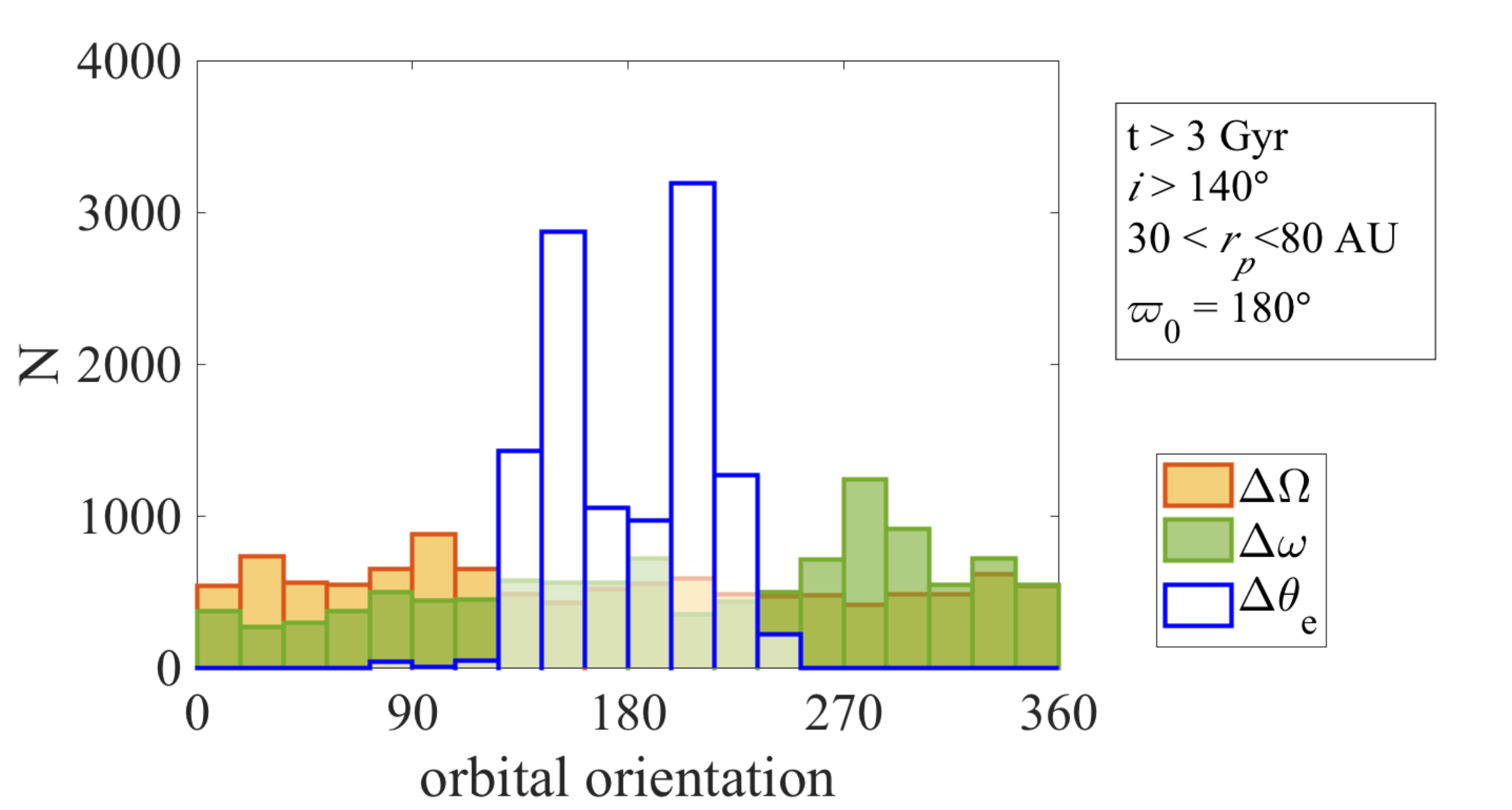}
\caption{Histogram of orbital orientation angles for the secular simulations illustrated in Figures \ref{f:Sec_epo} - \ref{f:Sec_iomega}, in which all particles were initialized anti-aligned with Planet Nine ($\Delta\varpi_0 = 180^\circ$). We select only particles with $30<r_p<80$ AU, and  plot in the \emph{\bf upper panel} low inclination ($i<40^\circ$) particles with $a>250$ AU. In the \emph{\bf middle panel}, we plot particles with $60^\circ<i<120^\circ$ and in the \emph{\bf bottom panel}, we plot particles with $i>140^\circ$ (there is no restriction on $a$ in the middle and lower panels, since only at larger semi-major axes can the orbital inclination be excited). The high inclination population ($60^\circ<i<120^\circ$) exhibits a strong clustering near $\sim 90^\circ$ and $\sim 270^\circ$ in $\Delta \Omega$ and $\Delta \theta_e$, and a mild clustering near $0^\circ$ and $180^\circ$ in $\Delta \omega$ for the low pericenter TNOs within the detection limit. Clustering near $\Delta\theta_e \sim 180^\circ$ can be seen in the low ($\lesssim 40^\circ$) and high ($\gtrsim 140^\circ$) populations. We note that the clustering in $\Delta\varpi \sim 180^\circ$ for the high inclination TNOs (in the middle panel) is missing due to the detectability cut.
 \label{f:hist_w180}}
\end{center}
\end{figure}

We summarize the orbital alignment in Figure \ref{f:hist_w180}, which shows histograms of $\Delta\theta_e$, $\Delta\omega$ and $\Delta\Omega$. 
We choose only particles with $t>3$Gyr and $30<r_p<80$ AU since these objects are more likely to be detectable. We break up the particles into three inclination bins in Figure \ref{f:hist_w180}: the upper panel selects particles with $a>250$ AU and $i<40^\circ$, the middle panel focuses on the high inclination ($60^\circ<i<120^\circ$) particles and the lower panel focuses on the retrograde particles with high inclination ($i>140^\circ$). For the low inclination population ($i<40^\circ$), there is a strong clustering around $\Delta \theta_e \sim 180^\circ$, and the clusterings in $\Delta\omega$ and $\Delta\Omega$ are weak. The high inclination population, shown in the middle panel, exhibits a clear deficit of particles around $\Delta \theta_e\sim 180^\circ$ and peaks near $\Delta \theta_e=90^\circ$ and $\Delta \theta_e=270^\circ$. 
We note that the clustering in $\Delta\varpi \sim 180^\circ$ for the high inclination TNOs (in the middle panel of Figure \ref{f:hist_w180}) is missing due to the detectability cut, but clustering in $\Delta \varpi \sim 180^\circ$ can be seen for high pericenter objects in Figure \ref{f:Sec_ipo}.
In addition, there is an excess of objects with $\Delta\Omega \sim 90^\circ$ and $270^\circ$, and with $\Delta\omega\sim 0$ and $180^\circ$, which illustrate clustering of high inclination orbits for the low pericenter TNOs.  The counter-orbiting particles ($i>140^\circ$) show a double-peaked clustering near $\Delta \theta_e\sim 150^\circ$ and $210^\circ$, with a slight deficit near $\Delta\theta_e\sim 180^\circ$ while $\Delta\omega$ and $\Delta\Omega$ are roughly uniform.

We note that we make the detectability cut at $r_p <80$ AU to facilitate comparison with observational results, in order to clarify the role of secular interactions in producing the observed clustering of the TNO orbits. Many dynamical effects, e.g., scatterings with Neptune, Planet Nine and mean motion resonances are neglected in the secular approach. 
Thus, we do not intend to reproduce the full dynamical interactions between the TNOs and Planet Nine using the secular methods. In addition, we note that including randomly initialized TNOs with $\omega$ and $\Omega$ uniformly distributed, the secular dynamics is similar to the exact coplanar case, where the there are two clustering of $\varpi$ around $0^\circ$ and $180^\circ$. The $0^\circ$ clustering is unstable if one considers the full dynamics including close encounters with Neptune and MMR with planet Nine \citep[e.g.,][]{Hadden18}.

\label{f:Sec_ranwhist}

\section{N-body Results}
\label{s:Nbd}
In this section, we apply the secular results of Section \ref{s:sec} to interpret N-body simulations of the evolution of trans-Neptunian objects (TNOs) orbiting under perturbations from Planet Nine, as well as to study the role of secular dynamics in sculpting the orbits of TNOs. We include a central $1 M_{\odot}$ star, $J_2$ moment corresponding to the inner three giant planets, but model both Neptune and planet Nine as fully-interacting standard massive particles. 

Similar to the near co-planar configuration included in \citet{Hadden18}, we set the the semi-major axis of Planet Nine $a_9 = 500$ AU, and the eccentricity of Planet Nine $e_9 = 0.6$. We set the inclination of Planet Nine to be $3^\circ$ for a near co-planar configuration\footnote{We do not use $i=0$ because we wish to have well-defined $\Omega_9$ and $\omega_9$ for Planet-Nine in our simulations, relative to which we can measure the TNO orientations.}, and we set the initial condition $\omega_9 = \Omega_9 = 0$. The initial condition of Neptune is set to agree with its configuration at the J2000 epoch: $a_N = 30.07$ AU, $e_N=0.0086$, $i_N = 1.77^\circ$, $\omega_N = -86.75^\circ$, $\Omega_N=131.72^\circ$, $\lambda_N=304.88^\circ$ \citep{Murray99}. We include 5000 test particles in our simulation, where the pericenters of the test particles are uniformly distributed within $30-50$ AU, and we draw the semi-major axes uniformly between $150-550$ AU. The inclination of the test particles are initially zero ($i_0 = 0^\circ$), while the non-coplanar Neptune torques the orbits of the TNOs, and Planet Nine further excites their inclinations. The argument of pericenter, longitude of ascending node and mean anomaly are all uniformly distributed between $0^\circ$ and $360^\circ$ for the TNOs. We use the Mercury package \citep{Chambers99} for the N-body simulation. We adopt the  ``hybrid'' Wisdom-Holman/Bulirsch-Stoer integrator \citep{Wisdom92, Press92}, with a time-step of $dt = 3000$ days, which is roughly $5\%$ of Neptune's orbital period.

\begin{figure}[ht]
\begin{center}
\includegraphics[width=3.2in]{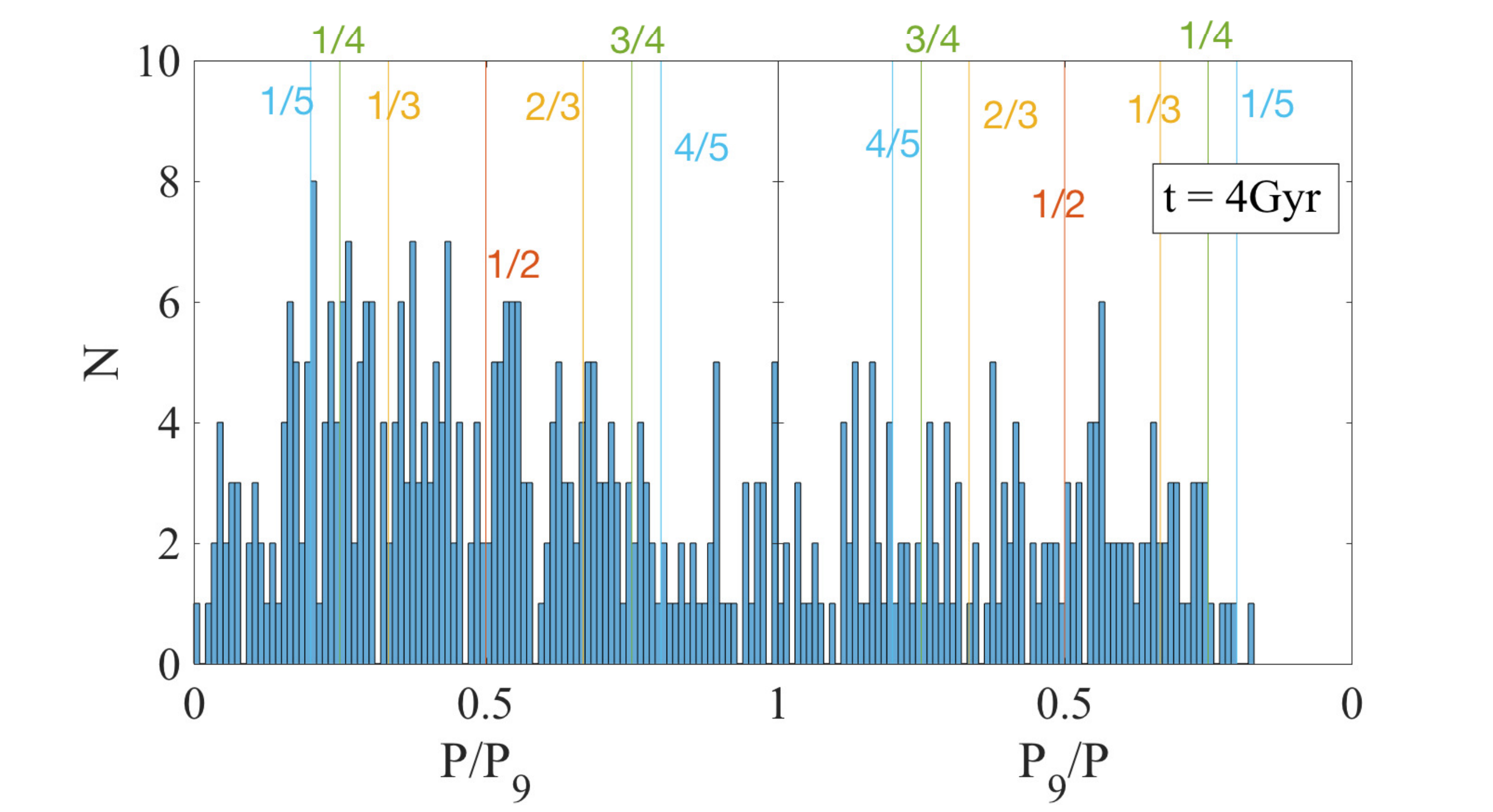} \\
\includegraphics[width=3.2in]{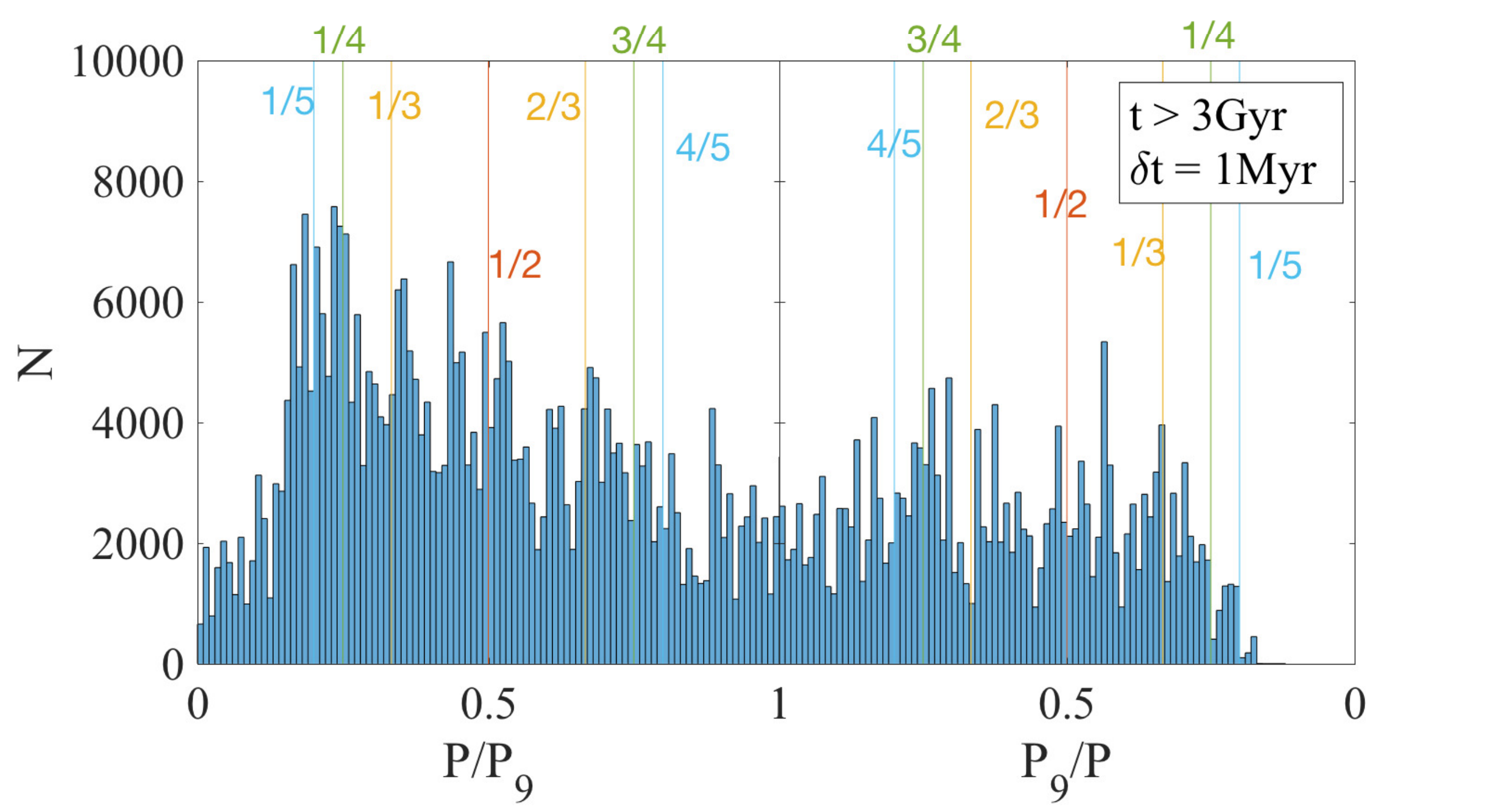} 
\caption{Period ratio of TNOs to Planet Nine for N-Body simulations in which Neptune is a point mass (instead of a $J_2$term).
Upper panel: Snapshot at $t=4$Gyr; Lower panel: All surviving particles for $t>3$Gyr, sampled at time-steps of $1$Myr. 
The left-half of each plot shows orbits \emph{interior} to Planet Nine, while those on the right are \emph{exterior}.
Contrary to the results of \citet{BatyginMorbi17} (where Neptune is treated by a $J_2$ term and using a coplanar set up), we find that many objects survive \emph{outside} of MMRs in the near coplanar configuration.
\label{f:NbdPratio}}
\end{center}
\end{figure}

After $3$ Gyr, $\sim12.5\%$ particles survived the simulation (bound to the Sun with $a<3000$ AU), and at $t=4$ Gyr, $8.7\%$ of the TNOs survived. The histogram of the TNO period ratios w.r.t. Planet Nine is shown in Figure \ref{f:NbdPratio}. 
In contrast to the co-planar case studied by \citet{BatyginMorbi17}, who treated Neptune as a $J_2$ term, many particles survive outside of the lower order mean motion resonance with Planet Nine, and some of them spend only a small fraction of time in high order mean motion resonances as illustrated in Figure \ref{f:res} in the appendix. Analyzing 100 survived TNOs, we find more than $\sim 80\%$ of them spend more than $\sim 80\%$ of time outside of MMRs. This is because scattering with the point mass Neptune removes TNOs from mean motion resonances with Planet Nine \citep{Hadden18}. Moreover, the misalignment between the TNO and Planet Nine's orbit helps to avoid orbit intersections and decreases the likelihood of collisions or close encounters with Planet Nine that eject the TNOs. The TNOs which survive outside of MMRs supports the importance of secular dynamics in the evolution of TNOs, in addition to the effects of the mean motion resonances highlighted by \citet{BatyginMorbi17}.

\begin{figure}[ht]
\begin{center}
\includegraphics[width=3.8in]{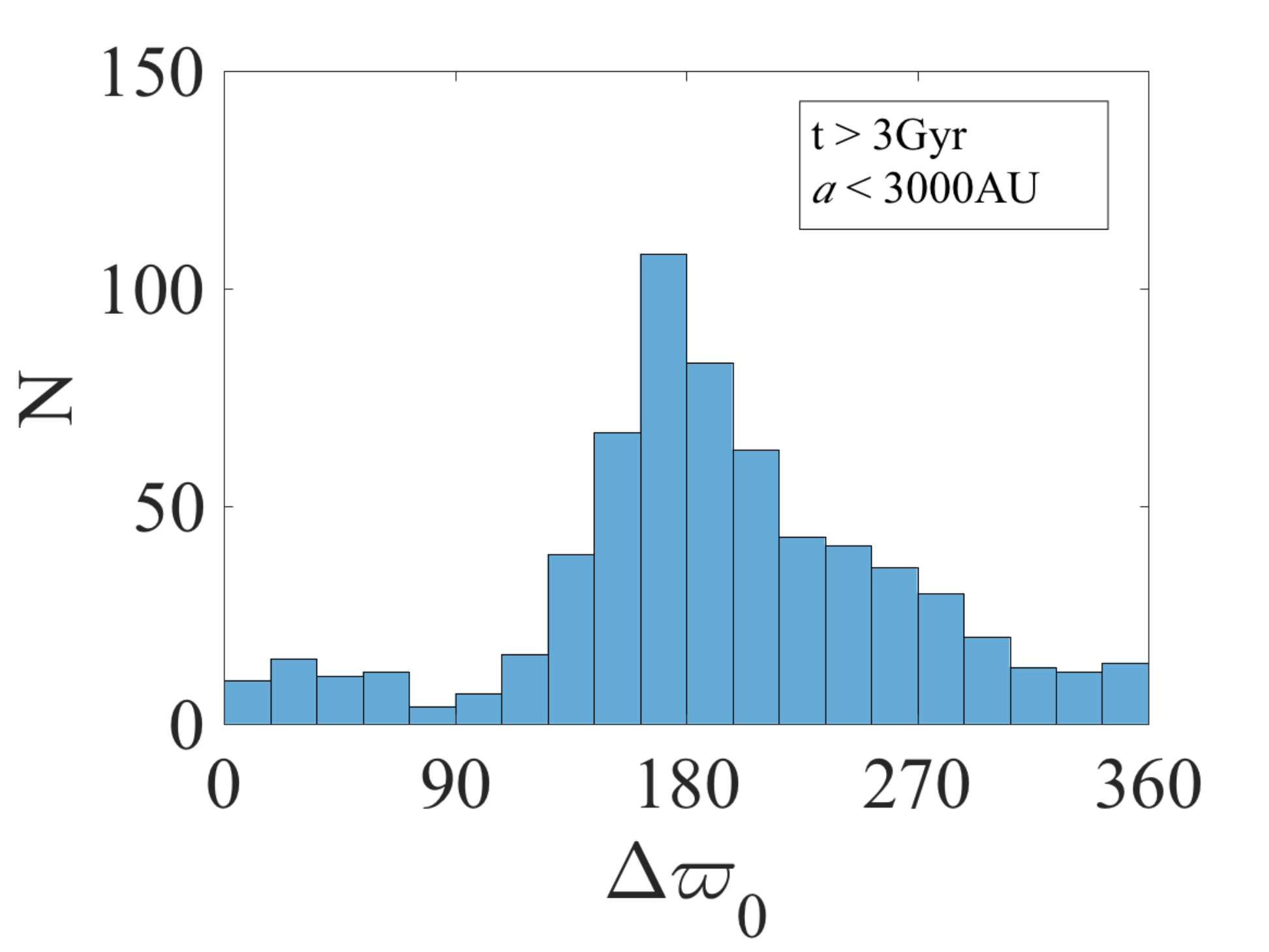} 
\caption{The histogram in the initial $\Delta\varpi$ of the survived particles (with $t_{\rm final}>3$Gyr). It shows that most of the survived ones start with $\Delta\varpi_0 \sim 180^\circ$, while the initially aligned populations are more likely ejected. \label{f:histNbodypo0}}
\end{center}
\end{figure}

The N-body results on the clustering of the TNO orbits in the near coplanar configuration have previously been shown in the literature \citep[e.g.][]{Batygin_inc16}, and we include the clustering of the TNO orbits in the Appendix (Figure \ref{f:Nbody} and \ref{f:Nbdhighi}) to allow detailed comparison with our secular results.

To illustrate the dependence of the surviving particles on their initial longitude of pericenter in more detail, we show the histogram of the initial longitude of pericenter for the surviving particles ($t_{\rm final} >3$Gyr) in Figure \ref{f:histNbodypo0}. There is a strong peak near $\Delta\varpi_0 \sim 180^\circ$. The initially anti-aligned population is more likely to survive due to the phase protection of the orbits in the libration region around $\Delta\theta_e \sim 180^\circ$. This keeps the TNO orbits away from being tangential with the orbit of Planet Nine, where the overlap of the mean motion resonances lead to chaotic evolution of TNOs, as discussed in \citet{Hadden18} for the coplanar case. Then, they are carried into the low pericenter distance Neptune scattering region and are often ejected \citep{Khain18}. The dominance of surviving particles with $\Delta\varpi_0 \sim 180^\circ$ also shows that secular investigations can reproduce the dynamical features of N-body simulations as discussed in Appendix B.

\begin{figure}[ht]
\begin{center}
\includegraphics[width=3.5in]{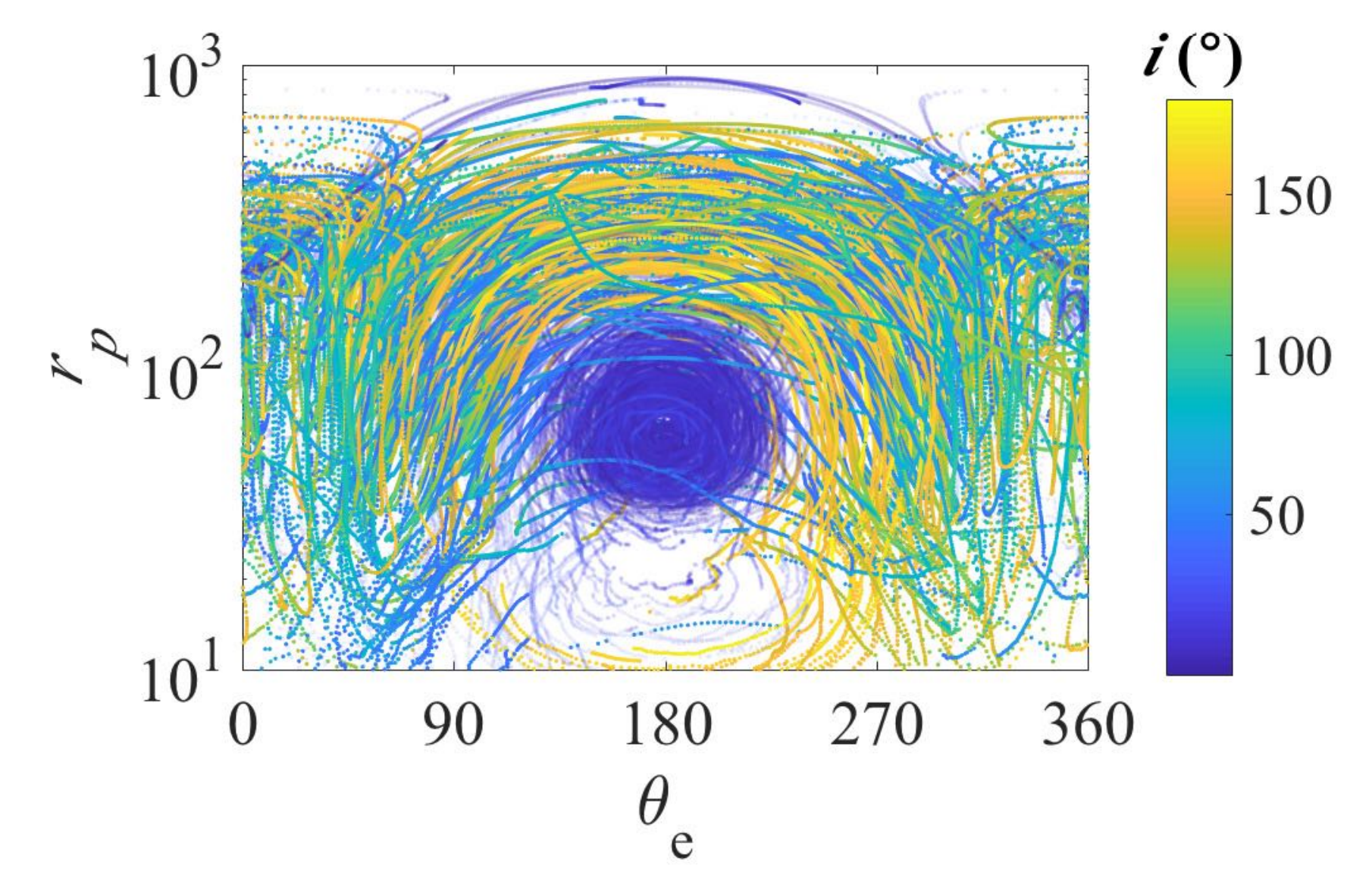} 
\caption{Evolution of particles for $t>3$Gyr, $a>300$ AU, $10<r_p<1000$ AU in the $r_p-\theta_e$ plane for the N-body simulations seen in Figure \ref{f:Nbody}. To highlight the high inclination evolution, we increase the transparency of the low inclination part of trajectories ($i<40^\circ$). The libration of the particles in $\theta_e$ is consistent with that caused by the secular resonances, as shown in Figure \ref{f:Sec_ethetae}. We show a wide range of pericenter distances here in order to illustrate the overall dynamics. \label{f:varpi-ap_Nbd}}
\end{center}
\end{figure}

Moreover, we can see libration of the geometrical longitude of pericenter during the evolution of the TNO orbits in the plane of $(\Delta\theta_e, r_p)$ (Figure \ref{f:varpi-ap_Nbd}). The libration in $\Delta\theta_e \sim 180^\circ$ when the inclination is low is similar to the coplanar case \citep{Hadden18}. This leads to the clustering in $\Delta\theta_e \sim 180^\circ$. When the inclination is higher, the trajectories can sometimes also librate in $\Delta\theta_e \sim 180^\circ$. This feature can also be seen in the secular results, as shown in Figure \ref{f:Sec_epo}.

\begin{figure}[ht]
\begin{center}
\includegraphics[width=3.8in]{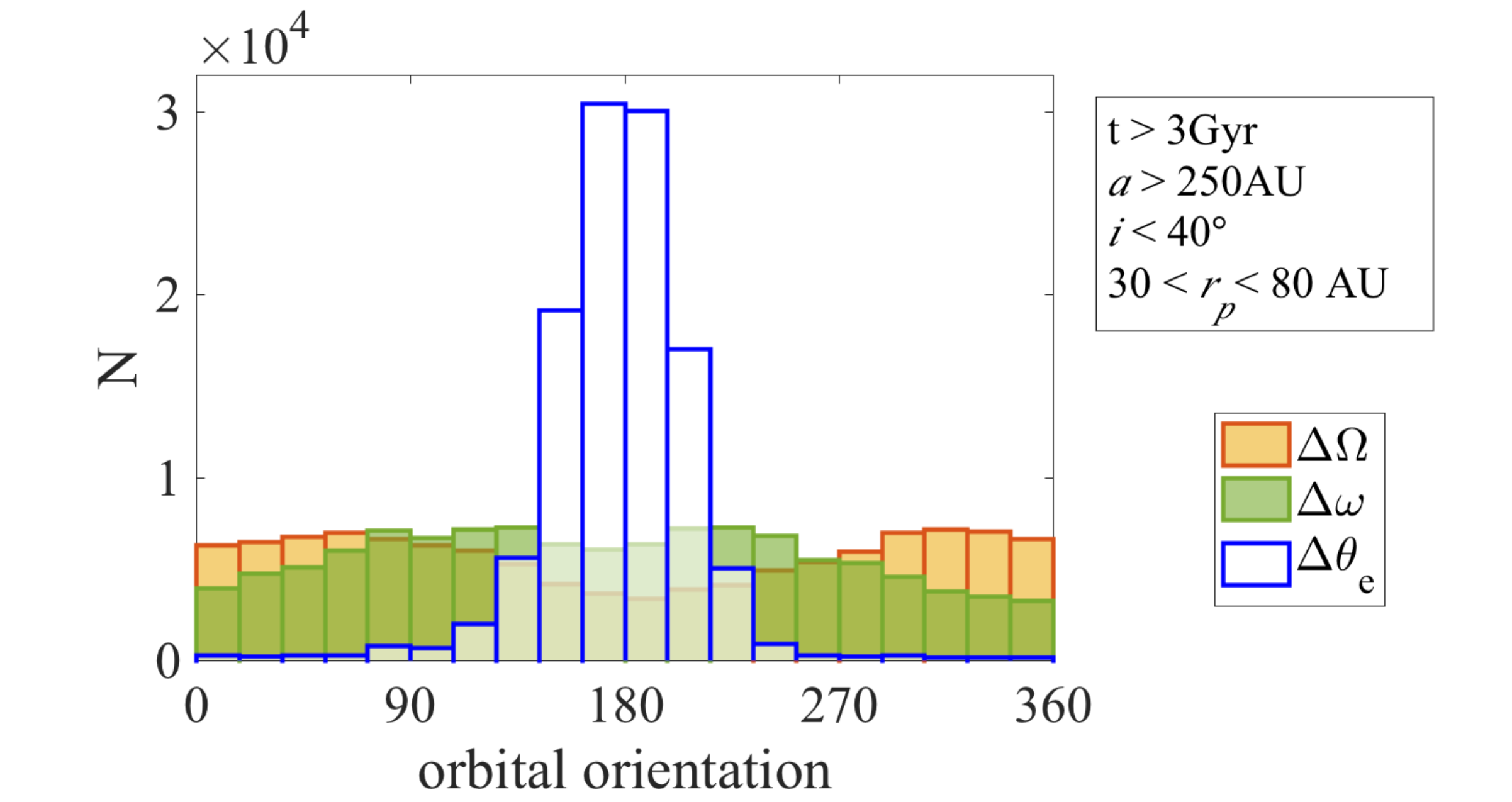} 
\includegraphics[width=3.8in]{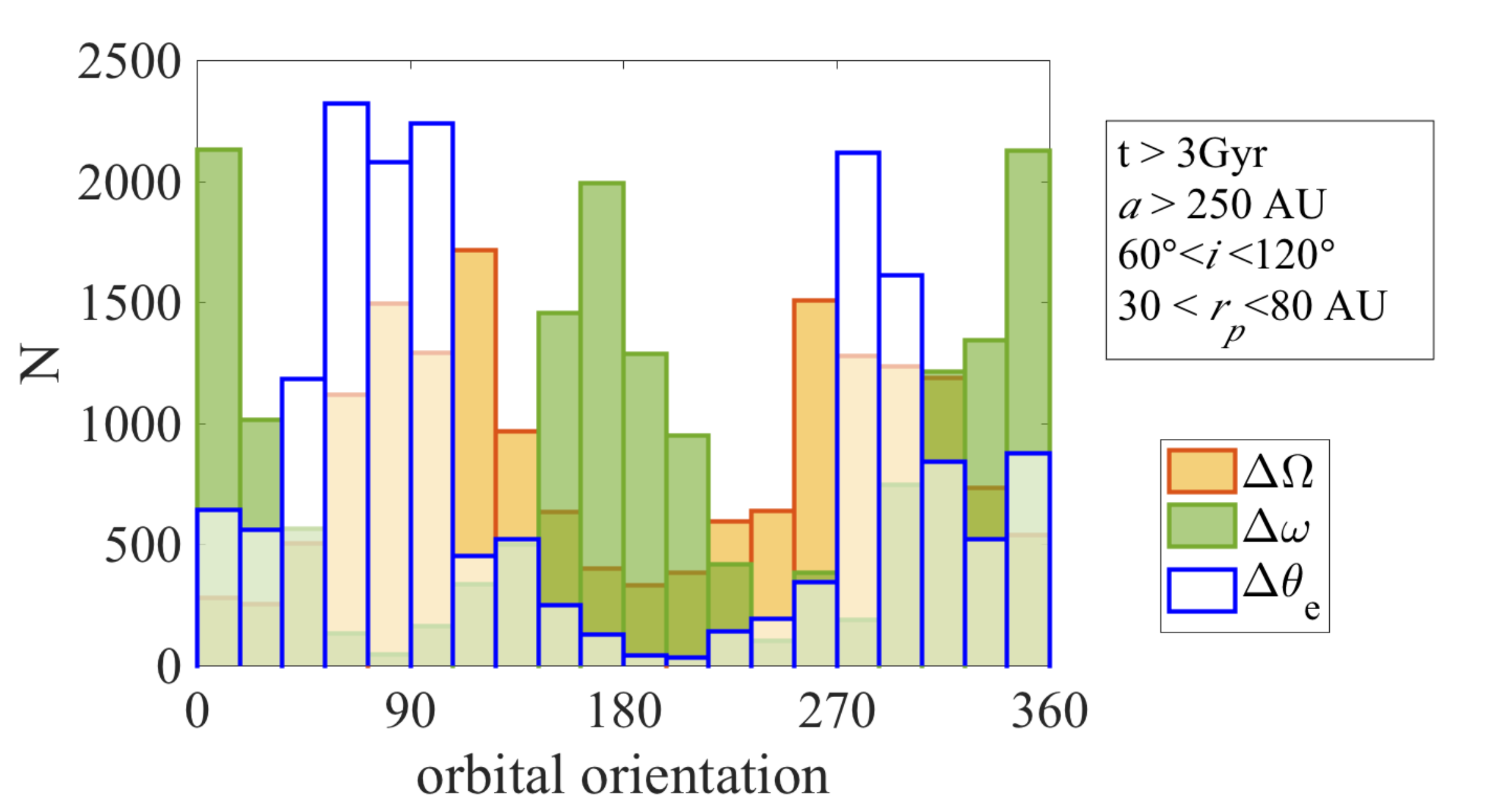}
\includegraphics[width=3.8in]{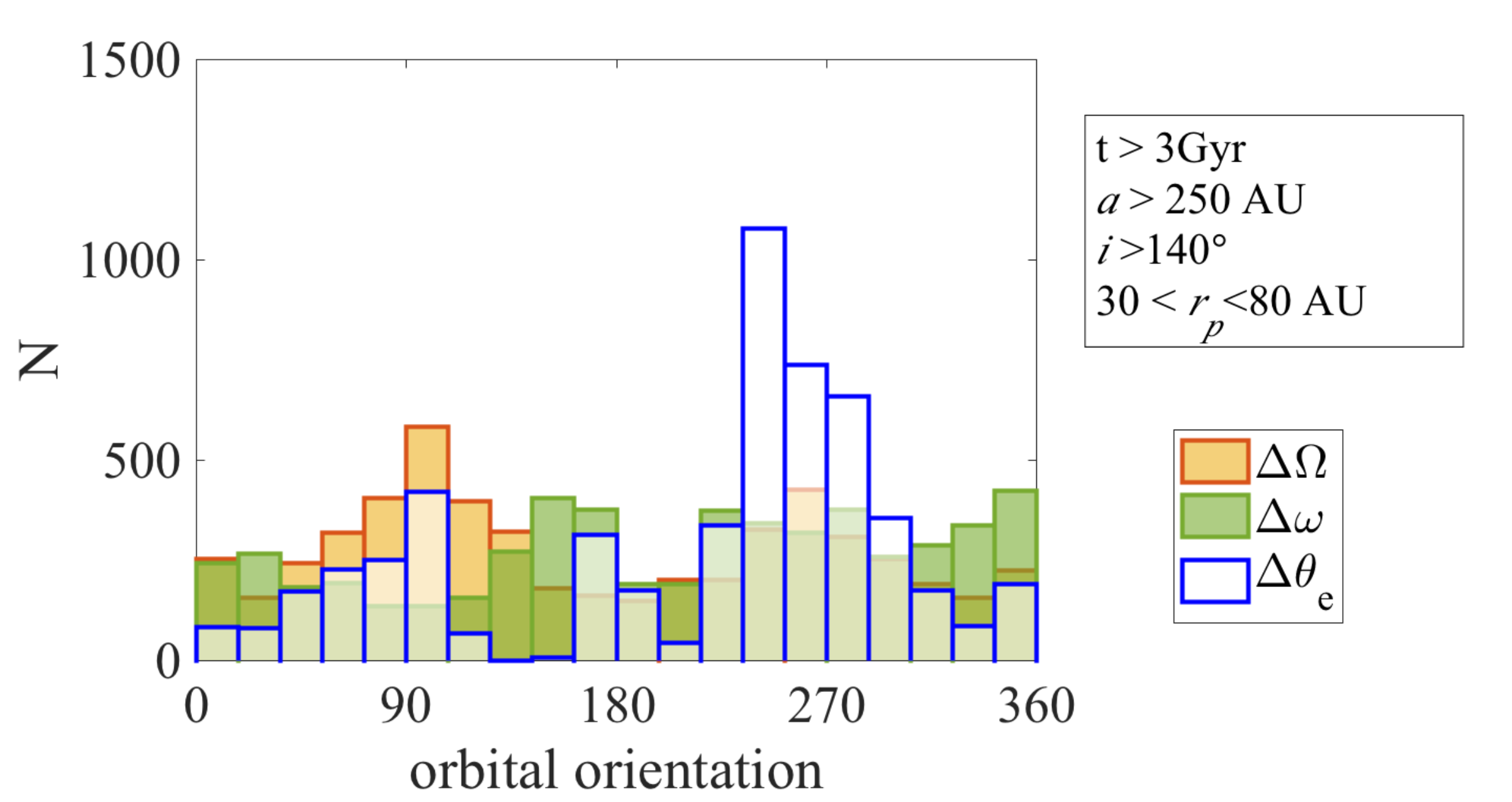}
\caption{Histogram of orbital orientation for $t>3$Gyr, $a>250$ AU, $30<r_p<80$ AU and $i<40^\circ$ (upper panel), $60^\circ<i<120^\circ$ (middle panel) and $i>140^\circ$ (upper panel), illustrating the strong clustering in the orbital orientations for the N-body simulations seen in Figure \ref{f:Nbody} and \ref{f:Nbdhighi} in the appendix. The clustering is similar to the secular results when the TNOs start with $\Delta\varpi_0 = 180^\circ$ in Figure \ref{f:hist_w180}. \label{f:histNbody}}
\end{center}
\end{figure}

To summarize the clustering features of the TNO orbits, we combine all the long lived ($t>3$Gyr) and small-pericenter particles ($30<r_p<80$ AU) in Figure \ref{f:histNbody}. The upper panels shows the histogram for particles with low inclination $i<40^\circ$. We select those with $a>250$ AU, since there is no clustering at smaller semi-major axes (due to the fast $J_2$ precession, as shown in Figure \ref{f:Nbody}). We plot histograms in the geometrical longitude of pericenter $\Delta\theta_e$, argument of pericenter $\Delta\omega$ and longitude of ascending node $\Delta\Omega$. There is a significant peak around $\Delta\theta_e\sim 180^\circ$, indicating strong clustering in the geometrical longitude of pericenter. However, the clustering in $\Delta\omega$ and $\Delta\Omega$ are weak. 

The middle panel shows the high inclination population, $60^\circ<i<120^\circ$. Similar to the secular results, we see clustering in $\Delta\omega \sim 0^\circ \& 180^\circ$, $\Delta\Omega \sim 90^\circ \& 270^\circ$ and $\Delta\theta_e \sim 90^\circ$ and $\Delta\theta_e \sim 270^\circ$ within the detection limit for the low pericenter TNOs. Only a small fraction of TNOs reached the counter orbiting configuration with $i>140^\circ$. Similar to the near anti-aligned population illustrated in Figure \ref{f:hist_w180}, there are two peaks around $\Delta\theta_e\sim135^\circ$ and $\Delta\theta_e\sim225^\circ$. The clustering in $\Delta\omega$ and $\Delta\Omega$ is weak for the near counter orbiting population. These N-Body clusterings are similar to the secular results in Figure \ref{f:hist_w180}. This indicates that the clusterings in the N-body results can be produced by pure secular interactions for the anti-aligned population.

\section{Orbital Flips and Dependence on Planet Nine}
\label{s:incl}

N-body simulations have shown that TNOs' orbits can flip with large amplitude and cross $90$ degrees \citep[e.g.,][]{Batygin_inc16, Lawler17, Shankman17a}. This might explain the origin of the detected retrograde TNOs \citep[e.g., ][]{Chen16}. The flip of the orbits is very similar to the near coplanar flip of the hierarchical ($a_{TNO} \ll a_{m9}$) three body interactions discovered in \citet{Li14a}, where a test particle's orbit can be flipped by nearly $180^\circ$ by a near co-planar perturber. Here, we investigate the flip in the much less hierarchical configurations ($a_{TNO} \sim a_{m9}$) in this section, and characterize the dependence of the flip on properties of the perturber, Planet Nine.

\subsection{Secular Investigation}


\begin{figure*}
\begin{center}
\includegraphics[width=6.in]{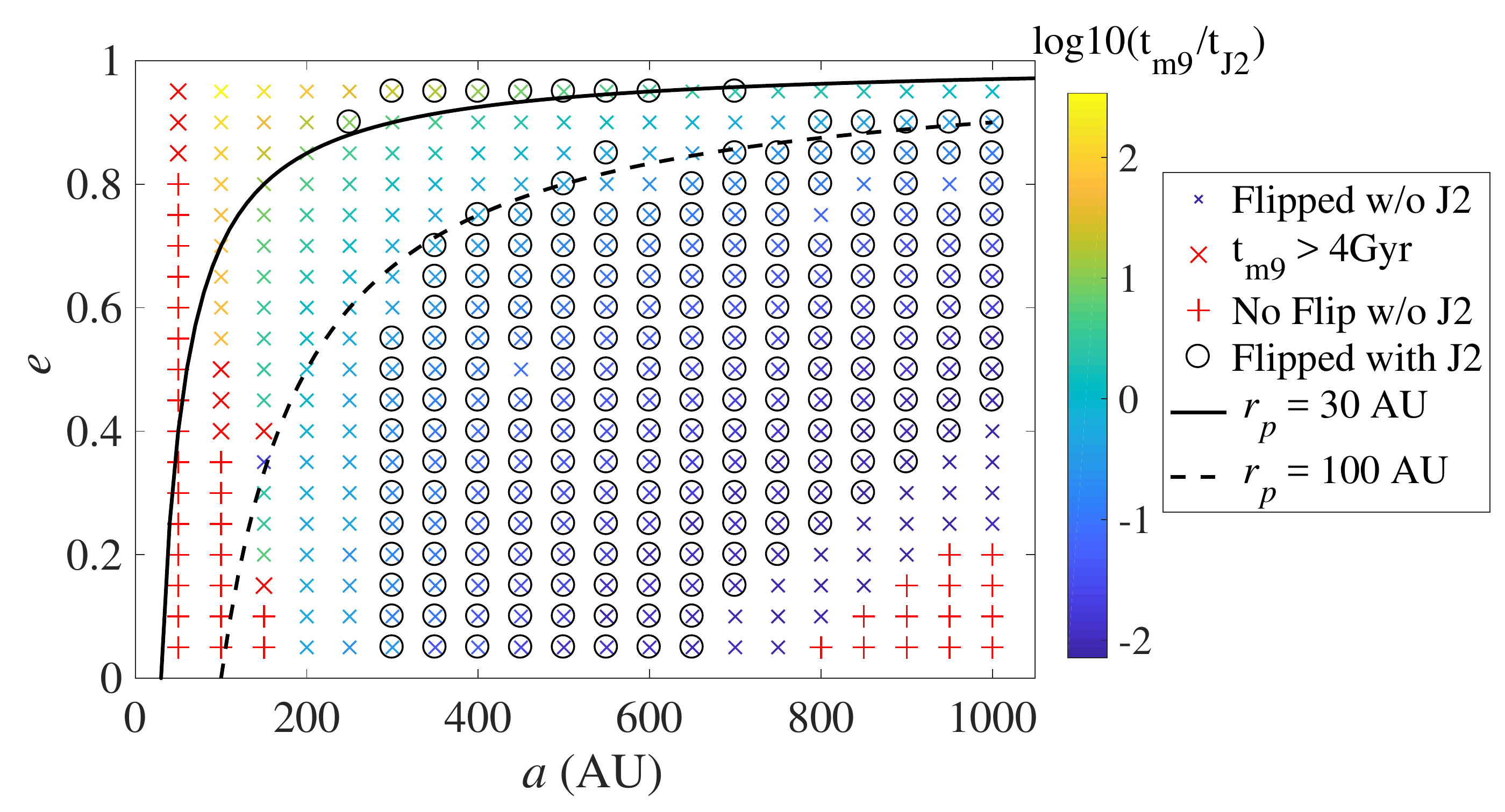} 
\caption{Flip condition in the secular approximation. The colors represent the ratio of the libration timescale of $\theta_e$ due to Planet Nine to the timescale of $J_2$ precession. The red pluses indicate the TNOs that do not flip in $4$Gyr even without the $J_2$ potential? and the circles denote the runs that still flip in the presence of the $J_2$ potential. The initial inclination of the TNOs are set to be $5^\circ$, and the pericenter of the TNOs are anti-aligned with that of Planet Nine ($\omega=\pi$, $\Omega = 0$) for illustration. The orbits are more likely to be flipped when the libration timescale of the longitude of pericenter vector is shorter than the $J_2$ precession timescale, with $a\gtrsim300$ AU. Flips in the region within $30\lesssim r_p\lesssim 100$ AU are suppressed due to $J_2$ precession. \label{f:flipsec}}
\end{center}
\end{figure*}

For simplicity, we start with the secular approximation.  To characterize the range of parameter space where the TNOs could flip, we investigate the flips in the plane of TNO semi-major axes and initial eccentricity using the secular integration in Figure \ref{f:flipsec}. Similar to the secular simulations discussed above, we set $a_9 = 500$ AU, $e_9 = 0.6$, $i_9=\omega_9=\Omega_9 =0$. We set the initial longitude of pericenter of the TNO to be $\varpi_0 = 180^\circ$ ($\omega_0 = \pi$, $\Omega_0 = 0$) here, since the surviving TNOs mostly have $\Delta\varpi_0\sim 180^\circ$ (as shown in Figure \ref{f:histNbodypo0}). The inclinations of the TNOs are all set to be $5^\circ$, slightly misaligned from the ecliptic plane. The parameter space that allows the TNOs to flip for the anti-aligned TNOs does not depend sensitively on the initial mutual inclination, as long as it is non-zero and near-coplanar.

To understand the flips better, we break down the problem into pieces, and consider the three-body interactions first, without the $J_2$-precession. Including only the central star, a TNO and the perturbing planet Nine, the orbit of the TNO can be easily flipped both when the TNO is inside the orbit of Planet Nine and when it is farther. For illustration, we represent the runs that could not flip using pluses and then use crosses to represent the runs that can flip when we ignore $J_2$ precession. 

Next, we marked by circles the runs which can flip in the presence of the $J_2$ precession. Including $J_2$ precession, the flip of the orbits can be suppressed. This is determined by the libration and $J_2$-precession timescales of the TNOs. To compare the timescales, we color-code the crosses using the ratio of the libration timescale of $\Delta\theta_e$ and the $J_2$-precession timescale. To obtain the timescales, we numerically calculated the libration timescale of the geometrical longitude of pericenter following the secular integration of the three-body interaction, and we calculated the $J_2$ precession due to the inner four giant planets analytically based on Equation (\ref{eqn:prec2}). 

When the libration timescale exceeds $4$Gyrs, we color the crosses in red. As expected, the runs can still flip when the libration timescale is shorter than the $J_2$ precession timescales in general. Most of the TNOs can still flip when they are farther from the inner giant planets with $a \gtrsim 300$ AU and $r_p \gtrsim 100$ AU. Note that we only focus on the anti-aligned configurations here, since these TNOs are more likely to survive. The TNO orbits are less likely to flip if their pericenters are aligned with that of Planet Nine, analogous to the hierarchical limit \citep{Li14a}. The general flip condition in the non-hierarchical limit is quite complicated, and is beyond the scope of our paper.

As illustrated in Figure \ref{f:flipsec}, the overall dynamics can be divided in the following three regions. \\
1) small semi-major axis region ($a\lesssim 150$ AU), where the precession timescale is much shorter than that of the $\Delta\theta_e$ libration timescale. Both the flip and the libration in $\Delta\theta_e$ can be suppressed. \\
2) large semi-major axis region ($a\gtrsim 150$ AU) within the two black lines inside $30\lesssim r_p<\lesssim 100$ AU in Figure \ref{f:flipsec}, where the two timescales are comparable. The flips are suppressed while the libration in $\Delta\theta_e$ still persists. \\
3) large semi-major axis region ($a\gtrsim 150$ AU), where the libration time scale is much shorter. Both the libration and the flip remain.

To illustrate the flips in more detail, we show examples of arbitrarily selected trajectories in these three regions below.

\begin{figure*}[ht]
\includegraphics[width=0.5\textwidth]{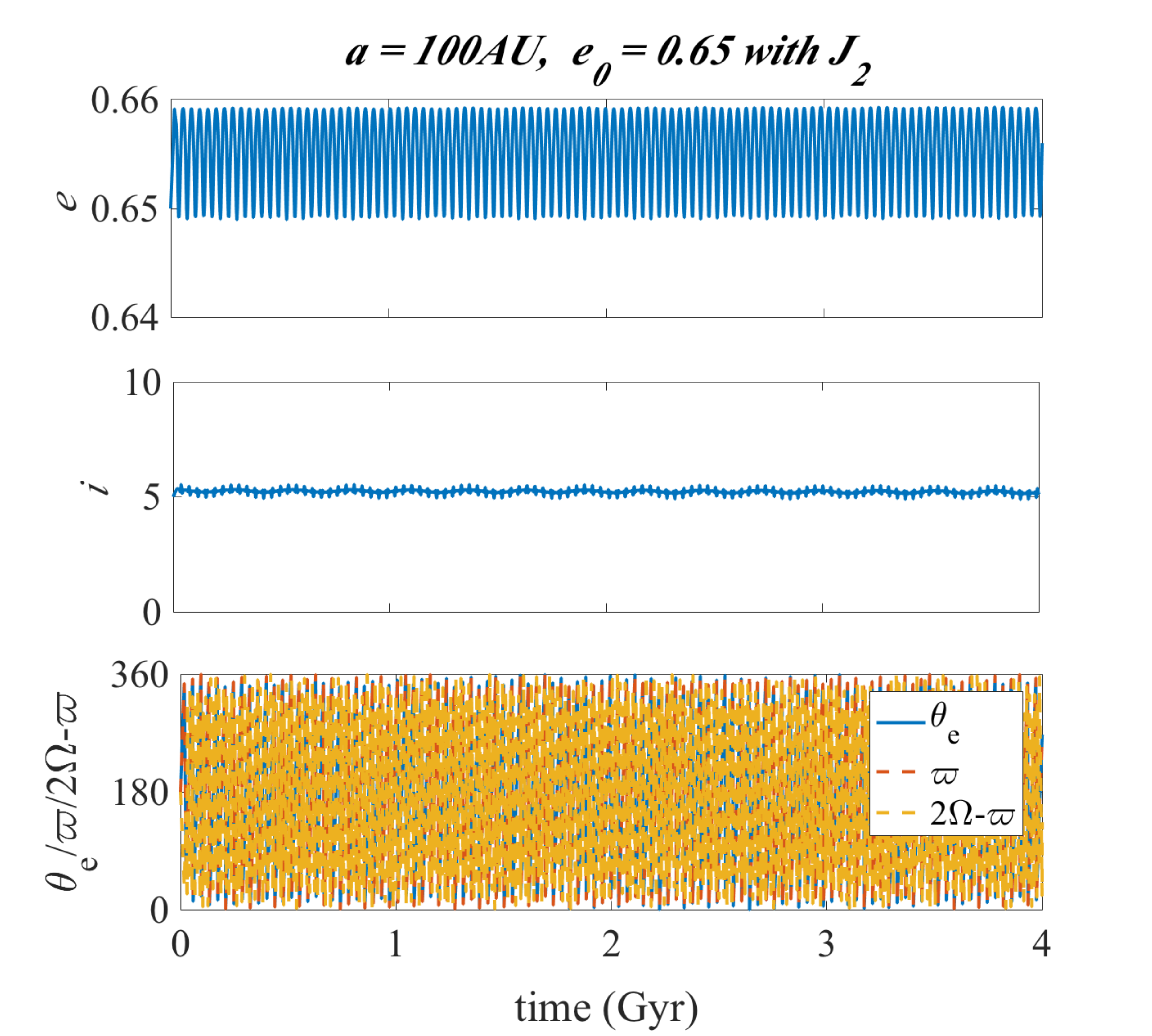}
\includegraphics[width=0.5\textwidth]{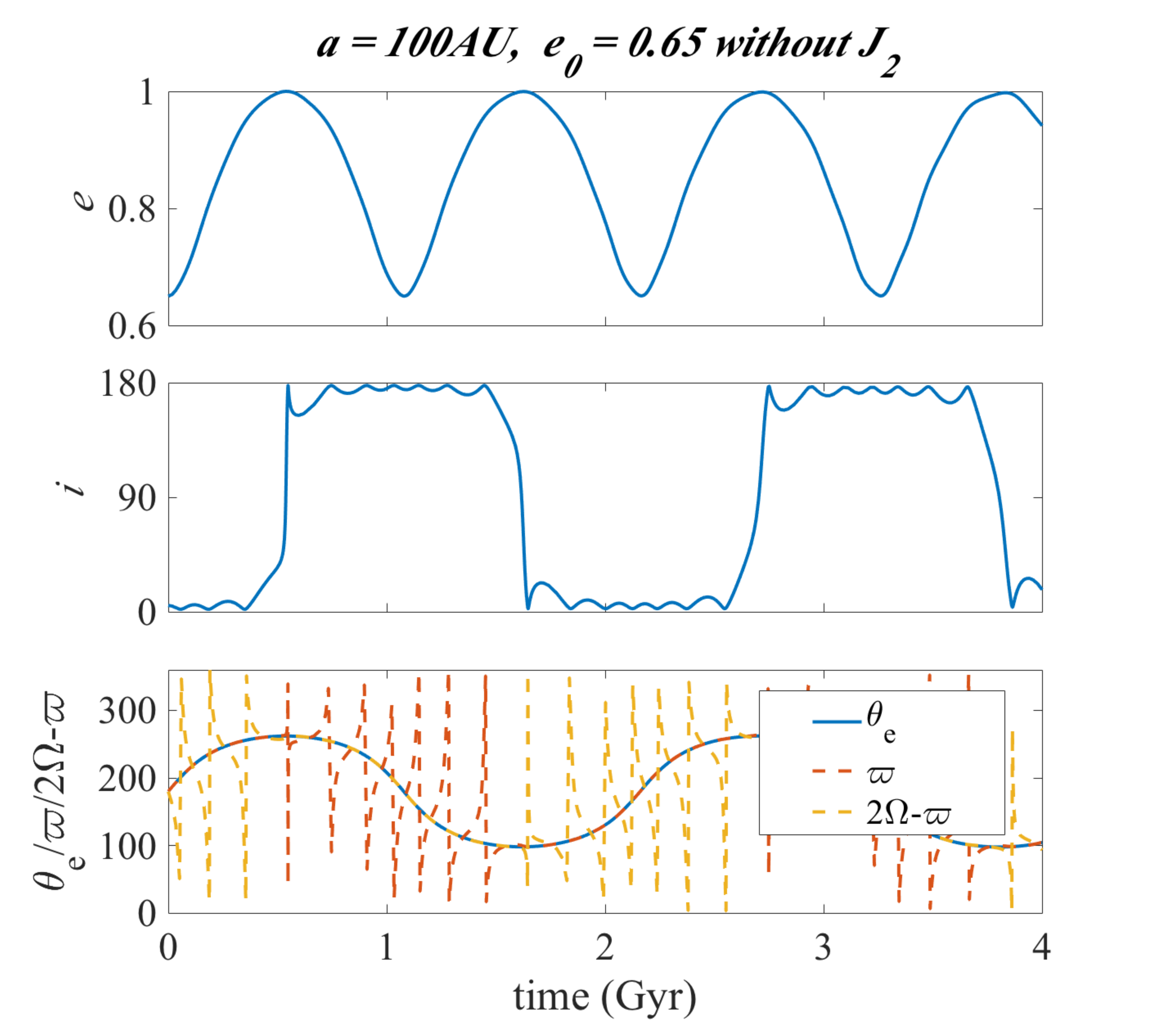}
\caption{Evolution of TNO in region 1), where $J_2$ precession dominates over the perturbation from Planet Nine. Without $J_2$ precession, $\Delta\theta_e$ librates around $180^\circ$, TNO eccentricity are excited, the orbit of TNO can be flipped and $\theta_e$ librates, similar to the hierarchical three-body dynamics \citep{Li14a}. Including $J_2$ potential, $\Delta\theta_e$ libration, eccentricity and inclinatino excitation are all suppressed. \label{f:secincl_traj1}}
\vspace{0.1cm}
\end{figure*}

Figure \ref{f:secincl_traj1} illustrates the evolution in region 1), where the semi-major axis of the TNO is small, and thus the $J_2$ term from the inner four giant planets dominants over the perturbation from the outer Planet Nine. The left panel shows the case with $J_2$-precession and the right panel shows the case without $J_2$-precession. Without $J_2$ precession, the flip is analogous to that in the hierarchical limit \citep{Li14a}. The precession timescale is much shorter comparing with the $\Delta\theta_e$ libration timescale (as shown in Figure \ref{f:flipsec}), and thus the libration of $\Delta\theta_e$ is suppressed when the $J_2$ term is included. The flip of the orbit is also suppressed in this case. 

\begin{figure*}[ht]
\includegraphics[width=0.5\textwidth]{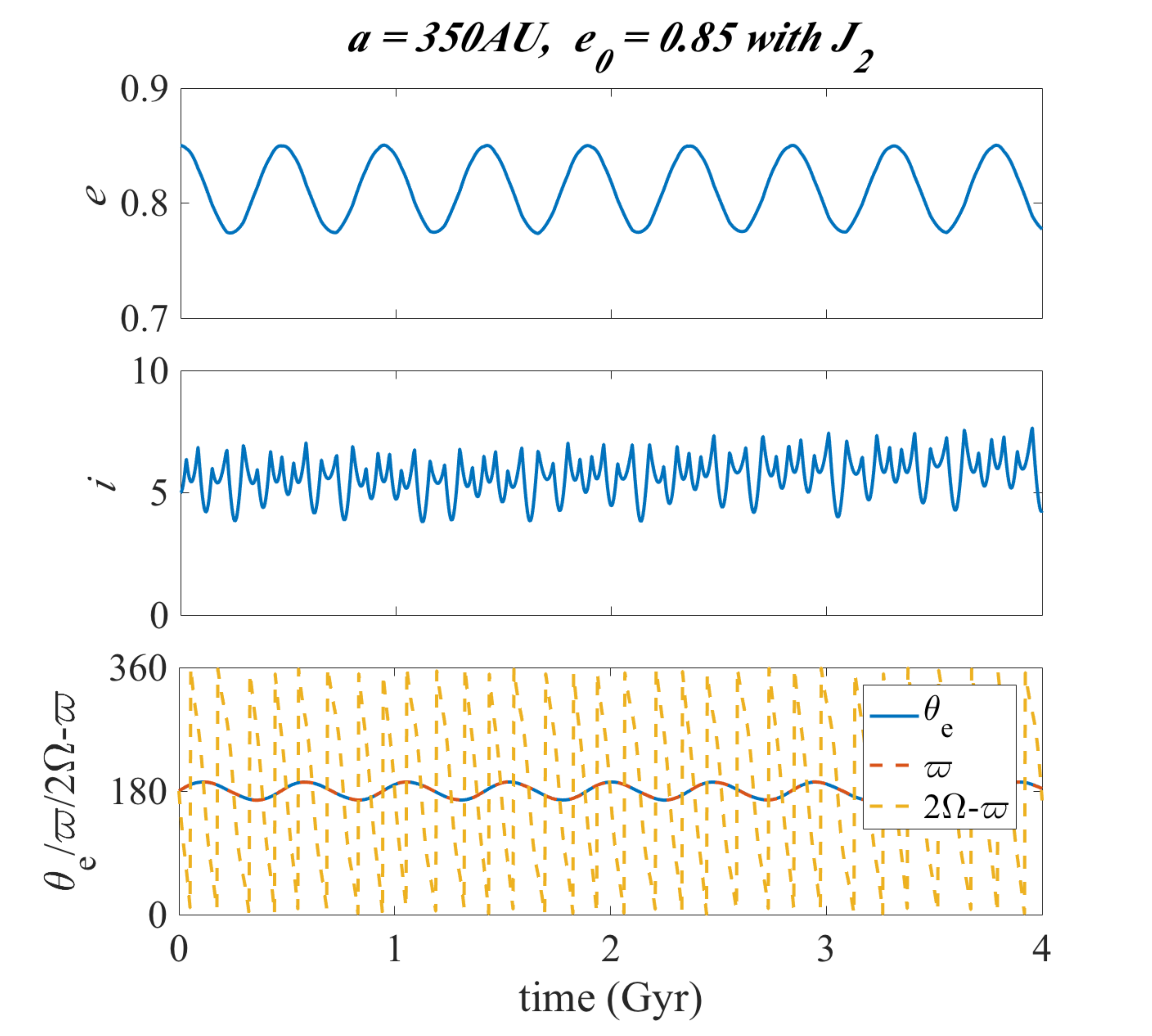}
\includegraphics[width=0.5\textwidth]{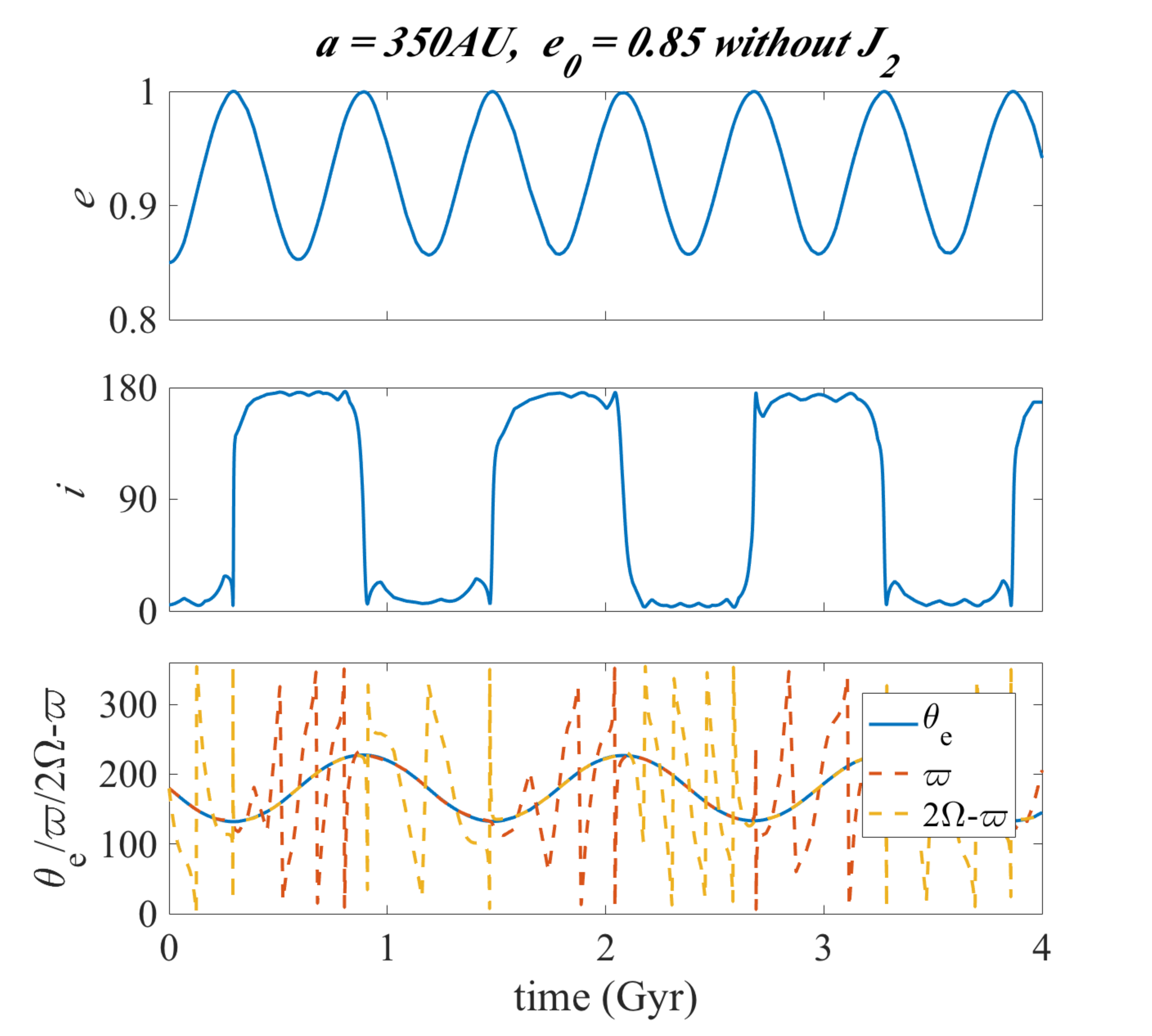}
\caption{Evolution of TNO in region 2), where $J_2$ precession timescale is similar to the libration timescale of $\Delta\theta_e$ due to perturbations from Planet Nine. The $\Delta\theta_e$ still librates with the presence of $J_2$ term, while the flip of the orbit is suppressed. \label{f:secincl_traj2}}
\vspace{0.1cm}
\end{figure*}

Figure \ref{f:secincl_traj2} illustrates the evolution in region 2), where the semi-major axis of the TNO is larger, and thus the $J_2$-precession from the inner four giant planets becomes comparable to the perturbation from the outer Planet Nine. Similar to Figure \ref{f:secincl_traj1}, the left panel shows the case with $J_2$ term and the right panel shows the case without the $J_2$ term. The $J_2$ precession timescale is similar to the $\Delta\theta_e$ libration timescale (as shown in Figure \ref{f:flipsec}). The libration of $\Delta\theta_e$ is not suppressed when the $J_2$ term is included, however the flip of the orbit is suppressed in this case. Orbits lying in this region are detectable with $r_p\lesssim 100$ AU, and they contribute to the alignment of the orbits around $\Delta\theta_e \sim 180^\circ$.

\begin{figure*}[ht]
\includegraphics[width=0.5\textwidth]{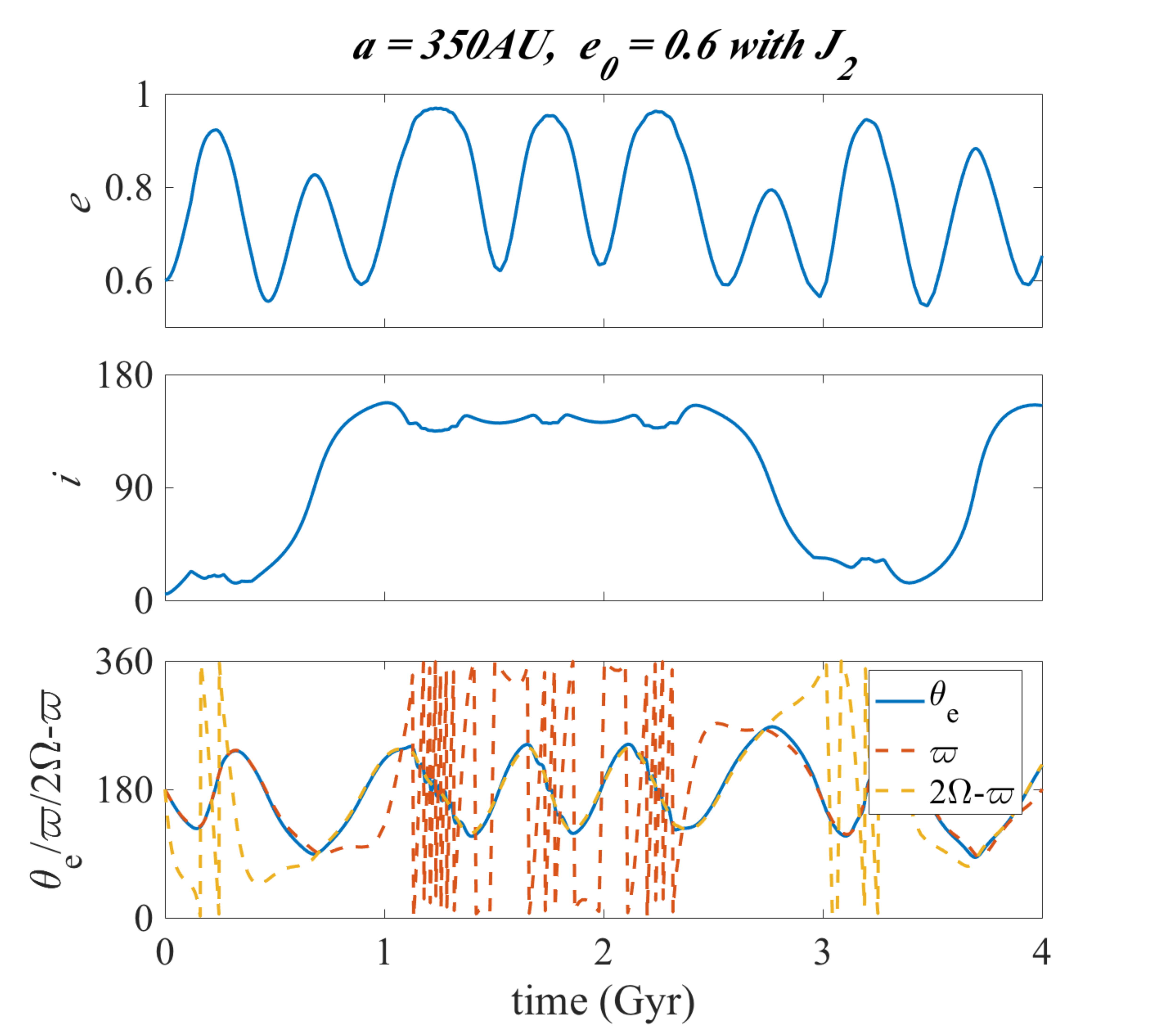}
\includegraphics[width=0.5\textwidth]{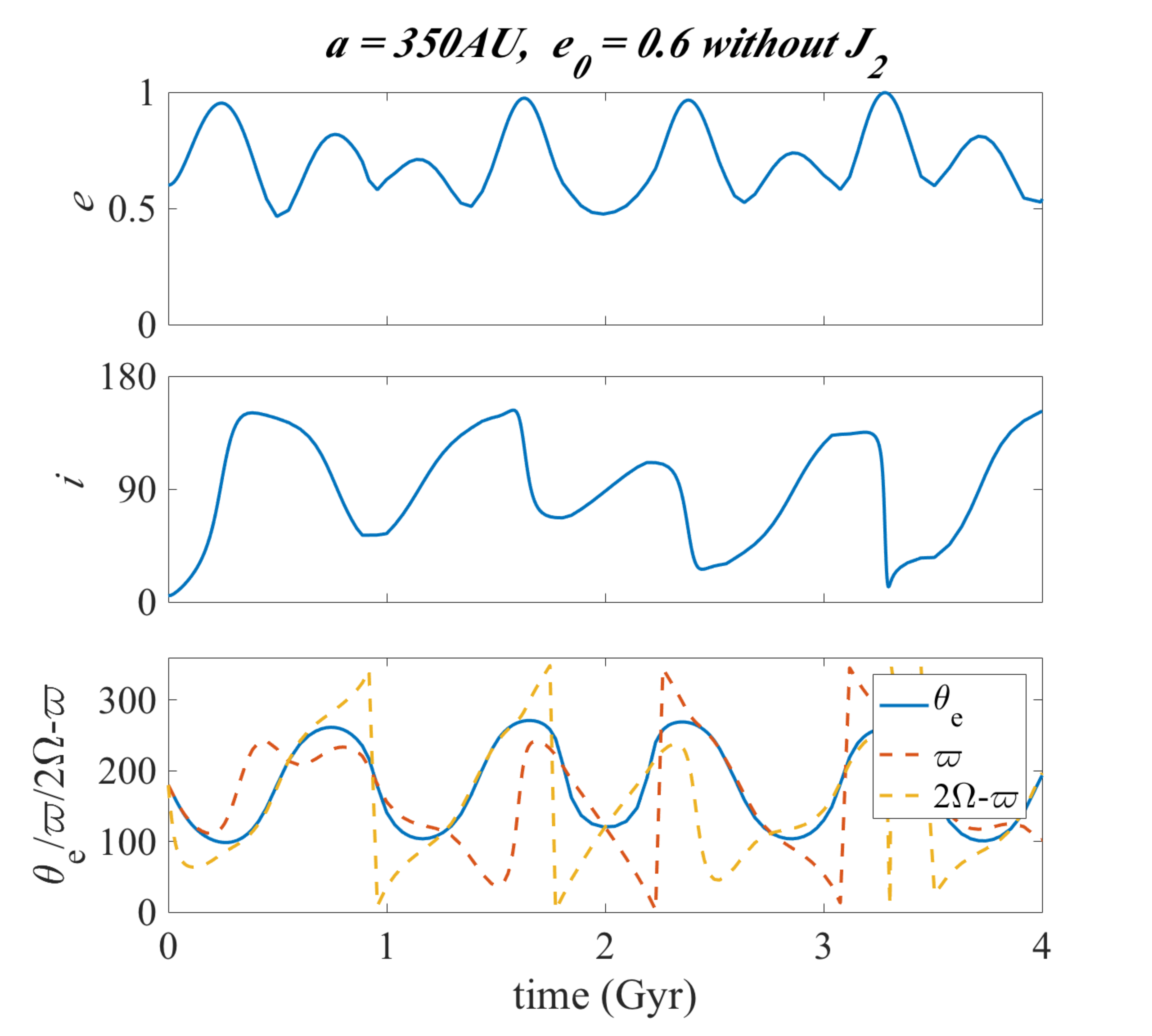}
\caption{Evolution of TNO in region 3), where $J_2$ precession timescale is longer than the libration timescale of $\Delta\theta_e$ due to perturbations from Planet Nine. Both the libration of $\Delta\theta_e$ and the flip of the orbit are not suppressed. \label{f:secincl_traj3}}
\vspace{0.1cm}
\end{figure*}

Finally, Figure \ref{f:secincl_traj3} illustrates the evolution in region 3), where the eccentricity of the TNO is lower, and the perturbation from the outer Planet Nine is more dominant. Again, the left panel shows the case with the $J_2$ term and the right panel shows the case without the $J_2$ term. The $J_2$ precession timescale is lower than the $\Delta\theta_e$ libration timescale (as shown in Figure \ref{f:flipsec}), and neither the libration of $\Delta\theta_e$ nor the flip of the orbit are suppressed when the $J_2$ term is included. Orbits lying in this region contribute to the flipped orbits. 

In addition, the TNO could stay in the high inclination regime for many cycles of low amplitude $2\Omega-\omega$ libration, and $\varpi$ circulates during this high inclination phase. This is similar to the N-body results presented in Figure \ref{f:Nbdhighi}, but this is different from the particles selected in Figure 10 of \citet{BatyginMorbi17}, where the particles fall back to low inclination regime quickly and $\varpi$ is still confined. We note that there are also particles which only spend one libration cycle at high inclination in our simulation, similar to Figure 10 of \citet{BatyginMorbi17}.

Interestingly, we notice that orbits within $30\lesssim r_p\lesssim 100$ AU and $a\gtrsim 200$ AU inside the solid and dashed black lines in Figure \ref{f:flipsec} do not flip and the geometrical longitude of pericenter $\Delta\theta_e$ still librates in the presence of $J_2$ term, since two timescales are comparable to each other. These TNOs lie within our selection criteria based on observational limits, and they lead to the alignment of the low inclination TNO orbits with $\Delta\theta_e\sim 180^\circ$, as shown in the N-body and secular simulations discussed in the previous sections. 

\begin{figure}[h]
\begin{center}
\includegraphics[width=3.4in]{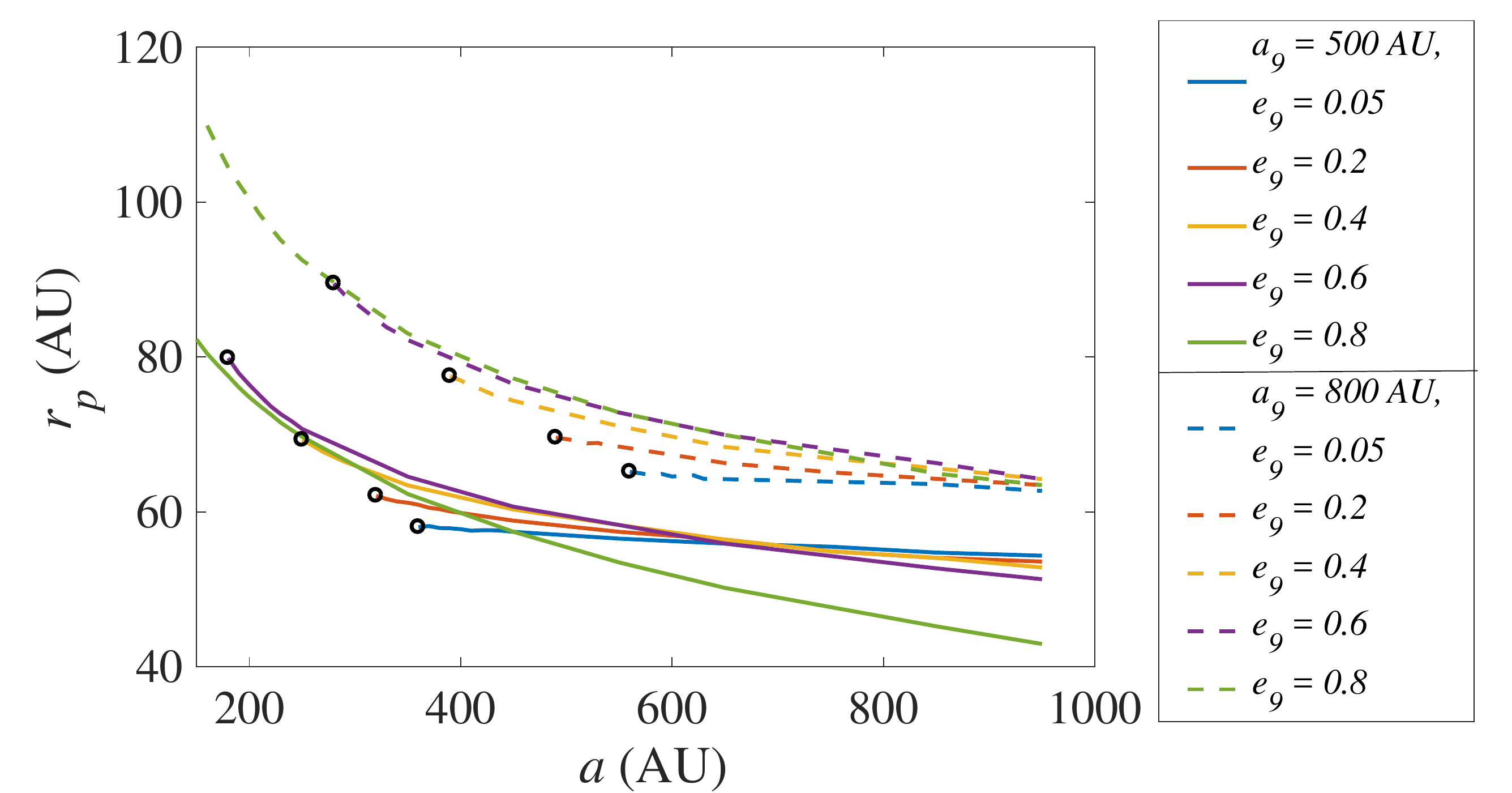} 
\caption{Fixed point pericenter distance v.s. TNO semi-major axis for different Planet Nine properties. The solid line represents the fixed points due to a $a=500$ AU Planet Nine, and the dash lines represent that due to a $a=800$ AU Planet Nine. The black circles represent the critical minimum semi-major axis when the libration regions appear. The libration region around $\Delta\varpi = 180^\circ$ arises in pericenter distance for closer TNOs when Planet Nine's orbit is more eccentric, and this libration region disappears when Planet Nine is circular and farther. \label{f:fixpt}}
\end{center}
\end{figure}

The parameter space of clustered low inclination TNOs correspond to the central libration region in the plane of $(\Delta\varpi,r_p)$ (e.g., Figure \ref{f:Sec_epo}), near $\Delta\varpi=180^\circ$. Thus, characterizing the dependence of the libration region on the properties of Planet Nine can help constrain the possible outer planet based on the detected clustering of low inclination orbits. 

In Figure \ref{f:fixpt} we show the corresponding pericenter distances of the fixed points as a function of semi-major axis, for different Planet Nine orbital parameters. The fixed point of the libration region locates at $\Delta\varpi = 180^\circ$, and we numerically calculate the pericenter distances of the fixed point at different semi-major axes, by searching for eccentricity corresponding to the maximum energy when $\Delta \varpi = 180^\circ$. 

The libration region exist around the fixed points. The secular resonances around $\Delta\varpi = 180^\circ$ disappears for low TNO semi-major axes (e.g., upper left panel of Figure \ref{f:Sec_epo}). We marked the minimum semi-major axes when the libration region appears by black circles. There is no fixed points when semi-major axis is small, in particular for wider and more circular Planet Nine's orbit.
In other words, when the orbit of Planet Nine is more circular, the anti-aligned libration appears only for TNO with larger semi-major axes. 
In addition, when Planet Nine is more distant, it also requires a more eccentric Planet Nine orbit to produce the alignment for the closer in TNOs. 
This is consistent with the N-body results shown in \citet{Brown16}, where Planet Nine favors a more eccentric orbit around $a\sim 500$ AU to produce the clustering.

\subsection{Inclination Distribution based on N-body results}
As shown in the previous section, the flip of the orbits depends on the existence of the libration region of $\Delta\theta_e\sim 180^\circ$, and this itself depends on the properties of Planet Nine. Thus, the retrograde TNO orbits and the signatures of the inclination distribution can provide valuable constraints on the properties of any possible outer planet. In this section, we illustrate the dependence of the inclination distribution using N-body simulations. First, we present the inclination distribution of the TNOs, under perturbations from a Planet Nine with $m_9 = 10 M_{\oplus}$, $a_9 = 500$ AU, $e_9 = 0.6$, $i_9 = 3^\circ$, similar to that included in \textsection \ref{s:Nbd}. We find that some high inclination objects have small pericenter distances ($r_p<30$ AU). Thus, to obtain a more accurate TNO inclination distribution, we re-run the N-body simulation including Jupiter, Saturn and Uranus as massive point particles, instead of their equivalent $J_2$-potential.

\begin{figure}[h]
\begin{center}
\includegraphics[width=3.4in]{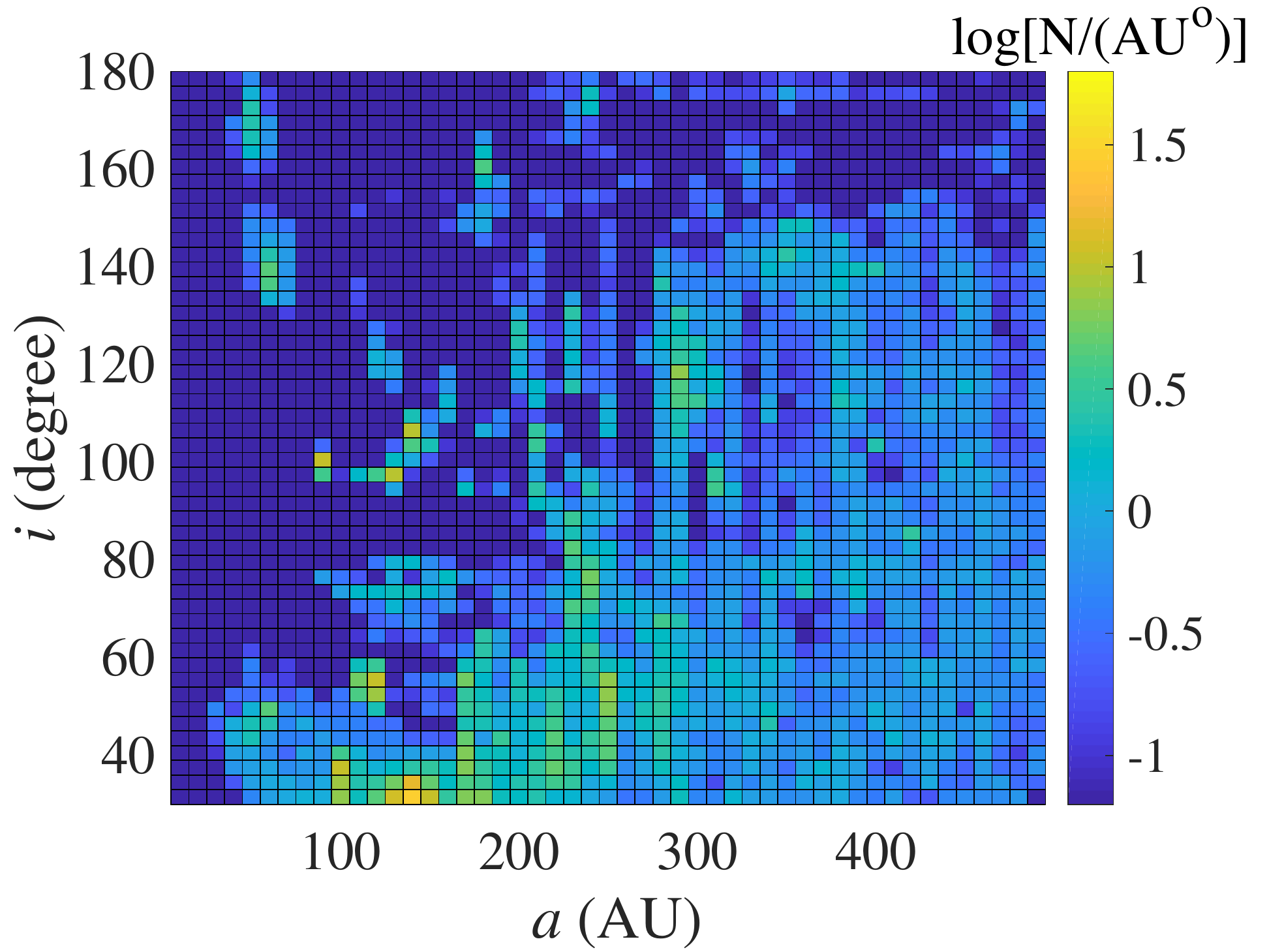}\\
\includegraphics[width=3.4in]{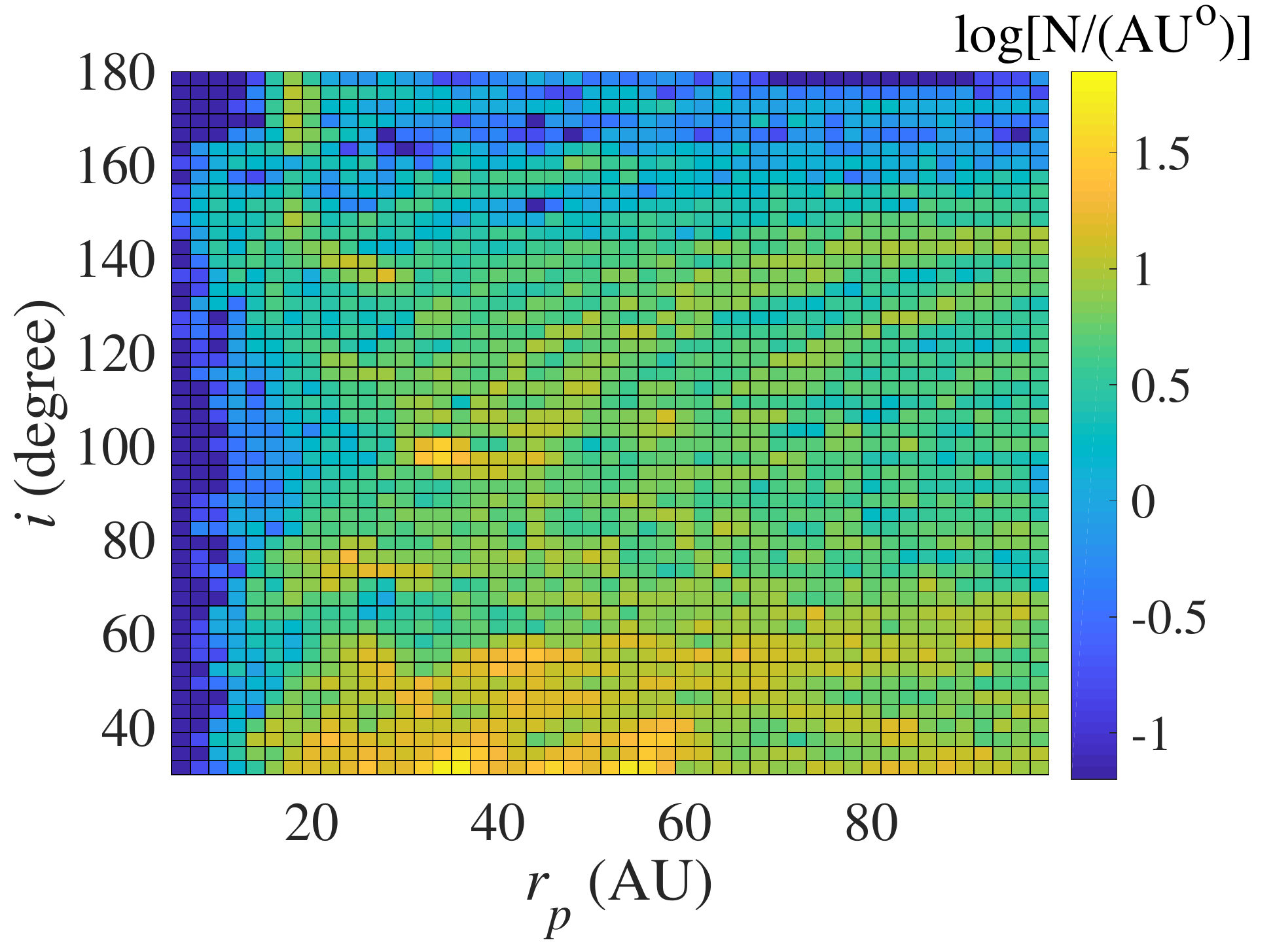}
\caption{Density map in the plane of inclination versus semi-major axis (upper panel), and in the plane of inclination versus pericenter distance (lower panel) for TNOs with $i>30^\circ$ $t>3$Gyr, based on N-body simulation including the four giant planets as point massive particles. The upper panel selects TNOs with $r_p<80$ AU. Inclination can only be flipped above $90^\circ$ when $a\gtrsim30$ AU. There is an over-density in the high inclination regime near $150^\circ$ when $a\lesssim 100$ AU, and the inclination distribution becomes more uniform when $a \gtrsim 300$ AU. When $r_p < 20$ AU, there is a deficit of TNOs near $\sim90^\circ$. \label{f:incdis}}
\end{center}
\end{figure}

In Figure \ref{f:incdis} we show the scatter in inclination as a function of semi-major axis from our N-body simulations. 
The colors represent the density of the TNOs (log of the number of TNOs per AU per degree). Particles are included with $t>3$ Gyr, plotted at snapshots taken every $1$Myr. To focus on the high inclination population, we selected TNOs with $i>30^\circ$, and to select those more likely to be observable, we require $r_p<80$ AU. 
We find that the TNOs become retrograde ($i>90^\circ$) when $a\gtrsim 30$ AU. This is closer than in the secular approximation since the semi-major axes of the particles can drift due to close encounters with Neptune and due to overlap of mean motion resonances, both of which are not captured in the secular approximation. 

In addition to the limiting semi-major axis for which TNO orbits can be flipped, the inclination distribution shows additional interesting features. As shown in the density map in Figure \ref{f:incdis}, within ${\it a} \lesssim 100$AU, there is an overdensity of inclined orbits around $150^\circ$. This is due to the nature of the flips when the semi-major axes is small, which crosses over $90^\circ$ fast, and then spends more time near $\sim 150^\circ$ before flipping back. As $\it a$ continues to increase ($a\gtrsim 300$ AU), the distribution of the inclination is more uniform.


The inclination distribution depends on the properties of Planet Nine. For instance, if Planet Nine is less massive ($\sim M_{\oplus}$), the perturbations on the TNOs are weaker, and the $J_2-$precession due to the inner planets dominates. Thus, it is difficult to flip the TNO orbits, and none of the TNO orbits is flipped over $90^\circ$.

\begin{figure}[h]
\begin{center}
\includegraphics[width=3.4in]{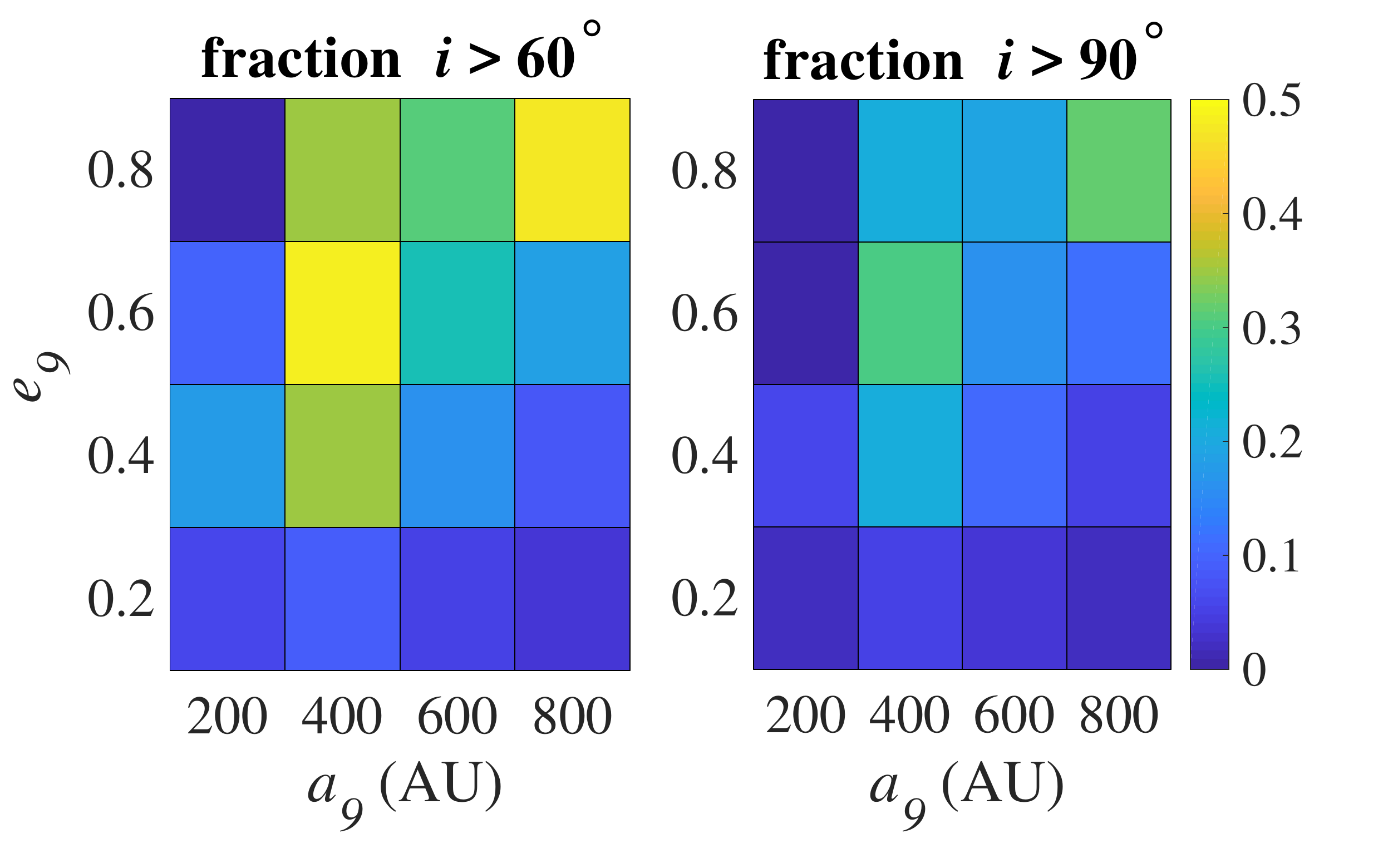}
\caption{Fraction of TNOs with inclination over $60^\circ$ (left panel) and over $90^\circ$ (right panel) at $4$Gyr, for different Planet Nine orbital parameters. It is more likely to flip the TNO orbits for more eccentric Planet Nine orbit with semi-major axis $\sim 400-800$ AU for a $10 {\rm M_{\oplus}}$ Planet Nine.\label{f:icrit}}
\end{center}
\end{figure}

Since the flips are associated with the libration of the geometrical longitude of pericenter ($\Delta\theta_e$) around $180^\circ$, and since this only occurs for Planet Nine orbits that have high eccentricity and are wide ($a\sim 400-800$ AU: see Figure \ref{f:fixpt}), we expect that the TNO orbits can flip only for such Planet Nine orbits. To test this, we performed more N-body simulations to calculate the fraction of TNOs that can have their inclination excited using different Planet Nine properties. We include a $10M_{\oplus}$ Planet Nine and varied the Planet Nine orbital parameters in these simulations. 

Figure \ref{f:icrit} shows the fraction of TNOs which have inclination over $60^\circ$ and $90^\circ$ at $4$Gyr. Specifically, it is more likely to flip the TNO orbits for an eccentric Planet Nine ($e\gtrsim 0.4$) with large semi-major axis of $\sim 400-800$ AU. This agrees with our expectation based on the appearance of the libration region around $\Delta\theta_e\sim 180^\circ$. 
The recent discovery and dynamical analysis of high inclination objects, e.g. 2015 ${\rm BP_{519}}$ \citep{Becker18}, supports the existence of possible outer planets. Future surveys on the inclination distribution of Centaurs can help further constrain parameters of possible outer planets \citep[e.g.,][]{Petit17, Lawler18}.

\section{Conclusion}
\label{s:dis}
A deeper understanding of the dynamics involved in interactions between TNOs and the hypothesized Planet Nine can help characterize the orbits of possible planets in the outer Solar System, and facilitate the observational detection of such planets. The secular interactions with Planet Nine play an important role in shaping the orbits of the TNOs.  We obtained the averaged Hamiltonian for a Planet Nine in the plane of the ecliptic but allowing TNOs small initial inclinations, and we compared our secular results with N-body simulations, extending beyond the secular effects in the exact coplanar configurations studied in the literature \citep{Beust16,BatyginMorbi17,Hadden18}. Our findings are listed as below:

\begin{itemize}
\item For systems in which Neptune is a fully interacting point mass and Planet Nine is slightly inclined ($\sim 3^\circ$), many TNOs survive \emph{outside} of MMR for $4$ Gyr.
\item The clustering of TNO orbits near $\Delta \varpi (\Delta \theta_e) \sim 180^\circ$ can be produced by secular interactions, similar to the coplanar results \citep{Beust16, BatyginMorbi17, Hadden18}. In addition, the clustering near $\Delta\varpi\sim 180^\circ$ is stronger if one selects only the low inclination ($<40^\circ$) TNO population. The clustering in $\Delta\omega$ and $\Delta\Omega$ for the low inclination population is weak.
\item For the high inclination population ($60^\circ<i<120^\circ$), there are clusters for the TNOs within the detection limit ($r_p < 80$ AU) near $\Delta\theta_e \sim 90^\circ$ and $270^\circ$, while the clusters in $\Delta\varpi$ are weaker. In addition, we can see clusters in $\Delta\omega$ and $\Delta\Omega$, around $\Delta\omega \sim 0^\circ$ and $180^\circ$ and around $\Delta \Omega \sim 90^\circ$ and $270^\circ$ for the low pericenter TNOs. The orbital alignment of the high inclination population can also be produced by secular effects.
%
%
\item Eccentric ($e\gtrsim 0.4$) Planet Nine beyond $a_2\gtrsim 400$ AU can facilitate flips of TNO orbits over $90^\circ$ due to secular interactions, analogous to the near coplanar flips in the hierarchical configurations \citep{Li14a}. During the flips, the geometrical longitude of pericenter ($\Delta\theta_e$) of TNOs librates. 
\item
$J_2-$precession caused by the inner giant planets can suppress the aforementioned flips TNO orbits, and lead to the clustering of low inclination TNOs around $\Delta\varpi (\Delta\theta_e)\sim180^\circ$ with $30<r_p<80$ AU. 
\end{itemize}

Configurations involving higher inclination Planet Nine may explain the clustering in $\Delta\omega$. This is consistent with the results of \citet{Batygin16} and \citet{Brown16}. A detailed analysis of the dynamics of a highly inclined Planet Nine is a promising future direction, which is beyond the scope of this paper. 

Future observations with more detailed orbital features of TNOs and in particular of the high inclination population will provide valuable information to constrain the properties of any outer planets in our Solar System. In addition to the orbital distribution of TNOs, it is likely that ecliptic comets can also help predict and/or constrain the properties of Planet Nine, for which \citet{Nesvorny17} suggests that the inclination distribution of Ecliptic Comets would be wider than the observed one for a range of Planet Nine configurations.

\section*{Acknowledgments}
GL thanks Konstantin Batygin and Andrea Milani for helpful discussions. The authors would like to thank the anonymous reviewer for constructive comments, which improved the quality of the paper.

\appendix
\section{A. Secular Non-hierarchical Three-body Interactions}
\label{APP:SEC1}
\begin{figure*}[ht]
\includegraphics[width=6.8in]{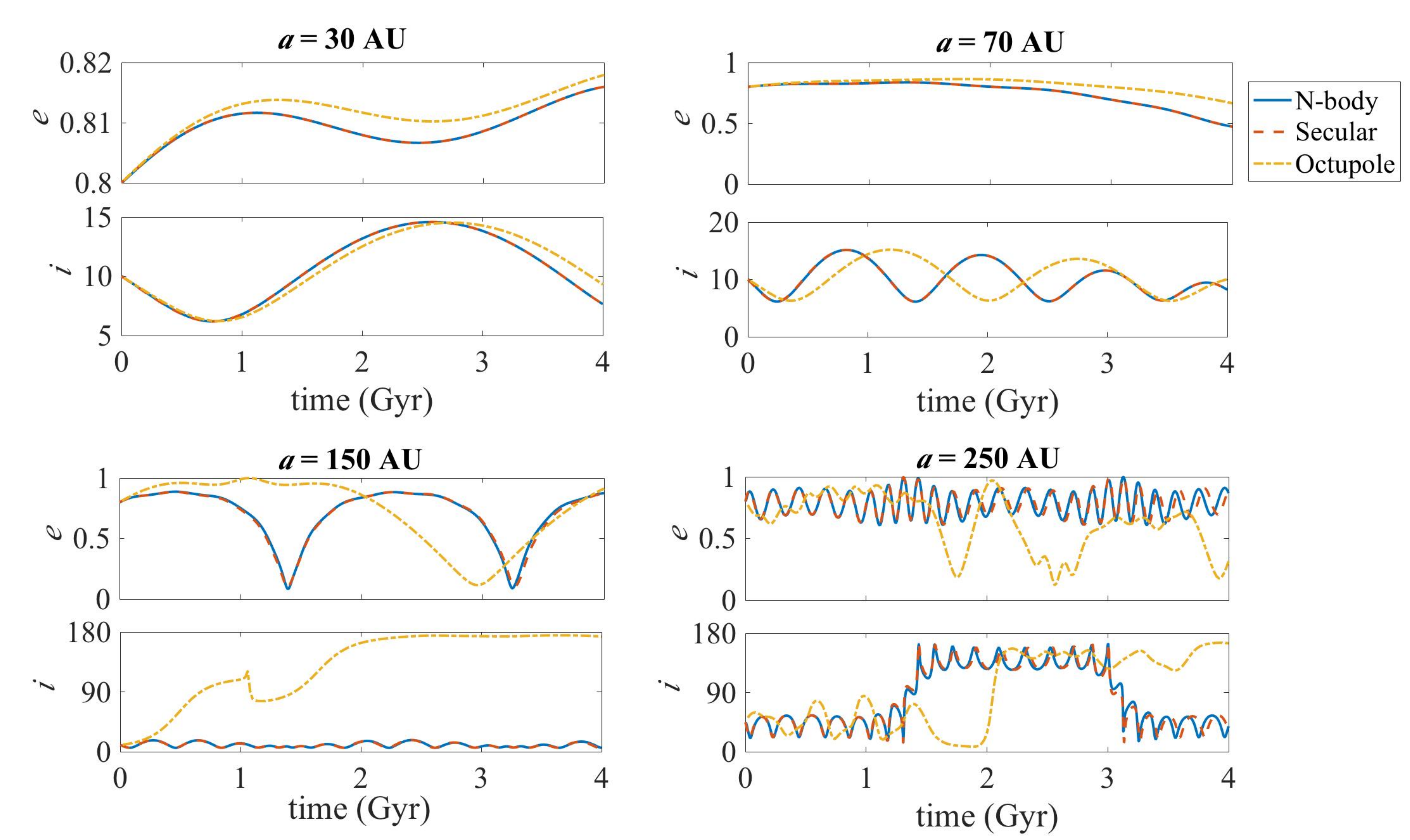}
\caption{Comparisons between N-body (solid-blue) and secular results (secular-averaging in dashed red lines, Octopole expansion in dot-dashed yellow lines). 
The secular direct-averaging results agree well with the N-body results, but the octupole results deviate from the N-body results significantly when $a \gtrsim 150$ AU ($\epsilon \gtrsim 0.28$).
\label{f:secNbd}}
\vspace{0.1cm}
\end{figure*}

To verify that the secular approach serves as a good approximation, we illustrate in Figure \ref{f:secNbd} results that use both direct secular averaging (dashed red lines) and N-body integrations (solid blue lines). We use the high accuracy adaptive time stepping integrator ``ias15'' \citep{Rein15} from the {\sc rebound} package to obtain the N-body trajectories, since the eccentricities of the particles approach close to unity in the three-body interactions without the $J_2$ terms. 
We also include results using the Hamiltonian at the octupole level of expansion in the semi-major axes ratio (dotted yellow lines), which characterize the Kozai-Lidov oscillation \citep{Lidov62, Kozai62}, and is a good approximation when the hierarchical parameter $\epsilon = (a/a_9)e_9/(1-e_9^2) < 0.1$ \citep[for a review, see][]{Naoz16}. 

As shown in Figure \ref{f:secNbd}, the hierarchical limit is still a good approximation when $a = 30$ AU ($a_9 = 500 AU$, $e_9 = 0.6$, $\epsilon = 0.056$), and the octupole results deviate from the N-body results when $a \gtrsim 70$ AU ($\epsilon \gtrsim 0.13$). The secular results all agree very well with the N-body results over the $4$ billion year evolution. When $a\gtrsim 350$ AU, the trajectories are chaotic.

\section{B. Some Details on N-body Results}
\label{APP:SEC2}
\subsection{B1. Presence of MMRs}
\begin{figure}[ht]
\begin{center}
\includegraphics[width=6.in]{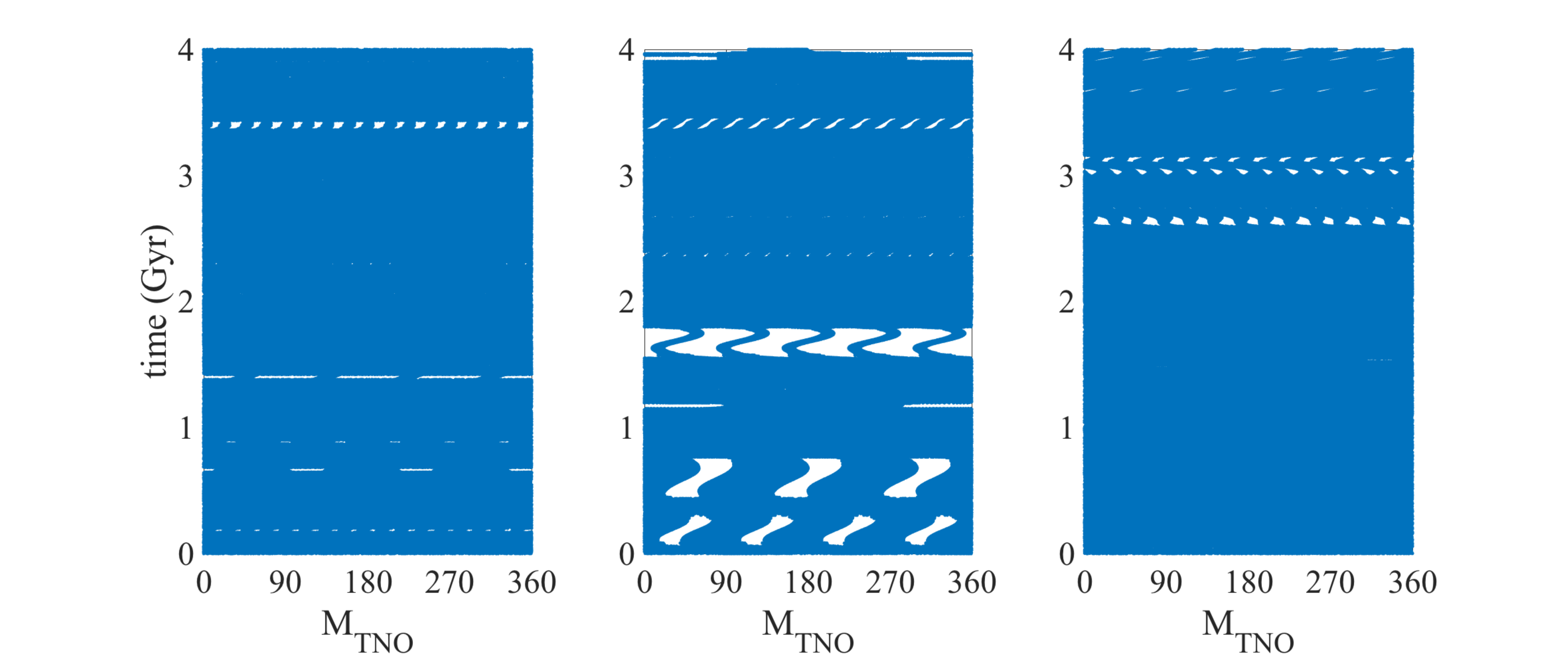} 
\caption{The mean anomaly of TNO when that of Planet Nine is zero versus time for three arbitrarily selected TNOs. Regular blank regions over time indicates the existence of mean motion resonances. For the three illustrated example, the TNOs in the left and the right panel only stay in MMR for a negligible fraction of time, while the TNO in the middle panels stay in MMR for $\sim 20-30\%$ of time. All of the TNOs show that pure secular interactions play an important role in their orbital dynmical evolution.
\label{f:res}}
\end{center}
\end{figure}

In this section, we present the detailed TNO orbital evolution starting in the near coplanar configuration as mentioned in Section \ref{s:Nbd} based on the N-body results. First, we illustrate the presence of high order MMR during the evolution of TNOs. As mentioned in the main text, Figure \ref{f:NbdPratio} shows that the TNOs could survive outside of the lower order mean motion resonances (MMR). To illustrate any presence of higher order MMR, we plot the mean anomaly of the TNOs when that of Planet Nine is zero in Figure \ref{f:res}. 

The evenly spaced blank region in Figure \ref{f:res} illustrate the presence of MMRs due to commensurability of between the TNO and Planet Nine. Two out of the three randomly selected TNOs spend only a negligible fraction of time in MMR, while the TNO in the middle panel spends around $\sim20-30\%$ time in MMR. All of them stay out of MMR for a large fraction of time, indicating the importance of pure secular interactions in their orbital evolution. Performing the similar analysis for 100 survived TNOs, we find that more than $\sim 80\%$ of them remain outside all MMRs for more than $\sim 80\%$ of the time.

In addition, we adopt a more systematic approach, where we cut the time series of $M_{TNO}$ at $M_9 = 0$ to 1Myr and 10Myr segments, and performed ks tests in each segment. If the ks test shows that the $M_{TNO}$ agrees with a uniform distribution, we mark the TNO to be out of MMR for the time segment. Using time segments of $1$Myr, $95\%$ of the TNOs spend more than $80\%$ of time outside of MMRs, and using time segments of $10$Myr, $85\%$ of the TNOs spend more than $80\%$ of time outside of MMRs. We note that this approach only gives a rough estimate and it is very sensitive on the segment widths.

\subsection{B2. Orbital Clustering}
\begin{figure}[ht]
\begin{center}
\includegraphics[width=3.2in]{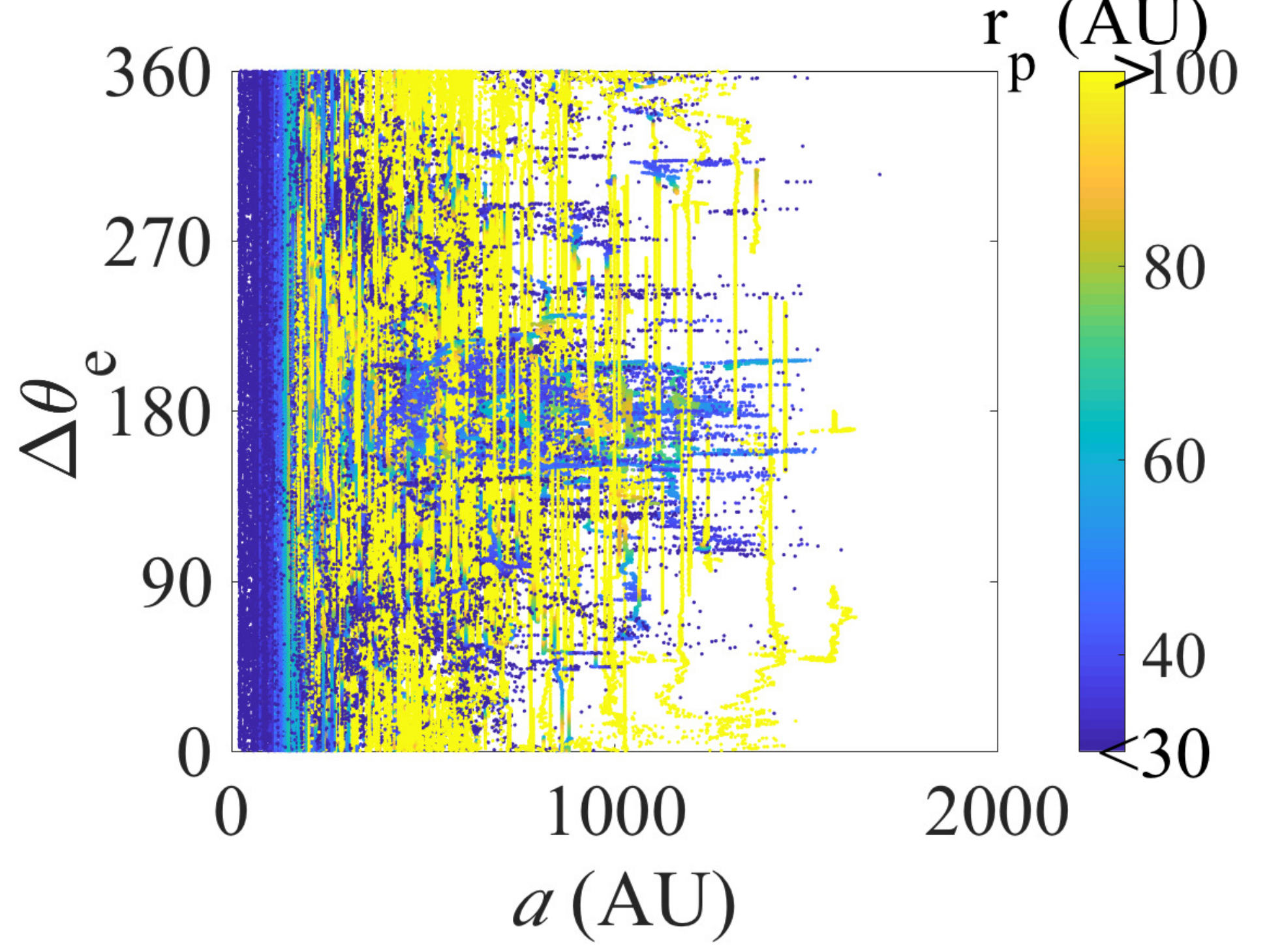} 
\includegraphics[width=3.2in]{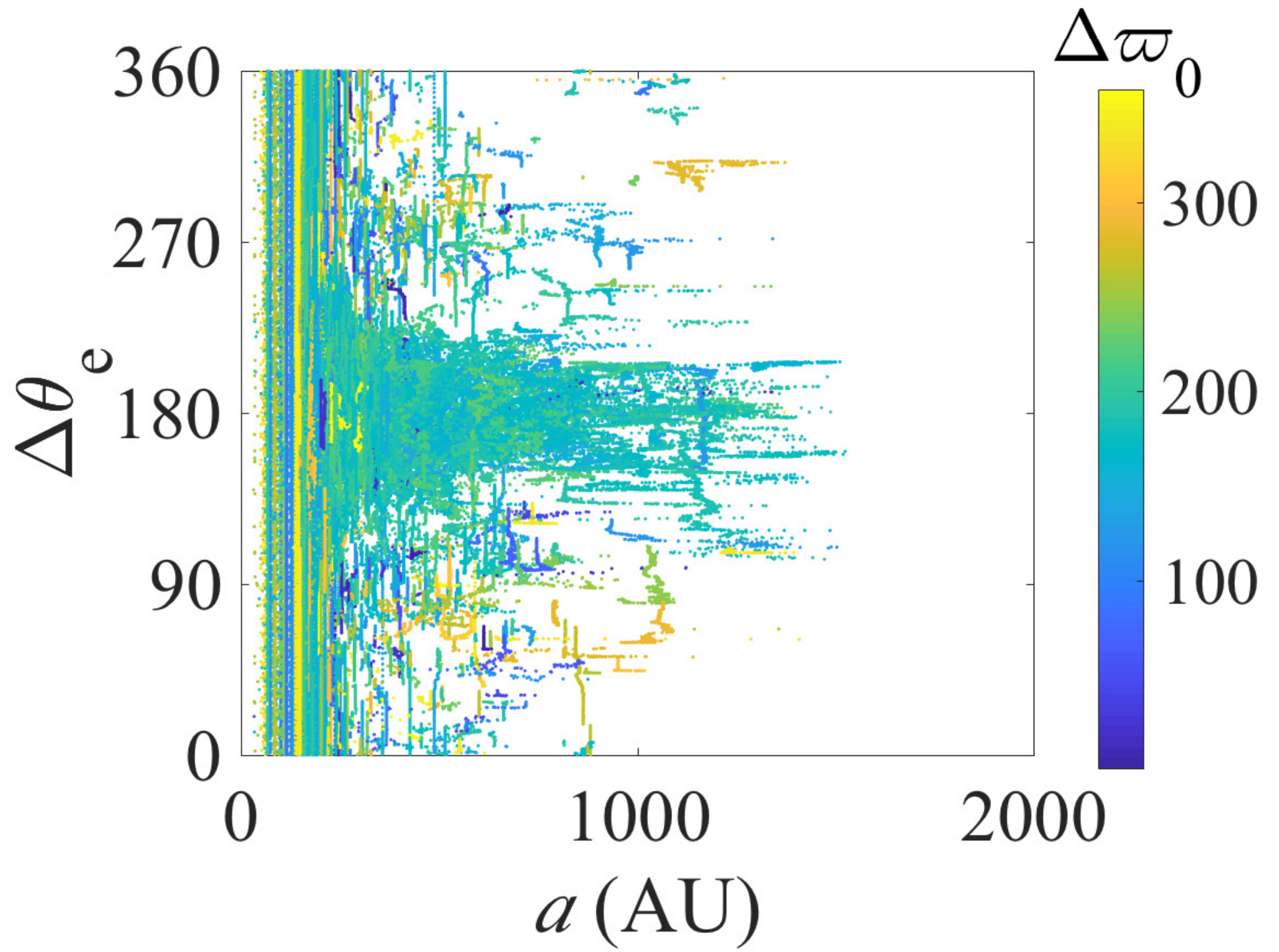} \\
\includegraphics[width=3.2in]{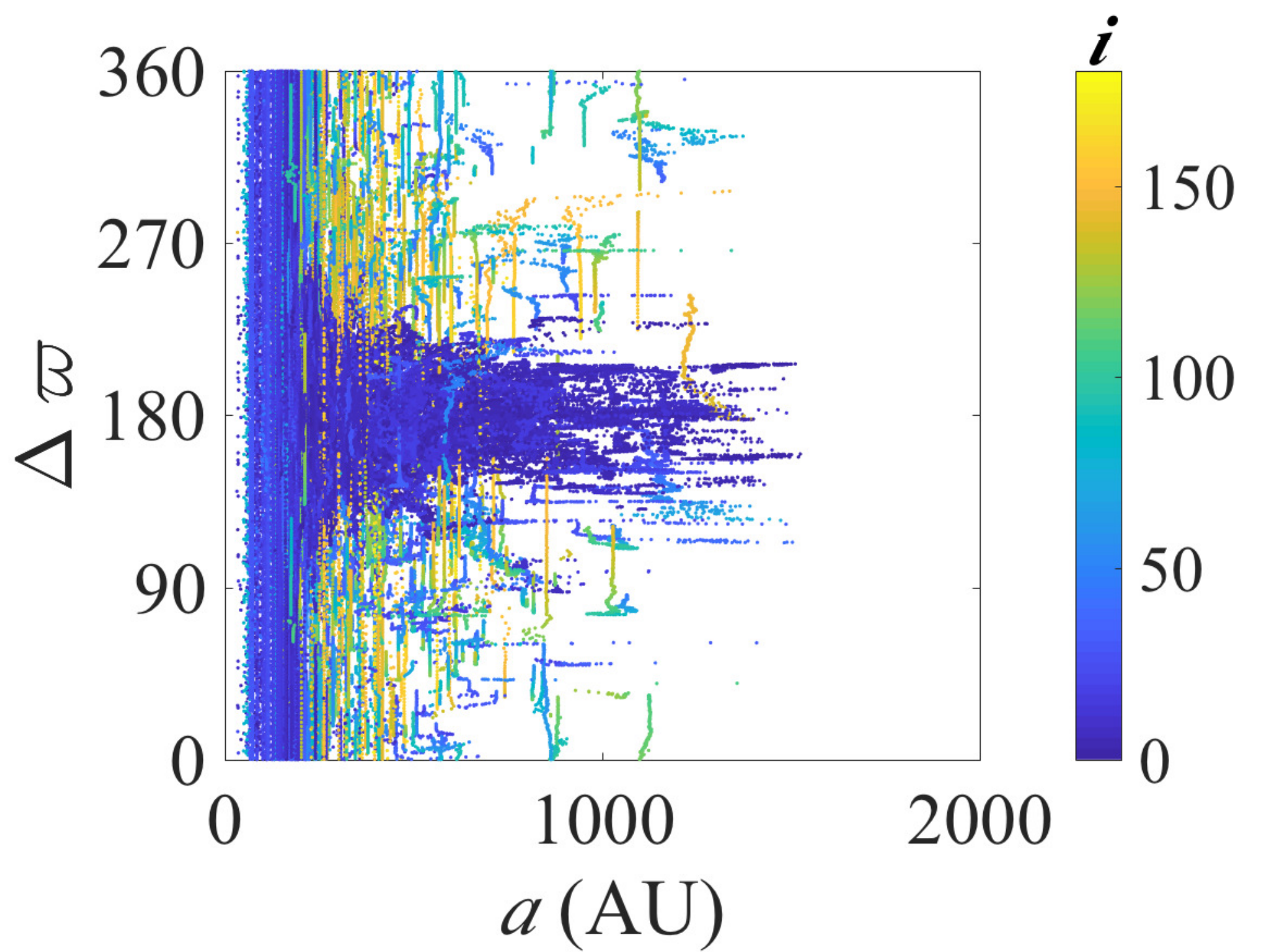} 
\includegraphics[width=3.2in]{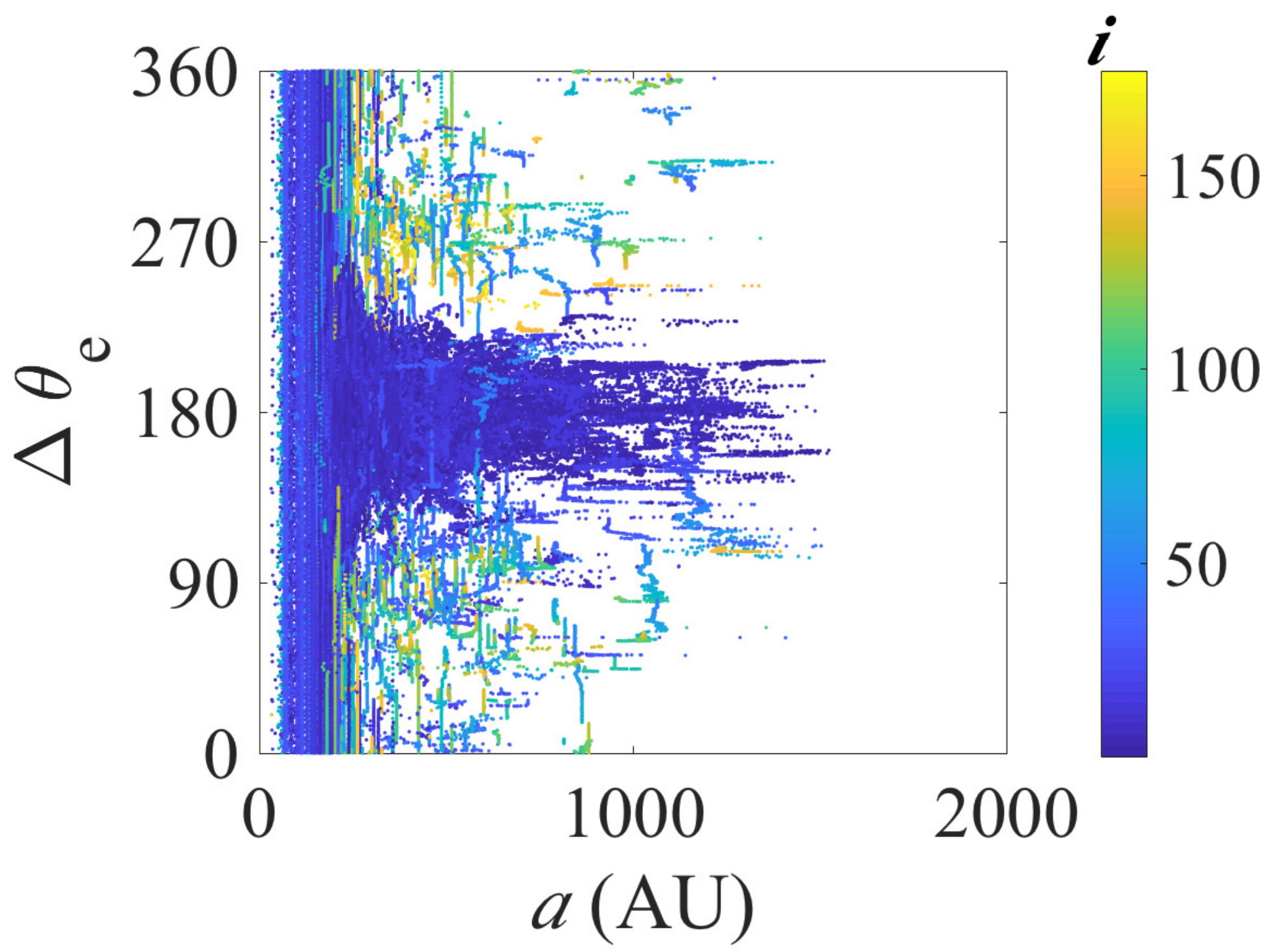} 
\caption{Alignment in the pericenter orientation of the TNOs under the perturbation of Planet Nine, Neptune and inner $J_2$ momentum caused by the inner three giant planets. Planet Nine is near coplanar with the ecliptic, and the initial test particle $\varpi$ are randomly distributed. Particles with $t>3$Gyr are selected and plotted with time-steps of 1Myr. Upper left panel: alignment in the geometrical longitude of pericenter $\Delta\theta_e$ color coded in pericenter. We include all survived TNOs with a wide range of pericenter to illustrate the overall dynamics; upper right panel: alignment in $\Delta\theta_e$ color coded in the initial longitude of pericenter, $\Delta\varpi_0$; lower left panel: alignment in $\Delta\varpi$ color coded in inclination; lower right panel: alignment in $\Delta\theta_e$ and color coded in inclination. The upper right, lower left and lower right panels only select particles with $30<r_p<80$ AU. There is a strong clustering in $\Delta\varpi\sim180^\circ$ and $\Delta\theta_e\sim180^\circ$ for the low inclination TNOs with initial $\Delta\varpi_0 \sim 180^\circ$. 
\label{f:Nbody}}
\end{center}
\end{figure}

\begin{figure}[ht]
\begin{center}
\includegraphics[width=3.2in]{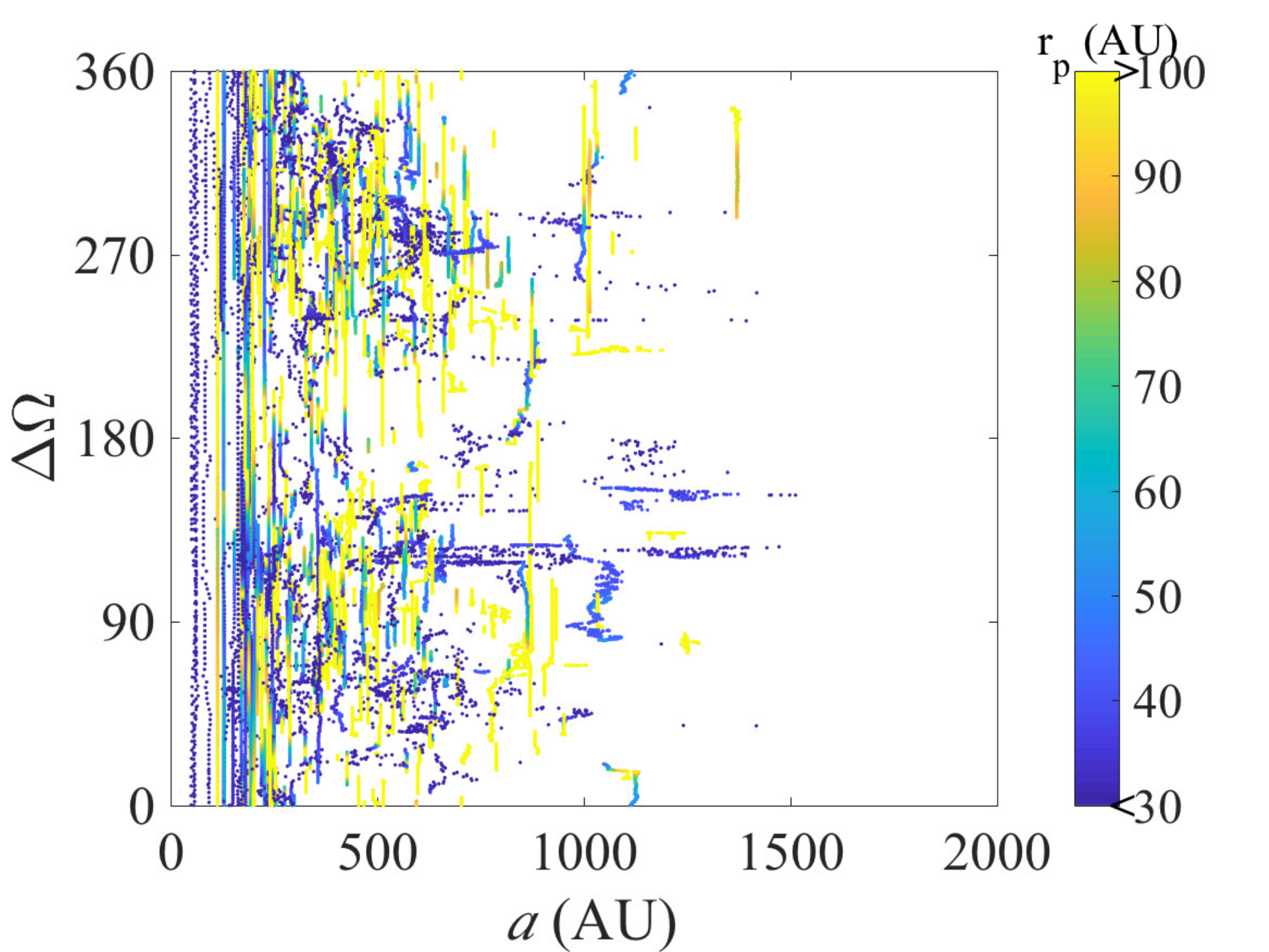} 
\includegraphics[width=3.2in]{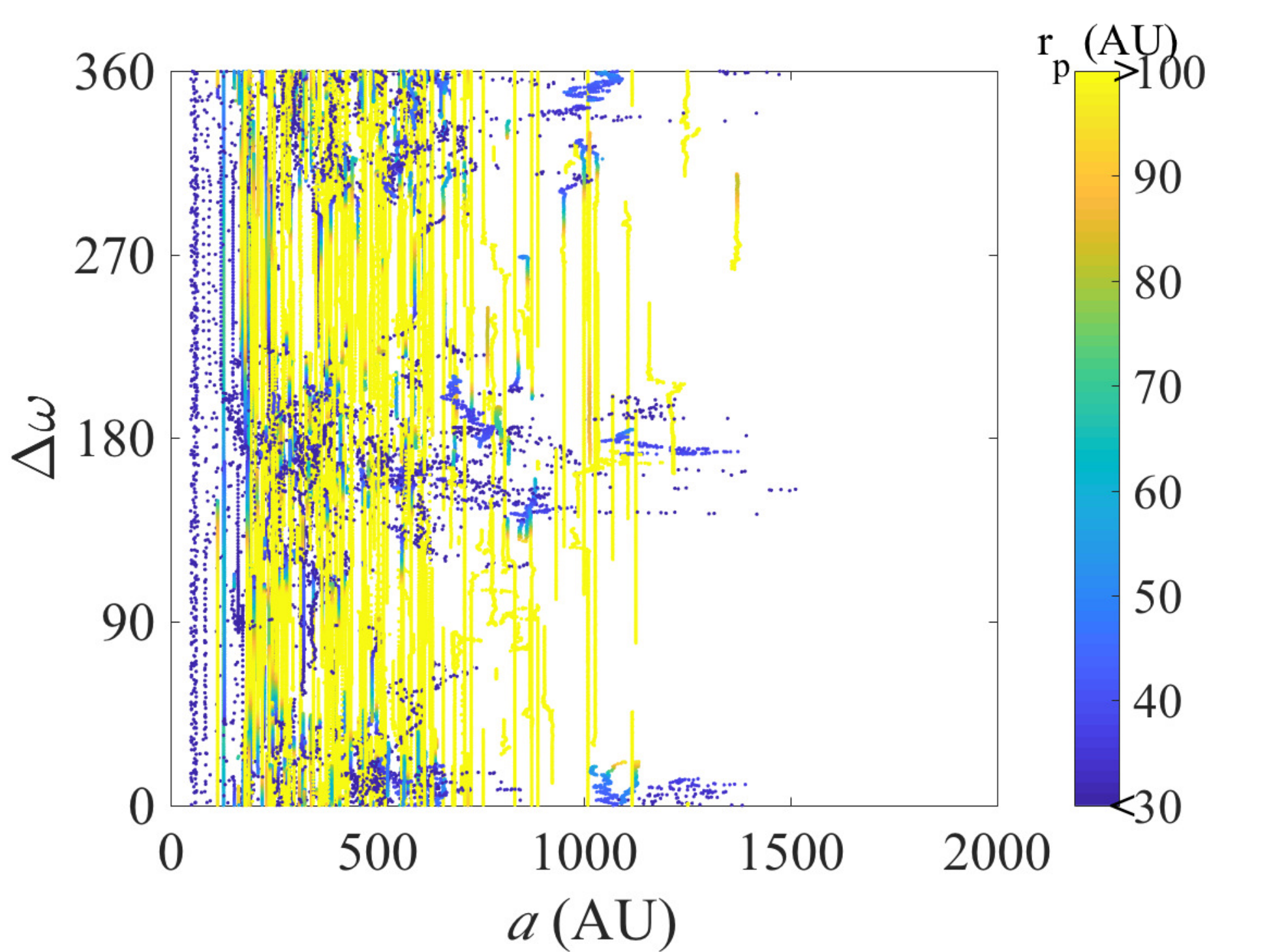}  \\
\includegraphics[width=3.2in]{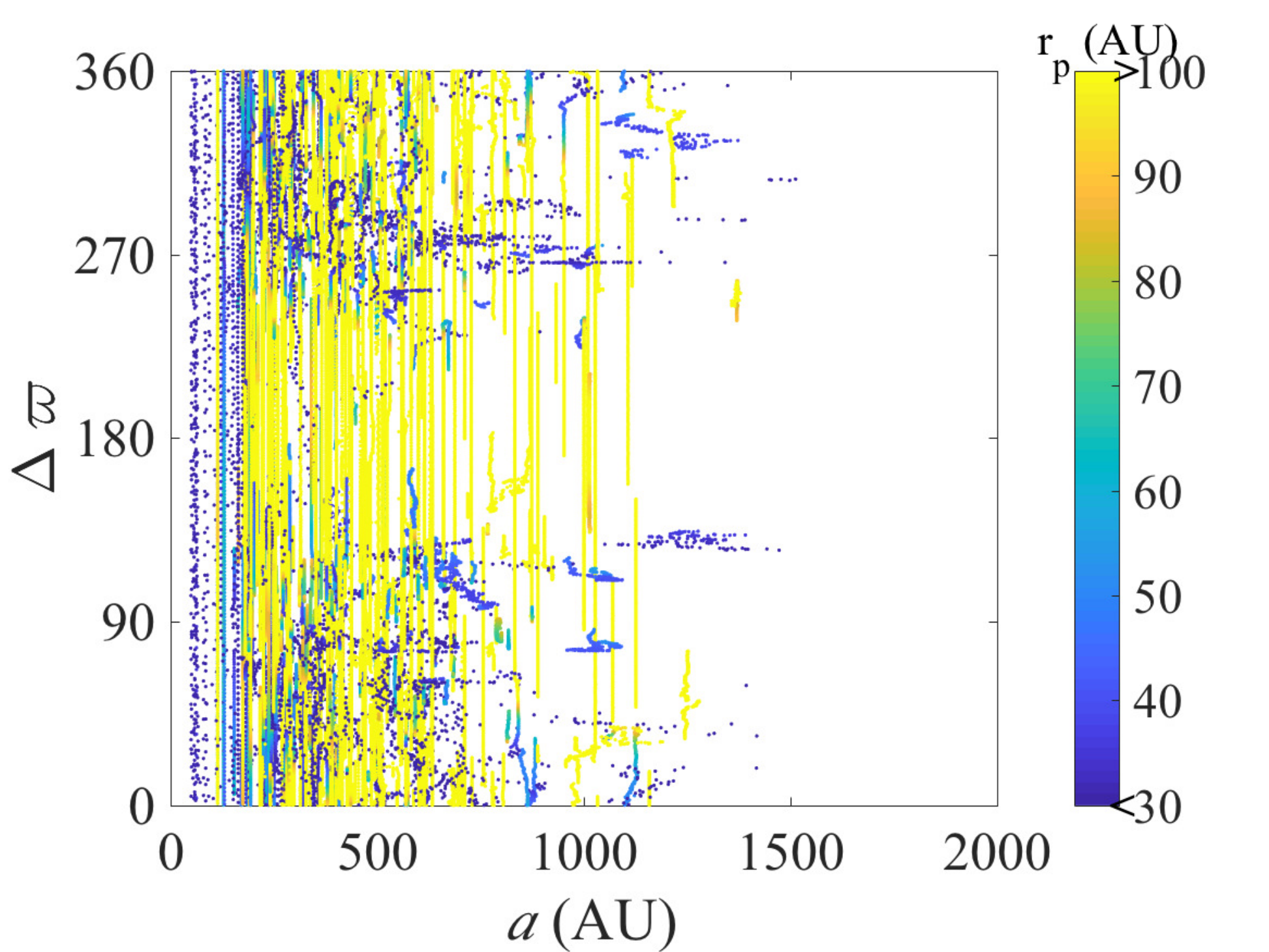}
\includegraphics[width=3.2in]{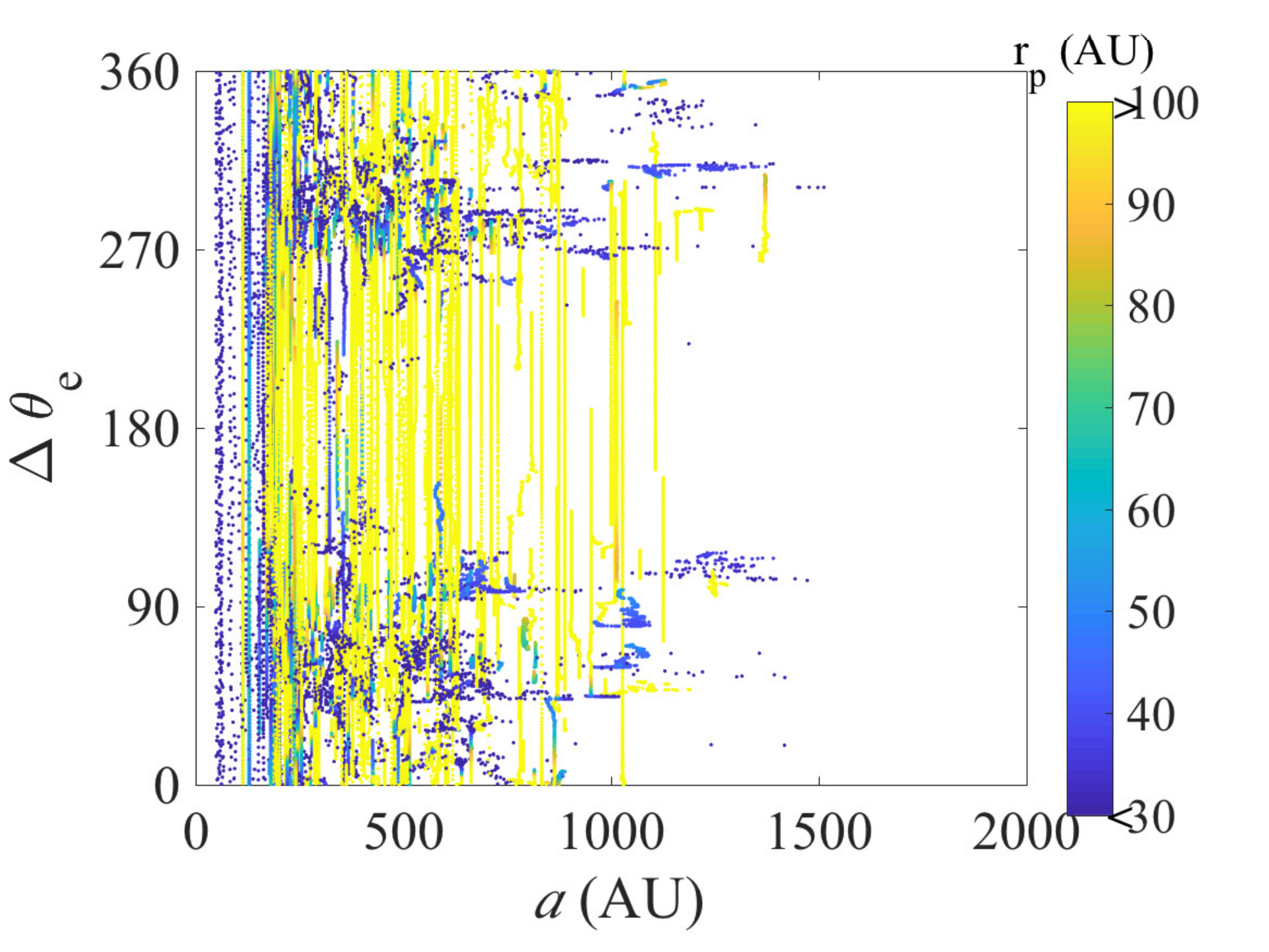}
\caption{Orbital alignment for high inclination particles ($60^\circ < i < 120^\circ$), and particles are selected if $t>3$ Gyr. There are alignments in $\Delta\Omega\sim 90^\circ \&~270^\circ$ (upper left panel), $\Delta\omega \sim 0^\circ \&~180^\circ$ (upper right panel), $\Delta\varpi\sim 90^\circ \&~\sim 270^\circ$ (lower left panel) and $\Delta\theta_e\sim 90^\circ \&~\sim 270^\circ$ (lower right panel) when the particles have small pericenter distances in the detection limit ($r_p\lesssim 80$ AU). Planet Nine lies $3^\circ$ inclined from the ecliptic, and the initial test particle $\varpi$ are randomly oriented (same as the configuration of Figure \ref{f:Nbody}). The distribution of TNOs is slightly more centrally peaked around $90^\circ$ and $270^\circ$ in $\Delta\theta_e$ comparing with $\Delta\varpi$. \label{f:Nbdhighi}}
\end{center}
\end{figure}

Next, we illustrate the clusterings of the TNO orbits in more detail, and compare with the secular results. The alignment of the test particles in pericenter ($\Delta\varpi$ and $\Delta\theta_e$) as a function of particle semi-major axes is shown in Figure \ref{f:Nbody}. 
We record the longitude of pericenter of the test particles every $1$ Myr for $t>3$ Gyr. 
To illustrate the dependence of the alignment on different parameters of the system, we include three different color-codes. 
The upper left panel of Figure \ref{f:Nbody} color-codes the TNOs in the pericenter distance, and it shows the orbital distribution of the TNOs in the plane of $(a,\Delta\theta_e)$. It illustrates that there is a clear alignment of the test-particles around the geometrical longitude of pericenter, $\Delta\theta_e \sim 180^\circ$ for small pericenter objects ($30<r_p<80$ AU), consistent with \citet{Brown16} and the observational results of  \citet{Trujillo14}. Particles with smaller pericenter distances, $r_p\lesssim 30$ AU, exhibit faster precession in the longitude of pericenter, suppressing any clustering, and the clustering is weak for longer pericenter distance $\gtrsim 80$ TNOs. In addition, the anti-aligned particles exhibit drifts in their semi-major axis at approximately constant pericenters. This is due to close encounters with Neptune when the particles are close to their pericenter, as shown in the coplanar case investigation by \citet{Hadden18}. 

Figure \ref{f:Nbody} shows that there is no clustering around $\Delta \theta_e \sim 0^\circ$, which differs from the secular case initialized with a uniform distribution of $\varpi$. 
This is because the TNOs with $\Delta \varpi_0 \sim 0$ are ejected due to instability caused by the overlap of mean motion resonances, as discussed in \citet{Hadden18}. This is illustrated in the upper right panel in Figure \ref{f:Nbody}, which is color coded in $\Delta\varpi_0$. It shows that most of the surviving TNOs that are clustered around $\Delta\theta_e\sim 180^\circ$ started with $\Delta\varpi_0\sim 180^\circ$. 
Meanwhile, there is a population of particles that exhibit drifts in pericenter distances at constant semi-major axis (most clearly seen as the multi-colored vertical paths in the top-left panel of Figure \ref{f:Nbody}), which arise due to secular effects. 

The lower panels of Figure \ref{f:Nbody} color-code the TNOs in inclination. The lower left one shows the alignment in $\Delta \varpi$. $\Delta\varpi$ clusters around $\sim180^\circ$, and the clustering is strongest when the test particle inclinations are low $\lesssim 40^\circ$. 
The majority ($\sim 90\%$) of the anti-aligned ($120^\circ<\varpi<240^\circ$), small pericenter ($30<r_p<80$ AU) particles maintain a low inclination throughout their $4$ Gyr evolution. On the other hand, the lower right panel shows the alignment in the geometrical longitude of pericenter ($\Delta\theta_e$), color-coded in inclination. In contrast to the result for $\Delta\varpi$, $\Delta\theta_e$ shows tight clustering, even for high inclination TNOs. This is because $\Delta\theta_e$ librates during the flips of the orbits, similar to the flips of the inner orbit in the hierarchical three-body interactions \citep{Li14a} (see more details in in section \ref{s:incl}). 

As shown in the secular results of Section 2, the high inclination population also shows interesting clustering in the orbital orientation. To illustrate this, we plot in Figure \ref{f:Nbdhighi} the alignment in different orbital orientations as a function of semi-major axis. All of the long lived ($t>3$Gyr) high inclination ($60<i<120^\circ$) particles are selected: There are 303 long lived TNOs that reached above $60^\circ$ and 255 of them reached retrograde configurations. The clustering in longitude of node, $\Delta\Omega$, is shown in the upper left panel, argument of pericenter, $\Delta\omega$, in the upper right, longitude of pericenter, $\Delta\varpi$, in the lower left, and the geometrical longitude of pericenter, $\Delta\theta_e$, in the lower right panel. 

Consistent with the secular results, there is a strong deficit of particles with $\Delta\varpi \sim 180^\circ$ and $\Delta\theta_e \sim 180^\circ$. In addition, there are clusters around $\Delta\varpi \sim 90^\circ$ and $\Delta\varpi \sim 270^\circ$, and the clusters around the geometrical longitude of pericenter is slightly tighter ($\Delta\theta_e \sim 90^\circ$ and $\Delta\theta_e \sim 270^\circ$) for the low pericenter TNOs. This is very different from the low inclination population. The clustering in $\Delta\omega \sim 0^\circ \& 180^\circ$ and $\Delta\Omega \sim 90^\circ \& 270^\circ$ are very similar to the secular results. The existence of such clustering in the future TNO detection can help constrain the orbital parameters of any outer planet.

\begin{figure*}[ht]
\includegraphics[width=\textwidth]{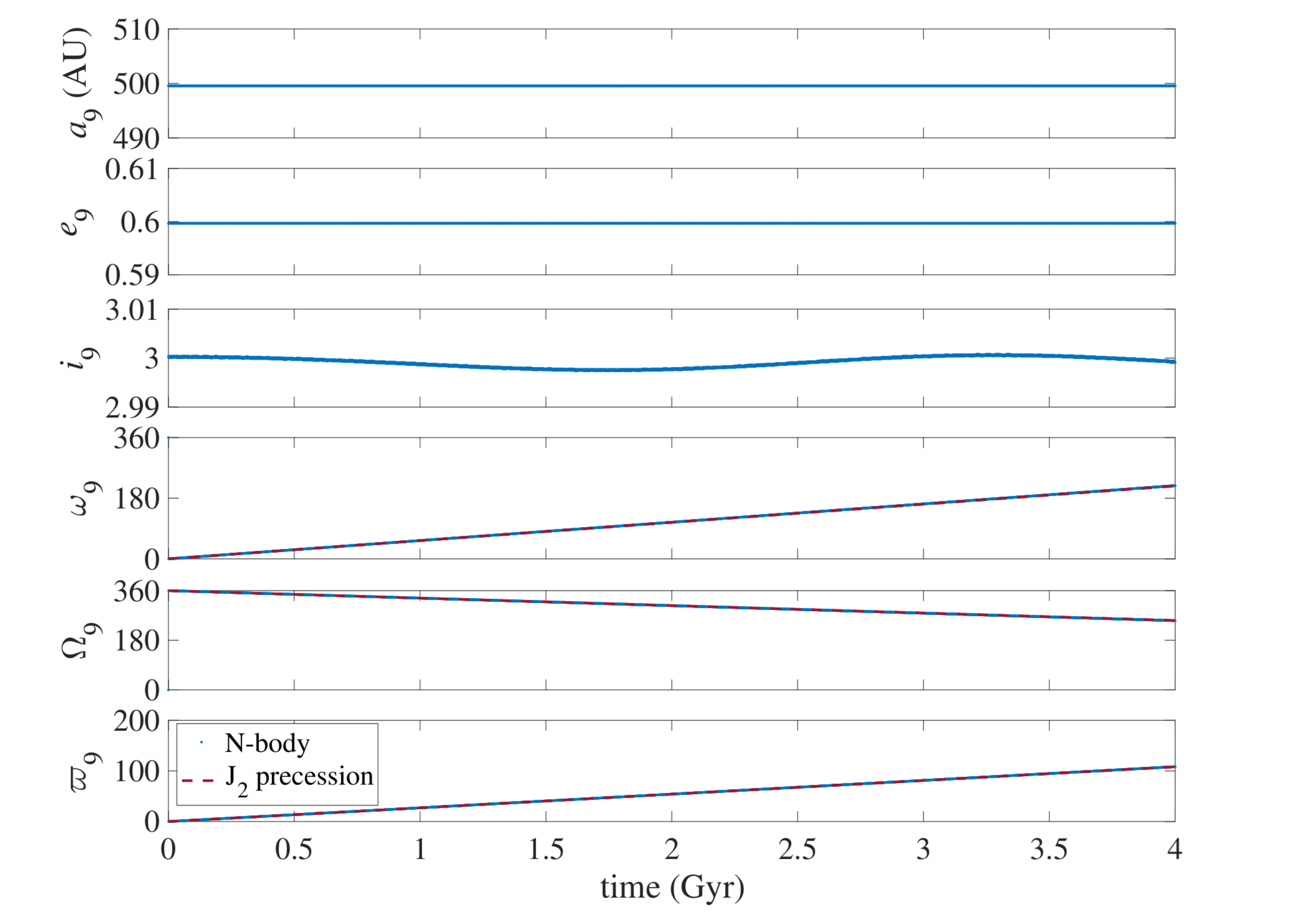}
\caption{Evolution of Planet Nine from the N-body simulation discussed in Section \ref{s:Nbd}. \label{f:P9_traj}}
\vspace{0.1cm}
\end{figure*}

\begin{figure*}[ht]
\includegraphics[width=\textwidth]{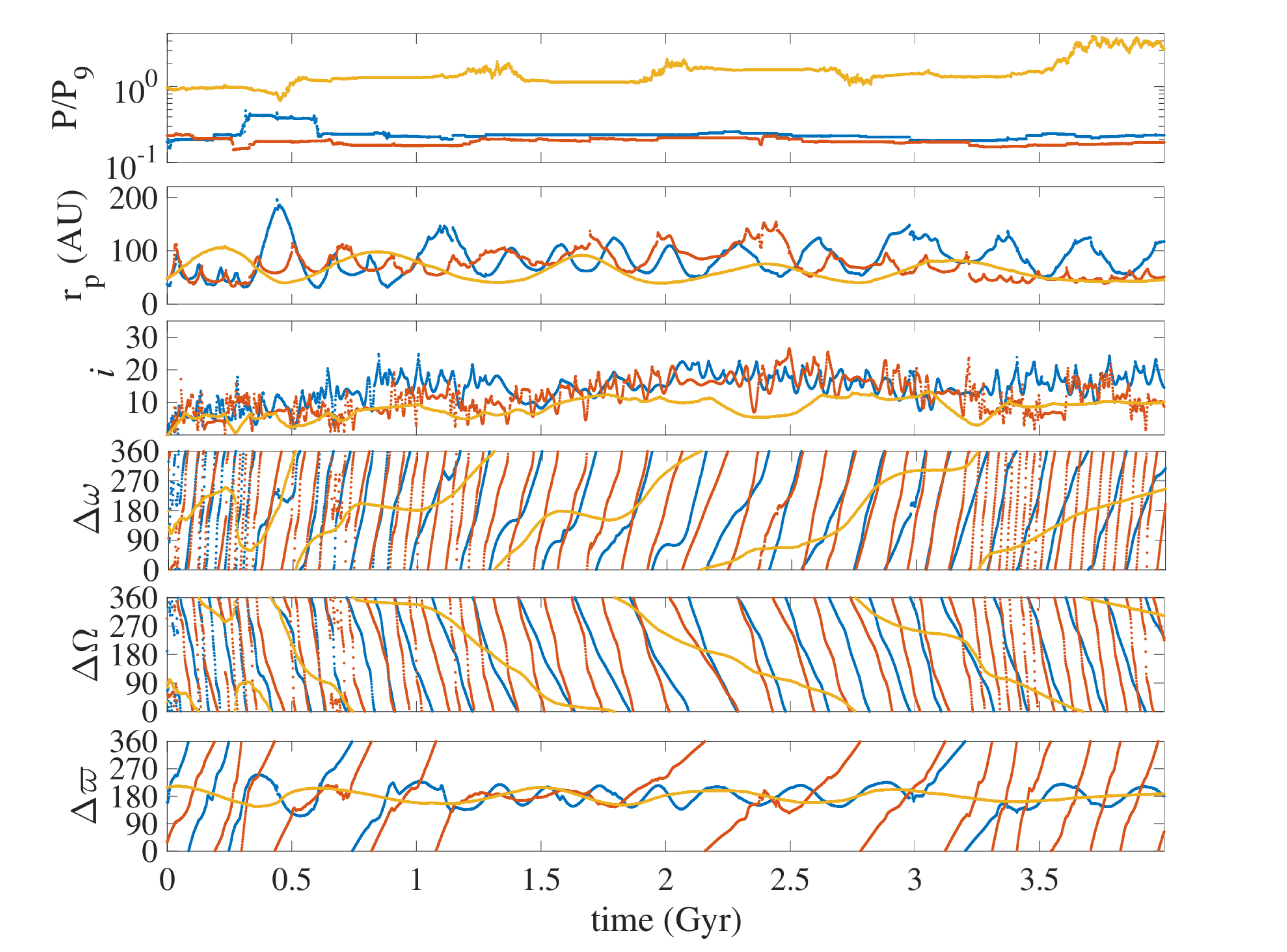}
\caption{Evolution of randomly selected particles with low inclination from the N-body simulation discussed in Section \ref{s:Nbd}. \label{f:lowi_traj}}
\vspace{0.1cm}
\end{figure*}

\begin{figure*}[ht]
\includegraphics[width=\textwidth]{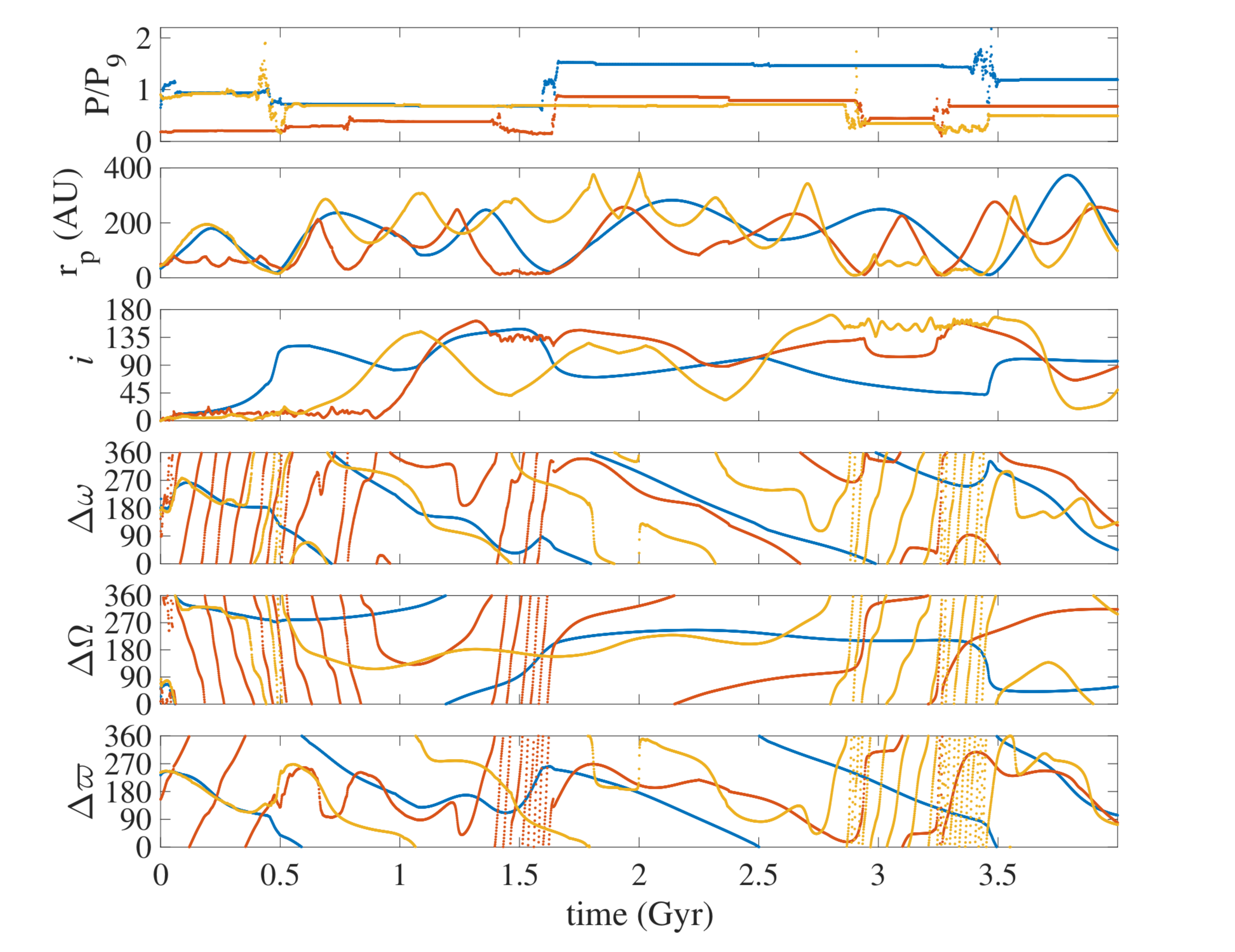}
\caption{Evolution of randomly selected particles with high inclination from the N-body simulation discussed in Section \ref{s:Nbd}. \label{f:highi_traj}}
\vspace{0.1cm}
\end{figure*}

To illustrate the evolution of the particles and Planet Nine, we plot Planet Nine and selected representative TNO orbital elements as a function of time in Figure \ref{f:P9_traj} - \ref{f:highi_traj}. 
Figure \ref{f:P9_traj} shows the evolution of Planet Nine, 
Figure \ref{f:lowi_traj} shows the evolution of some examples of low inclination objects, 
and Figure \ref{f:highi_traj} shows the evolution of some high inclination objects.

The semi-major axis and eccentricity of Planet Nine show little variation, as illustrated in Figure \ref{f:P9_traj}. 
The inclination of Planet Nine has low-amplitude, long-timescale variations due to interactions with Neptune. 
The red dashed lines in the lower three panels represent the precession due to the $J_2$ component by the inner giant planets. 
The $J_2$ precession agrees well with the changes of $\omega$, $\Omega$ and $\varpi$ as a function of time.

Figure \ref{f:lowi_traj} shows the trajectories of low inclination TNOs. 
The semi-major axes of the particles are perturbed by Planet Nine and Neptune, which causes large period-ratio variations. 
The eccentricity of the particles also varies throughout the $4$ Gyr of integration, which allows the pericenter distance to oscillate. The inclinations stay low ($\lesssim30^\circ$) within the $4$ Gyr integration. $\Delta\varpi$ librates for the particles with large semi-major axis, as illustrated by the particle represented by the yellow line. 
At smaller semi-major axes, $\Delta\varpi$ starts to circulate. 
$\Delta\omega$ and $\Delta\Omega$ generally circulate within $4$Gyr. 

Figure \ref{f:highi_traj} shows the trajectories of some representative high inclination TNOs. 
The semi-major axes of the particles are perturbed by Planet Nine and Neptune, which cause large period ratio variations, similar to the low inclination case. The eccentricity of the particles also varies throughout the $4$ Gyr of integration, while the inclination can be excited to high values. 
Only about $\sim 20\%$ of the trajectories flip back to low inclination $\lesssim20^\circ$ within the 4 Gyr integration. 
Different from the low inclination particles, $\Delta\varpi$ of the particles does not always librate around $180^\circ$ even when the semi-major axes are large. 
Decreases in $\Delta\varpi$ can be seen when $\Delta\varpi$ does not librate, which explains the over-density for $\Delta\varpi\sim 90^\circ$ compared with $\Delta\varpi\sim 270^\circ$, since the long-lived particles starting near $\Delta\varpi\sim 180^\circ$ tend to reach $\sim 90^\circ$ earlier and slightly more often than $\sim 270^\circ$. There are libration regions in $\Delta\omega$ and $\Delta\Omega$ as their inclinations increase, which lead to clusterings in $\Delta\omega$ and $\Delta\Omega$ for high inclination objects ($60^\circ<i<120^\circ$). 



\bigskip




\bibliographystyle{hapj}
\bibliography{msref.bib}

\end{document}